\documentclass[12pt,a4paper]{article}
\usepackage{amsmath,amssymb,epsfig}
\usepackage{epsfig}
\renewcommand\topfraction{.95}   
\renewcommand\bottomfraction{.6} 
\renewcommand{\theequation}{\thesection.\arabic{equation}}

\newlength{\extraspace}
\newlength{\extraspaces}
\newcommand{\be}{\begin{equation}
\addtolength{\abovedisplayskip}{\extraspaces}
\addtolength{\belowdisplayskip}{\extraspaces}
\addtolength{\abovedisplayshortskip}{\extraspace}
\addtolength{\belowdisplayshortskip}{\extraspace}}
\newcommand{\ee}{\end{equation}}
 
\newcommand{\ba}{\begin{eqnarray}
\addtolength{\abovedisplayskip}{\extraspaces}
\addtolength{\belowdisplayskip}{\extraspaces}
\addtolength{\abovedisplayshortskip}{\extraspace}
\addtolength{\belowdisplayshortskip}{\extraspace}}
\newcommand{\ea}{\end{eqnarray}}

\newcommand{\bas}{\begin{eqnarray*}
\addtolength{\abovedisplayskip}{\extraspaces}
\addtolength{\belowdisplayskip}{\extraspaces}
\addtolength{\abovedisplayshortskip}{\extraspace}
\addtolength{\belowdisplayshortskip}{\extraspace}}
\newcommand{\eas}{\end{eqnarray*}}
 
\newcounter{subequation}[equation]
\makeatletter

\expandafter\let\expandafter
\reset@font\csname reset@font\endcsname

\def\subeqnarray{\arraycolsep1pt
    \def\@eqnnum\stepcounter##1{\stepcounter{subequation}%
        {\reset@font\rm(\theequation\alph{subequation})}}
\jot5mm     \eqnarray}

\def\subarray{\arraycolsep1pt
    \def\@eqnnum\stepcounter##1{\stepcounter{subequation}%
        {\reset@font\rm(\alph{subequation})}}
\jot5mm     \eqnarray}

\makeatother

\newcommand{\newsection}[1]{
\vspace{15mm}
\pagebreak[3]
\addtocounter{section}{1}
\setcounter{equation}{0}
\setcounter{subsection}{0}

\setcounter{footnote}{0}
\addcontentsline{toc}{section}
{\protect\numberline{\arabic{section}}{#1}}
 
\begin{flushleft}
{\large\bf \thesection. #1}
\end{flushleft}
\nopagebreak
\medskip
\nopagebreak}
 
\newcommand{\newsubsection}[1]{
\vspace{1cm}
\pagebreak[3]
\addtocounter{subsection}{1}

\addcontentsline{toc}{subsection}
{\protect\numberline{\thesection.\arabic{subsection}}{#1}}
 
\noindent{ \bf \thesection.\arabic{subsection} #1}
\nopagebreak
\vspace{2mm}
\nopagebreak}

\newcommand{\newappendix}[1]{
\vspace{15mm}
\pagebreak[3]
\addtocounter{section}{1}
\setcounter{equation}{0}
\setcounter{subsection}{0}

\addcontentsline{toc}{section}
{\protect\numberline{\thesection}{#1}}

\renewcommand{\theequation}{\Alph{section}.\arabic{equation}}
\begin{flushleft}
{\large\bf \Alph{section}: #1}
\end{flushleft}
\nopagebreak
\medskip
\nopagebreak}

\newcommand{\NP}[1]{Nucl.\ Phys.\ {\bf #1}}
\newcommand{\PL}[1]{Phys.\ Lett.\ {\bf #1}}

\newcommand{\PR}[1]{Phys.\ Rev.\ {\bf #1}}

\newcommand{\N}{\mathbb{N}}

\newcommand{\Z}{\mathbb{Z}}
\newcommand{\R}{\mathbb{R}}

\newcommand{\1}{\mbox{1\hspace{-.8ex}1}}
\newcommand{\bra}{\langle}
\newcommand{\ket}{\rangle}
\newcommand{\ra}{\rightarrow}

\newcommand{\is}{ &\! =\! & }
\newcommand{\nonum}{\nonumber \\[1.5mm]}
\newcommand{\sspace}{\makebox[1cm]{ }}

\newcommand{\nspace}{\!\!\!\!\!\!\!\!\!\!}

\renewcommand{\th}{{\theta}}
\newcommand{\eps}{\epsilon}
\newcommand{\lb}{\lambda}
\newcommand{\om}{\omega}

\newcommand{\dd}{{\partial}}

\newcommand{\cD}{{\cal D}}
\newcommand{\cE}{{\cal E}}

\newcommand{\cL}{{\cal L}}

\newcommand{\cM}{{\cal M}}
\newcommand{\cO}{{\cal O}}

\newcommand{\cR}{{\cal R}}

\newcommand{\betabf}{\mbox{\boldmath{$\beta$}}}

\newcommand{\gn}{{\rm g}_{\sl N}}

\newcommand{\n}{{\sc n}}
\newcommand{\dk}{ k\frac{\dd}{\dd k} }

\newcommand{\gbab}{\bar{g}_{\alpha\beta}}

\begin{document}

\begin{titlepage}

\mbox{}
\vspace{5mm}

\begin{center}
\mbox{{\large \bf The Asymptotic Safety Scenario in Quantum Gravity}}\\[6mm]
\mbox{{\large \bf -- An Introduction -- }} 
\vspace{1.5cm}

{{\sc M. Niedermaier}%
\footnote{Membre du CNRS.} 
\\[4mm]
{\small\sl Laboratoire de Mathematiques et Physique Theorique}\\
{\small\sl CNRS/UMR 6083, Universit\'{e} de Tours}\\
{\small\sl Parc de Grandmont, 37200 Tours, France}
\vspace{1.5cm}
}

{\bf Abstract}
\vspace{-3mm}

\end{center}

\begin{quote}
The asymptotic safety scenario in quantum gravity is reviewed, 
according to which a renormalizable quantum theory of the gravitational 
field is feasible which reconciles asymptotically safe couplings 
with unitarity. 
All presently known evidence is surveyed: (a) from the $2+\eps$ expansion, 
(b) from the perturbation theory of higher derivative gravity theories and 
a `large $N$' expansion in the number of matter fields,
(c) from the 2-Killing vector reduction, and (d) from truncated flow equations
for the effective average action. Special emphasis is given to 
the role of perturbation theory as a guide to `asymptotic safety'.  
Further it is argued that as a consequence of the scenario the 
selfinteractions appear two-dimensional in the extreme ultraviolet. 
Two appendices discuss the distinct roles of the ultraviolet 
renormalization in perturbation theory and in the flow equation 
formalism.   
\end{quote}
\vfill

\setcounter{footnote}{0}
\end{titlepage}


\thispagestyle{empty}


\tableofcontents

\setlength\paperwidth  {210mm}   %
\renewcommand\bottomfraction{.6} 
\nopagebreak

\renewcommand\topfraction{.95}   
\setlength\paperheight {297mm}   

\newpage

\setcounter{equation}{0}
\newsection{Survey of the scenario and evidence for it}

The quest for a physically viable theory of quantized gravitation is 
ongoing; in part because the physics it ought to describe is unknown,
and in part because different approaches may not `approach' the 
same physics. The most prominent contenders are string theory 
and loop quantum gravity, with ample literature available on either 
sides. For book-sized expositions see for example \cite{GSW, Polchinski,
Rovelli}. The present review surveys a circle of ideas 
which differ in several important ways from these approaches;
we refer to \cite{Gravirept} for a more detailed account with 
a slightly different emphasis. 

\newsubsection{Survey of the scenario}

In brief, the scenario delineates conditions under which a 
functional integral based quantum theory of gravity can be 
viable beyond the level of an effective field theory: first a 
physics premise (``antiscreening'') is made about the selfinteraction 
of the quantum degrees of freedom in the ultraviolet. Second, 
the effective diminution of the relevant degrees of freedom in the 
ultraviolet (on which morally speaking all approaches 
agree) is interpreted as universality in the statistical physics sense
in the vicinity of an ultraviolet renormalization 
group fixed point. Third, the resulting picture of microscopic geometry 
is fractal-like with a local dimensionality of two. 

The concrete implementation of these ideas has begun only 
recently and led to a number of surprising results to be reviewed here. 
Part of the physics intuition, on the other hand, dates back to an 
1979 article by S.~Weinberg~\cite{Weinberg}, see also
\cite{GomisWeinb}. Motivated by the analogy to the asymptotic freedom 
property of nonabelian gauge theories, the term ``asymptotic safety'' 
was suggested in~\cite{Weinberg}, indicating that physically motivated 
running couplings should be ``safe'' from divergencies at all scales. 
Following this suggestion we shall refer to the above circle of ideas as the 
``asymptotic safety scenario'' for quantum gravity. For convenient 
orientation we display the main features in overview: 
\medskip

{\bf Asymptotic safety scenario -- main ideas:}
\vspace{-1mm}

\begin{itemize}
\item The gravitational field itself is taken as the prime carrier 
of the relevant classical and quantum degrees of freedom; its  
macro- and micro-physics are related through a renormalization flow. 
\item As the basic physics premise stipulate that the physical 
degrees of freedom in the ultraviolet interact predominantly 
antiscreening. 
\item Based on this premise benign renormalization properties in the 
ultraviolet are plausible. The resulting ``Quantum Gravidynamics'' 
can then be viewed as a peculiar quasi-renorm\-aliz\-able 
field theory based on a non-Gaussian fixed point. 
\item In the extreme ultraviolet the residual interactions appear 
two-dimensional. 
\end{itemize}
The first point is shared by the effective field theory framework
for quantum gravity, the others are needed to go beyond it in 
a functional integral based approach. The rationale for 
trying to so is twofold: first, the effective field framework
gives rise only to very few universal corrections which 
are quantitatively too small to be phenomenologically 
interesting. Second, once the physics premise  underlying a
regularized functional integral for gravity has been made a 
``UV completion'' based simply on removal of the regulator, 
if feasible, is physically well-motivated and computationally 
seamless.

A strategy centered around a functional integral picture was 
indeed adopted early on~\cite{Misner} but is now mostly abandoned.
A functional integral over geometries of course has to differ in 
several crucial ways from one for fields on a fixed geometry. This 
lead to the development of several formulations (canonical, covariant
\cite{deWitt1, deWitt2, deWitt3}, proper time~\cite{Teitelb1, Teitelb2} 
and covariant Euclidean~\cite{Hawking, GibbonsHawking}). 
As is well-known the functional integral picture is also 
beset from severe technical 
problems~\cite{tHVelt74, DesNieuw74,GibbonsHawking,Mottola}. 
Nevertheless this should not distract attention from the fact that 
a functional integral has a physics content which 
differs from the physics content of other approaches. 
For want of a better formulation we shall refer to this 
fact by saying that a functional integral picture  
``takes the degrees of freedom of the gravitational field 
seriously also in the quantum regime''.

Let us briefly elaborate on that. 
Arguably the cleanest intuition to `what quantizing gravity 
might mean' comes from the functional integral picture. 
Transition or scattering amplitudes for nongravitational 
processes should be affected not only by one geometry solving 
the gravitational field equations, but by a `weighted superposition' 
of `nearby possible' off-shell geometries. The rationale behind this intuition 
is that all known (microscopic) matter is quantized that way, and using an 
off-shell matter configuration as the source of the Einstein field equations 
is in general inconsistent, unless the geometry is likewise
off-shell. Moreover, relativistic quantum field theory suggests that the 
matter-geometry coupling is effected not only through averaged or 
large scale properties of matter. For example nonvanishing connected 
correlators of a matter energy momentum tensor should be a legitimate source 
of gravitational radiation as well, as should be the Casimir energy, 
see~\cite{Ford1,Fullingetal}. 
Of course this doesn't tell in which sense the geometry 
is off-shell, nor which class of possible geometries ought to be 
considered and be weighed with respect to which measure. 
Rapid decoherence, a counterpart of spontaneous symmetry 
breaking, and other unknown mechanisms may in addition mask 
the effects of the superposition principle. Nevertheless the 
argument suggests that the degrees of freedom of the gravitational 
field should be taken seriously also in the quantum regime,
roughly along the lines of a functional integral.

Doing so one readily arrives at the effective field theory 
description of quantum gravity, see \cite{burgess} for a recent 
review. It is a commonly accepted criterion that a theory 
of quantum gravity, even one evading a functional integral 
over geometries, should match whatever universal results 
are obtained from an effective field theory framework. 
The issue at stake thus is the extent to which  different 
``UV completions'' of the effective field theory description have a 
physics content different from the latter and from each other. Of 
course in the absence of empirical guidance the `true' physics of 
quantum gravity is unknown; so for the time being it will be important 
to try to isolate differences in the physics content of the various 
``UV completions''. By physics content 
we mean here qualitative or quantitative results for the 
values of ``quantum gravity corrections'' to {\it generic} physical 
quantities in the approach considered. Generic physical quantities should
be such that they in principle capture the entire invariant content of 
a theory. In a conventional field theory S-matrix elements by and large 
have this property, in canonical general relativity Dirac observables 
play this role~\cite{Anderson, Torreobs, Dittrich}. In quantum 
gravity, in contrast, no agreement has been reached on the nature 
of such generic physical quantities. 

The present scenario proposes a UV completion proper,  one which is based 
on the very same physics principles that make the effective field theory 
description so credible and which renders the crucial interpolating 
regime computationally accessible. We share the viewpoint expressed by Wilczek 
in \cite{QFTWilc}: ``Whether the next big step will require a sharp 
break from the principles of quantum field theory, or, like the 
previous ones, a better appreciation of its potentialities, 
remains to be seen''. Here we center the discussion around the above 
four main ideas, and, for short, call a quantum theory of gravity based on them
{\bf Quantum Gravidynamics}. For the remainder of Section 1.1 we now 
discuss a number of key issues that arise.

In any functional integral picture one has to face the crucial 
{\bf renormalizability problem}. Throughout we shall be concerned 
exclusively with (non-)renormalizability in the ultraviolet. 
The perspective on the nature of the impasse entailed 
by the perturbative nonrenormalizability of the Einstein--Hilbert action
(see Bern~\cite{Bern} for a review), however, has changed 
significantly since the time it was discovered by 't Hooft and 
Veltmann~\cite{tHVelt74}. First, the effective field theory framework 
applied to quantum gravity provides unambiguous answers some lowest 
order corrections despite the perturbative nonrenormalizability of the 
`fundamental' action, as stressed by Donoghue (see~\cite{Bjerrum-BohrD,burgess} 
and references therein). 
The role of an a-priori microscopic action is 
moreover strongly deemphasized when a Kadanoff--Wilson view 
on renormalization is adopted. We shall give a quick reminder on this 
framework in Appendix A. Applied to gravity it means that the 
Einstein--Hilbert action should {\it not} be considered as the 
microscopic (high energy) action, rather the renormalization flow 
itself will dictate to a 
certain extent which microscopic action to use, and whether 
or not there is a useful description of the extreme
ultraviolet regime in terms of `fundamental' (perhaps nonmetric)
degrees of freedom. The extent to which this is true hinges 
on the existence of a fixed point with a renormalized trajectory emanating
from it. The fixed point guarantees universality in the statistical
physics sense. If there is a fixed point any action on a renormalized trajectory 
describes identically the same physics on all energy scales lower than 
the one where it is defined. Following the trajectory back (almost) into 
the fixed point one can in principle extract unambiguous answers for 
physical quantities on all energy scales.

Compared to the effective field theory framework the main advantage
of genuine renormalizability lies not primarily in the gained energy 
range in which reliable computations can be made, but rather that one has a 
chance to properly identify `large' quantum gravity effects
at low energies (assuming they exist). The effective field theory framework 
rests on a {\bf decoupling assumption}: there exists a potentially
process-dependent scale $M_{\rm eff}$ such that the low energy 
degrees of freedom ($E/M_{\rm eff} \ll 1$) relevant for the process 
obey an approximately autonomous dynamics.  Based on this assumption 
some unambiguously defined low energy effects of quantized gravity  
can be identified, but are found to be suppressed by the powers of 
energy scale/Planck mass expected on dimensional grounds. 
However in the presence of massless degrees of freedom the 
decoupling assumption may fail (mediated e.g.~by anomalies \cite{AMM} or 
by spontaneous symmetry breaking) and the extent to which 
it is valid in `quantized' gravity is a {\it dynamical problem}. 
In a theory of quantum gravidynamics this dynamical problem 
can be investigated: the effect of high energy (Planck scale) 
processes can in principle be computationally propagated 
through many orders of magnitudes down to accessible energies,
where they may leave a detectable low energy imprint. 

Note that the nature of the `fundamental' degrees of freedom 
is of secondary importance in this context. From the 
viewpoint of renormalization theory it is the universality class 
that matters not the particular choice of dynamical variables. Once 
a functional integral picture has been adopted even nonlocally and 
nonlinearly related sets of fields or other variables may describe 
the same universality class -- and hence the same physics.

Generally, the arena on which the renormalization group acts is a space 
of actions or, equivalently, a space of (regularized) measures. A typical 
action has the form: $\sum_{\alpha} u_{\alpha} P_{\alpha}$, where 
$P_{\alpha}$ are interaction 
monomials (including kinetic terms) and the $u_{\alpha}$ are scale dependent 
coefficients. The subset $u_i$ which cannot be removed by field 
redefinitions are
called {\it essential} parameters, or couplings. Usually one makes them 
dimensionless by taking out a suitable power of the scale parameter
$\mu$, ${\rm g}_i(\mu) = \mu^{-d_i} u_i(\mu)$. In the following 
the term essential coupling will always refer to these dimensionless 
variants. We also presuppose the principles according to which a 
(Wilson-Kadanoff) renormalization flow is defined on this arena.
For the convenience of the reader a brief glossary is included 
Section 1.4. In the context of Quantum Gravidynamics some key notions 
(coarse graining operation, unstable manifold and continuum limit) 
have a somewhat different status which we outline below.

Initially all concepts in a Wilson-Kadanoff renormalization 
procedure refer to a choice of {\bf coarse graining} operation. 
It is part of the physics premise of a functional 
integral approach that there is a 
physically relevant distinction between coarse grained and fine
grained geometries. On a classical level this amounts to 
the distinction, for example, between a perfect fluid solution of 
the field equations and one generated by its $10^{30}$ or so molecular 
constituents. A sufficiently large set of Dirac observables 
would be able to discriminate two such spacetimes. One can also 
envisage a vacuum counterpart of this distinction and view the 
coarse graining scale as analogous to an `intrinsic clock' variable 
in general relativity. Whenever we 
shall refer later on to ``coarse grained'' versus ``fine grained'' 
geometries we have a similar picture in mind for the 
ensembles of off-shell geometries entering a (regularized) functional 
integral. For example, the value of integrated curvature invariants like 
$(\nabla_{\alpha} R_{\beta \gamma\delta \rho} \nabla^{\alpha} R^{\beta
  \gamma\delta \rho})^2$ may provide a rough measure for the 
coarseness. Tested proposals for an intrinsic coarse graining 
scale for geometries are however presently not available.

As a substitute one can define the coarse graining with respect 
to a {\bf state-dependent dynamically adjusted background metric}. 
Let $\bar{g}_{\alpha\beta}$ be an initially prescribed 
background metric and suppose that it has been used to 
define a ``background covariant'' notion of coarse graining, 
e.g.~by referring to the spectrum of a covariant differential 
operator built from $\bar{g}$. The coarse graining can then be 
used to construct the functional integral averages 
$\bra \;\; \ket_{\bar{g}}$ subject to suitable boundary 
conditions that encode information about the state vector.  
Eventually one obtains a functional $\cO \mapsto \bra 
\cO \ket_{F(\bar{g})}$ (``a state'', roughly in the algebraic 
sense) which depends parameterically on the background 
$\bar{g}$ via a functional $F(\bar{g})$. In a second step one 
then selfconsistently adjusts the background metric to one solving 
\be 
\bra q_{\alpha \beta} \ket_{F(\bar{g}^*)} 
\stackrel{\displaystyle{!}}{=} \bar{g}_{\alpha\beta}^*\,.
\label{adjust}
\end{equation}
Here $F$ is defined via a stationarity condition referring to 
the full quantum dynamics and hence implicitly to the 
underlying state, see section 1.2. Equation (\ref{adjust}) 
can thus be viewed as selecting a class of state dependent 
backgrounds $\bar{g}^*_{\alpha\beta}$ such that the average 
of the quantum metric in the state co-determined by 
$\bar{g}_{\alpha\beta}^*$ coincides with $\bar{g}_{\alpha\beta}^*$. 
For definiteness we formulated (\ref{adjust}) in terms of 
the metric, assuming in particular that 
$\bra q_{\alpha \beta} \ket_{F(\bar{g})}$ is well-defined. 
This assumption is dispensible, however, as one could rephrase 
the above construction with whatever (nonlocal) composites 
or observables $\cO$ one decides to work with: given a family of 
$\bar{\cO}$'s containing the information about the background 
metric $\bar{g}$ deemed relevant, one can for any 
initially prescribed set of their values define a coarse graining 
operation relative to it and use the coarse graining to 
construct the functional averages $\bra\;\;  \ket_{F(\bar{\cO})}$,
depending parametrically on the $\bar{\cO}$ values. 
In a second step one can stipulate the counterpart of 
(\ref{adjust}), i.e.~$\bra \cO \ket_{F(\bar{\cO})} = \bar{\cO}$,
which dynamically adjusts the values $\bar{\cO}$ to
the selfconsistent ones $\bar{\cO}^*$. In formulating 
(\ref{adjust}) we assumed that the `infrared problem' has been 
solved, in particular that the full averages used for the 
adjustment contain information also about the infrared 
degrees of freedom and are well-defined. The same adjustment 
could, however, be done using scale dependent Wilsonian averages
at some scale infrared cutoff scale $\mu$, see Section 1.2.

With respect to a given coarse graining operation one can ask whether 
the flow of actions or couplings has a {\bf fixed point}. The existence 
of a fixed point is the raison d'\^{e}tre  for the universality properties 
(in the statistical field theory sense) which eventually are `handed down' 
to the physics in the low energy regime. By analogy with 
other field theoretical systems one should probably 
not expect that the existence (or nonexistence) of a (non-Gaussian) 
fixed point will be proven with mathematical rigor in the near future. 
From a physics viewpoint, however, it is the high degree of universality 
ensued by a fixed point that matters,
rather than the existence in the mathematical sense. For example 
nonabelian gauge theories appear to have a (Gaussian) 
fixed point `for all practical purposes', while their 
rigorous construction as the continuum limit of a lattice theory 
is still deemed a `millennium problem'. In the case of 
quantum gravity we shall survey in Section 1.3 various pieces of 
evidence for the existence of a (non-Gaussian) fixed point.

Accepting the existence of a (non-Gaussian) fixed point as a 
working hypothesis one is led to determine the structure of its 
{\bf unstable manifold}. Given a coarse graining operation and 
a fixed point of it, the stable (unstable) manifold 
is the set of all points connected to the fixed point 
by a coarse graining trajectory terminating at it (emanating from it). 
It is not guaranteed though that the space of actions can in the 
vicinity of the fixed point be divided into a stable and an 
unstable manifold; there may be trajectories which develop 
singularities or enter a region of coupling space deemed unphysical 
for other reasons and thus remain unconnected to the fixed point. 
The stable manifold is the innocuous part
of the problem, it is the unstable manifold which is crucial 
for the construction of a continuum limit. 
By definition it is swept out by flow lines emanating 
from the fixed point, the so-called {\it renormalized 
trajectories}. Points on such a flow line correspond to 
actions or measures which are called {\it perfect} in that they 
can be used to compute continuum answers for physical quantities 
even in the presence of an ultraviolet (UV) cutoff, like one which 
discretizes the base manifold \cite{Hasenf3}. In practice the unstable 
manifold is not known and renormalized trajectories have to be identified 
approximately by a tuning process. What is easy to determine is 
whether in a given expansion ``sum over coupling times interaction monomial''
a coupling will be driven away from the value the corresponding coordinate 
has at the fixed point after a sufficient number of coarse graining steps 
(in which case it is called {\it relevant}) or will move 
towards this fixed point value (in which case it is called {\it irrelevant}).  
Note that this question can be asked even for trajectories which 
are not connected to the fixed point. The dimension of the unstable 
manifold equals the number of independent 
relevant interaction monomials that are `connected' to the 
fixed point by a (renormalized) trajectory.

In quantum gravity traditionally the Einstein-Hilbert action is 
taken as the microscopic action. Perturbatively this action is
not connected to a fixed point, not even to the perturbative Gaussian 
one. The question whether or not the situation improves in a 
nonperturbative formulation has been mostly addressed in discretized
formulations, see \cite{Hamber,AmbjornDT} and references therein. 
The discretized action used then may no longer have a 
naive (classical) continuum limit reproducing the Einstein-Hilbert 
action, but it is still labelled by two bare parameters.    
Conceptually one can assign to the discretized two-parametric measure
a microscopic action in the above sense by requiring that combined 
with the regularized continuum measure \cite{BernBlauMott} it reproduces 
approximately the same correlation functions. The microscopic action 
defined that way would presumably be different from the Einstein-Hilbert 
action but it would still contain only two tunable parameters. 
Presupposing again the existence of a fixed point, this type of 
construction relies on the hope that the non-naive discretization 
procedure adopted gets all but two coordinates of the unstable 
manifold automatically right. We refer to \cite{Hamber,AmbjornDT} 
for the numerical evidence.

In the present context a counterpart of these constructions 
starting from a perturbatively (weakly or strictly) microscopic 
renormalizable action (see Section 2) would seem more promising. The 
tuning to the unstable manifold then is more complicated, but 
perturbation theory (or other expansion techniques) can be used 
as a guideline, both analytically and for the extrapolation of 
numerical results.

Typically the unstable manifold is indeed locally a manifold,
though it may have cusps. Although ultimately it is only 
the unstable manifold that matters for the construction of a
continuum limit, relevant couplings which blow up somewhere 
inbetween may make it very difficult to successfully identify 
the unstable manifold. In practice, if the basis of interaction 
monomials in which this happens is deemed natural and a change 
of basis in which the pathological directions could simply be 
omitted from the space of actions is very complicated, the 
problems caused by such a blow up may be severe. An important
issue in practice is therefore whether in a natural basis 
of interaction monomials the couplings are `safe' from such 
pathologies and the space of actions decomposes in the 
vicinity of the fixed point neatly into a stable and an 
unstable manifold. This regularity property is one 
aspect of ``asymptotic safety'', as we shall see below.

A second limitation appears in infinite dimensional situations. 
Whenever the coarse graining operates on an infinite set 
of potentially relevant interaction monomials convergence issues
in the infinite sums formed from them may render formally 
equivalent bases inequivalent. In this case the geometric 
picture of a (coordinate independent) manifold breaks down 
or has to be replaced by a more refined functional analytic 
framework. An example of a field theory with an infinite set 
of relevant interaction monomials is QCD in a lightfront 
formulation~\cite{PerryWilson} where manifest Lorentz and gauge invariance 
is given up in exchange of other advantages. In this case it is 
thought that there are hidden dependencies among the associated 
couplings so that the number of independent relevant couplings is 
finite and the theory is eventually equivalent to 
conventional QCD. Such a {\it reduction of couplings} is nontrivial 
because a relation among couplings has to be preserved 
under the renormalization flow. In quantum gravity related 
issues arise to which we turn later.

As an interlude let us mention the special role of 
{\bf Newton's constant} in a diffeomorphism invariant theory 
with a dynamical metric. 
Let $S[q,{\mathrm{matter}}]$ be any local action,
where $q = (q_{\alpha\beta})_{1 \leq \alpha ,\beta \leq d}$ is the 
(`quantum') metric entering the (regularized) functional integral
and the ``matter'' fields are not scaled when the metric is. 
Constant rescalings of the metric then give rise to a variation of 
the Lagrangian which vanishes on shell:
\be 
\frac{d}{d \om^2} S[\om^2 q,{\mathrm{matter}}]\Big|_{\om =1} = 
\int\! dx \sqrt{q}\, q_{\alpha\beta} 
\frac{\delta S[q,{\mathrm{matter}}]}{\delta q_{\alpha\beta}}\,. 
\label{iom}
\end{equation}
As a consequence one of the coupling parameters which in the absence 
of gravity would be {\it essential} (i.e.\ a genuine coupling) becomes
{\it inessential} (i.e.\ can be changed at will by a redefinition of the 
fields). The running of this parameter, like that of a wave function 
renormalization constant, has no direct significance.  
If the pure gravity part contains the usual 
Ricci scalar term, $\sqrt{q} R(q)$, the parameter that becomes 
inessential may be taken as its prefactor $Z_N$. Up to a dimension 
dependent coefficient it can be identified with the 
inverse of Newton's constant $Z_N^{-1} \sim G_{\rm Newton}$.
It is also easy to see that in a background field formalism $\om$ 
sets the overall normalization of the spectral/momentum values. 
Hence in a theory with a dynamical metric the three (conceptually distinct)
inessential parameters: overall scale of the metric, the inverse of 
Newton's constant, and the overall normalization of the spectral/momentum
values are in one-to-one correspondence; see section 2.1 for details. 
For definiteness let us consider the running of Newton's constant here.    

Being inessential, the quantum field theoretical running of $G_{\rm Newton}$ 
has significance only relative to the running coefficient of some reference 
operator. The most commonly used choice is a cosmological constant term 
$\tilde{\Lambda} \int\! dx \sqrt{q}$. Indeed
\be 
G_{\rm Newton} \tilde{\Lambda}^{\frac{d-2}{d}} =: {\rm const}\, \tau(\mu)^{2/d}\,,
\label{iom2}
\end{equation}  
is dimensionless and invariant under constant rescalings of the 
metric \cite{Kawai2}. The associated essential coupling $\tau(\mu)$   
is in the present context assumed to be asymptotically safe,
i.e.~$\sum_{\mu_0 \leq \mu \leq \infty} \tau(\mu) < \infty$, 
$\lim_{\mu \ra \infty} \tau(\mu) = \tau_*$, where here $0< \tau_* < \infty$. 
Factorizing it into the dimensionless Newton constant ${\rm g}_N \sim 
\mu^{d-2} G_{\rm Newton}$ and $\lb(\mu) = 
\mu^{-d} {\rm g}_N(\mu) \tilde{\Lambda}/2$, 
there are two possibilities: One is that the scheme choices 
are such that both ${\rm g}_N$ and $\lb$ behave like asymptotically 
safe couplings, i.e.~satisfy (\ref{safegs}) below. This is advantageous 
for most purposes. The second possibility is realized when a 
singular solution for the flow equation for ${\rm g}_N$ is 
inserted into the flow equation for $\lb$. This naturally occurs 
when $G_{\rm Newton}$, viewed as an inessential parameter, is 
frozen at a prescribed value, say $[G_{\rm Newton}]^{1/(d-2)} = 
M_{\rm Pl} \approx 1.4 \times 10^{19}$ GeV, which amounts to working 
with Planck units. Then the ${\rm g}_N$ flow is trivial, 
${\rm g}_N(\mu) \sim (\mu/M_{\rm Pl})^{d-2}$, but the flow equation 
for $\lb$ carries an explicit $\mu$-dependence \cite{PercacciPN}. By and 
large both formulations are mathematically equivalent, see section 2.1. 
For definiteness we considered here the cosmological constant term 
as a reference operator, but many other choices are possible.    
In summary, the dimensionless Newton constant can be treated either as 
an inessential parameter (and then frozen to a constant value) or as a 
quasi-essential coupling (in which case it runs and assumes a finite 
positive asymptotic value).

The unstable manifold of a fixed point is crucial for 
the construction of a continuum limit. The fixed point itself describes a
strictly scale invariant situation. More precisely the situation 
at the fixed point is by definition invariant under the chosen coarse graining 
(i.e.\ scale changing) operation. In particular any dependence on an 
ultraviolet cutoff must drop 
out `at' the fixed point, which is why fixed points are believed to be 
indispensable for the construction of a scaling limit. 
If one now uses a different coarse graining operation the location of the 
fixed point will change in the given coordinate system provided by the 
essential couplings. One aspect of universality is that all field theories 
based on the fixed points referring to different coarse graining operations 
have the same long distance behavior.

This suggests to introduce the notion of a {\bf continuum limit} as 
an `equivalence class' of scaling limits in which the physical quantities 
become strictly independent of the UV cutoff, largely independent of the 
choice of the coarse graining operation, strictly independent of the 
choice of gauge slice and, ideally, invariant under 
local reparameterizations of the fields.

In the framework of statistical field theories one distinguishes between
two construction principles, a {\it massless} scaling limit and a 
{\it massive} scaling limit. In the first case all the actions/measures 
on a trajectory emanating from the fixed point describe a scale invariant 
system, in the second case this is true only for the action/measure at 
the fixed point. 
In either case the unstable manifold of the given fixed point 
has to be at least one dimensional. Here we shall exclusively be 
interested in the second construction principle. Given a coarse 
graining operation and a fixed point of it with a nontrivial unstable 
manifold a scaling limit is then constructed by 
`backtracing' a renormalized trajectory emanating from the 
fixed point. The number of parameters needed to specify a 
point on the unstable manifold gives the number of possible scaling limits 
-- not all of which must be physically distinct, however.

In this context it should be emphasized that the number of relevant 
directions in a chosen basis is {\it not} directly related to the 
predictive power of the theory. A number of authors have argued 
in the effective field theory framework that even theories with an infinite 
number of relevant parameters can be predictive 
\cite{Kubo, Atance, Bjerrum-BohrD}. 
This applies all the more if the theory under consideration is based on 
a fixed point, and thus not merely effective. One reason lies 
in the fact the number of independent relevant directions connected 
to the fixed point might not be known. Hidden dependencies would 
then allow for a (genuine or effective) reduction of couplings 
\cite{Zimmermann, OehmeZimmermann, PerryWilson, anselmi2, Atance}. For quantum 
gravity the situation is further complicated by the fact 
that {\it generic} physical quantities are likely to be related 
only nonlocally and nonlinearly to the metric. What matters for the 
predictive power is not 
the total number of relevant parameters but how the observables depend 
on them. To illustrate the point imagine a (hypothetical) case 
where $n^2$ observables are injective functions of $n$ relevant 
couplings each:
\be 
\cO_i({\rm g}_1, \ldots , {\rm g}_n)\,,\quad i =1, \ldots, n^2\,.
\end{equation}
Then $n$ measurements will determine the couplings, leaving $n^2-n$ 
predictions. This gives plenty of predictions, for any $n$, and it remains 
true in the limit $n \ra \infty$, despite the fact that one then has 
infinitely many relevant couplings. This example may be seen as 
a mathematical abstraction of the reason why effective field 
theories (or renormalizable ones with a UV cutoff kept) 
are predictive. The $\cO_i$'s may depend on additional couplings, 
but if this dependence is quantitatively sufficiently suppressed 
the situation is qualitatively as in the example.

Initially infinitely many essential couplings arise when 
a perturbative treatment of Quantum Gravidynamics is based 
on a $1/p^2$ type propagator. Perturbation theory can be seen 
as a  degenerate special case of the general framework described 
before. Depending on the structure of the coupling flow the 
associated {\bf perturbative Gaussian fixed point} 
does or does not reflect a Gaussian fixed point proper. In the case 
of gravity, as first advocated by 
Gomis and Weinberg \cite{GomisWeinb}, the use of a $1/p^2$ type 
graviton propagator in combination with higher derivative terms 
avoids the problems with unitarity that occur in other 
treatments of higher derivative theories. Consistency requires that 
quadratic counterterms (those which contribute to the propagator)
can be absorbed by field redefinitions. This can be seen to be the case 
\cite{anselmi1} either in the absence of a cosmological constant term or 
when the background spacetime admits a metric with constant curvature. 
The price to pay for the $1/p^2$ type propagator is that all nonquadratic
counterterms have to be included in the bare action, so that independence 
of the UV cutoff can only be achieved with infinitely many essential 
couplings, but it can be \cite{GomisWeinb}.  
In order to distinguish this from the familiar notion of perturbative 
renormalizability with finitely many couplings we shall 
call such theories (perturbatively) {\it weakly renormalizable}. 
The above results then show the existence of a ``weakly renormalizable''  
but ``propagator unitary'' Quantum Gravidynamics based on a perturbative 
Gaussian fixed point.

The beta functions for this infinite set of 
couplings are presently unknown. If they were known, expectations are that
at least a subset of the couplings would blow up at some finite 
momentum scale $\mu = \mu_{\mathrm{term}}$ and would be unphysical for 
$\mu > \mu_{\mathrm{term}}$. In this case the computed results for 
physical quantities (``reaction rates'') are likely to blow up 
likewise at some (high) energy scale $\mu = \mu_{\mathrm{term}}$.

This illustrates Weinberg's concept of asymptotic safety. To quote 
from~\cite{Weinberg}: 
``A theory is said to be {\bf asymptotically safe} if the essential 
coupling parameters approach a fixed point as the momentum scale 
of their renormalization point goes to infinity''. Here
`the' essential couplings ${\mathrm{g}}_i$ are those which are useful 
for the absorption of 
cutoff dependencies, i.e.\ not irrelevant ones. The momentum scale 
is the above $\mu$, so that the condition amounts to having 
nonterminating trajectories for the ${\mathrm{g}}_i$'s with a finite 
limit: 
\be 
\sup_{\mu_0 \leq \mu \leq \infty} {\mathrm g}_i(\mu) < \infty \,,
\sspace 
\lim_{\mu \ra \infty} {\mathrm{g}}_i(\mu) = {\mathrm{g}}_i^* < \infty\,, 
\label{safegs}
\end{equation}
for some $i$-independent $\mu_0$. 
In other words in an asymptotically safe theory the above blow up in the 
couplings and hence in physical observables does not occur. 
We suggest to call couplings satisfying (\ref{safegs}) 
{\it asymptotically safe}. 
As a specification one should add~\cite{Weinberg}: ``Of course the 
question whether or not an infinity in coupling constants betokens 
a singularity in reaction rates depends on how the coupling constants 
are parameterized. We could always adopt a perverse definition 
(e.g.\ $\tilde{{\mathrm{g}}}(\mu) = 
({\mathrm{g}}_* - {\mathrm{g}}(\mu))^{-1}$) 
such that reaction rates are finite even at an infinity of the 
coupling parameters. This problem can be avoided if we define 
the coupling constants as coefficients in a power series expansion of 
the reaction rates themselves around some physical renormalization 
point''.

A similar remark applies to the signs of coupling constants.
When defined through physical quantities certain couplings or coupling 
combinations will be constrained to be positive. For example in a 
(nongravitational) effective field theory this constrains the couplings 
of a set of leading power counting irrelevant operators to be 
positive~\cite{Adams,Shoreobs}. 
In an asymptotically safe theory similar constraints are expected 
to arise and are crucial for its physics viability.

Note that whenever the criterion for asymptotic safety 
is met {\it all} the relevant couplings lie in the unstable manifold of the 
fixed point (which is called the ``UV critical surface'' 
in~\cite{Weinberg}, p.802, 
a term now usually reserved for the surface of infinite correlation length). 
The regularity property described earlier is then 
satisfied and the space of actions decomposes in the vicinity
of the fixed point into a stable and an unstable manifold.

Comparing the two perturbative treatments of Quantum Gravidynamics 
described earlier one sees that they have complementary advantages 
and disadvantages: higher derivative theories based on a $1/p^4$ 
propagator are strictly renormalizable with couplings that are 
presumed to be asymptotically safe; however unphysical propagating 
modes are  present. Defining higher derivative gravity perturbatively 
with respect to a $1/p^2$ propagator has the advantage that all propagating 
modes are physical, but initially infinitely many essential couplings are 
needed, a subset of which is presumed to be not asymptotically 
safe. From a technical viewpoint the challenge of Quantum Gravidynamics 
lies therefore not so much in achieving renormalizability 
but to {\bf reconcile asymptotically safe couplings} with the 
{\bf absence of unphysical propagating modes}.

Even in the above perturbative formulations one can see heuristically 
how this might be feasible: both descriptions should be related through 
a reduction of couplings, i.e.~the infinite set of couplings in 
the $1/p^2$ formulation should be thought of as having hidden 
dependencies such that a nonredundant set corresponds to the 
finitely many safe couplings in the $1/p^4$ formulation. 
The proper computational implementation presumably requires 
new (perturbative or nonperturbative) techniques. 

Assuming that this can be achieved certain qualitative features such a 
gravitational functional integral must have can be inferred without actually 
evaluating it. One is the presence of anti-screening configurations,
the other is a dimensional reduction phenomenon in the ultraviolet.

In nonabelian gauge theories the {\bf anti-screening} phenomenon 
can be viewed as the physics mechanism underlying their 
benign high energy behavior (as opposed to abelian gauge theories, 
say), see e.g.~\cite{PeskinSchroeder} for an intuitive discussion. 
It is important not to identify ``anti-screening'' with its most 
widely known manifestation, the sign of the dominant contribution 
to the one-loop beta function. In an exact continuum formulation 
of a pure Yang-Mills theory, say, the 
correlation functions do not even depend on the gauge coupling. Nevertheless 
they indirectly do know about ``asymptotic freedom'' through their 
characteristic high energy behavior. The phenomenon is also state-dependent: 
it is the Yang-Mills vacuum that acts like a specific polarizable medium. 
In the functional integral measure this in principle comes about through the 
dominance of certain configurations/histories which one might 
also call ``anti-screening''.

By analogy one would expect that in a gravitational functional 
integral that allows for a continuum limit a similar mechanism is 
responsible for its benign ultraviolet 
behavior (as opposed to the one expected by power counting considerations 
with respect to a $1/p^2$ propagator, say). As in the Yang-Mills case
a certain class of states will act like a polarizable, predominantly 
``antiscreening'' medium.  Importantly, since a preferred ground state 
is unlikely to exist in quantum gravity, one can take advantage 
of the ensued ambiguity to select the class of states appropriately. 
In a functional integral the state dependence can be encoded in 
boundary terms for the microscopic action, so that a corresponding 
{\it ambiguity} in the definition of the functional integral will result. 
Some insight into the nature of the gravitational antiscreening 
mechanism can be gained from a hamiltonian formulation 
of the functional integral but a proper understanding of the 
interplay between the class of states, the dominant 
geometries/histories, and the renormalization properties in the 
ultraviolet remains to be found. Nevertheless it is clearly 
legitimate to utilize the beforementioned ambiguities
so as to faciltate the construction of a continuum limit. 
For simplicity we shall refer to such an adjustment as the 
implementation of an ``anti-screening constraint''. 

In a discretized functional integral the dominance 
of antiscreening configurations/histories would by definition be 
responsable for the benign ultraviolet properties associated with a 
a non-Gaussian fixed point. Conversely understanding the nature of 
these antiscreening geometries/histories might help 
to design good discretizations. A discretization of the gravitational 
functional integral which allows for a continuum limit might 
also turn out to exclude or dynamically disfavor configurations that 
are taken into account 
in other, off-hand equally plausible, discretizations. Compared 
to such a naive discretization it will look as if a constraint 
on the allowed configurations/histories has been imposed. 
A useful analogy is the inclusion of a causality constraint in the 
definition of the (formal Euclidean) functional integral originally 
proposed by Teitelboim~\cite{Teitelb1, Teitelb2}, and recently put 
to good use in the framework of dynamical triangulations~\cite{Ambjorn1}. 
Just as the inclusion of a good causality constraint is justified 
retroactively, so would be the inclusion of a suitable 
``antiscreening'' constraint.

A second qualitative property of a gravitational functional 
integral where  the continuum limit is based on a non-Gaussian 
fixed point is a  {\bf dimensional reduction of the residual 
interactions in the UV}. 
There are several arguments for this phenomenon which we  
describe in section 3. Perhaps the simplest one is based on  
the large anomalous dimensions at a non-Gaussian fixed point 
and runs as follows: (We present here a model-independent 
variant~\cite{MNnotes} of the argument used in~\cite{LR1}). 
Suppose that the unkown microscopic 
action is local and reparameterization invariant. The only term containing 
second derivatives then is the familiar Einstein-Hilbert term 
$\int \! dx \sqrt{q} 
R(q)$ of mass dimension $2\!-\!d$ in $d$ dimensions, if the metric is taken 
dimensionless. As explained before the dimensionful running prefactor 
$Z_N$ ($N$ for ``Newton'') multiplying it 
plays a double role, once as a wave function renormalization constant and 
once as a quasi-essential coupling ${\rm g}_N(\mu)$. Both aspects 
are related as outlined before; in particular 
\be 
Z_N(\mu) = \frac{\mu^{d-2}}{\gn(\mu)}\,.
\label{N1}
\end{equation}
Here $\gn$ is a dimensionless coupling which is treated as 
``quasi-essential'' and whose running may also depend on all 
the other couplings 
(gravitational and non-gravitational) made dimensionless by 
taking out a suitable power of $\mu$. The short distance behavior of the 
propagator will now be governed by the ``anomalous dimension''
$\eta_N = - \mu \dd_\mu \ln Z_N(\mu)$, by the usual field theoretical 
arguments, say, via the Callan-Symanzik equation for the effective 
action. On the other hand the flow equation for ${\mathrm{g}}_N$ 
can be expressed in terms of $\eta_N$ as 
\be 
\mu \dd_\mu \gn = [d-2 + \eta_N(\gn,{\mathrm{other}})]\,\gn\,,
\label{N2}
\end{equation}
where we schematically indicated the dependence on the other dimensionless 
couplings. {\it If} this flow equation now has a nontrivial fixed point 
$\infty > \gn^* >0$, the only way how the right hand side can 
vanish is for $\eta_N(\gn^*, {\mathrm{other}}) = 2\!-\!d$, irrespective of 
the detailed behavior of the other couplings as long as no blow-up occurs. 
This is a huge anomalous dimension. For a graviton ``test propagator''
(see below) the key property of $\eta_N$ is that it gives rise to a high 
momentum behavior of the form $(p^2)^{-1 + \eta_N/2}$ modulo logarithms, 
or a short distance behavior of the form $(\sqrt{x^2})^{2-d -\eta_N}$ modulo 
logarithms. Keeping only the leading part the vanishing power at $\eta_N =
2-d$ translates into a logarithmic behavior, $\ln x^2$, 
formally the same as for massless (scalar) propagators in a 
two-dimensional field theory. We shall comment on potential pitfalls 
of such an argument below.

In accordance with this argument a $1/p^4$ type propagator goes hand in 
hand with a non-Gaussian fixed point for ${\rm g}_N$ in two other 
computational settings: in strictly renormalizable higher derivative 
theores (see section 2.2) and in the $1/N$ expansion 
\cite{Tomboulis1, Tomboulis2, Smolin}. In the latter case a  
nontrivial fixed point goes hand in hand with a graviton propagator 
whose high momentum behavior is of the form $1/(p^4 \ln p^2)$, in four 
dimensions, and formally $1/p^d$ in $d$ dimensions.

The fact that a large 
anomalous dimension occurs at a non-Gaussian fixed point was first 
observed in in the context of the $2+\eps$ expansion~\cite{Kawai2, Kawai3}
and then noticed in computations based on truncated flow equations 
\cite{LR1}. The above variant of the argument shows that 
no specific computational 
information enters. It highlights what is special about the Einstein--Hilbert 
term (within the class of local gravitational actions): 
it is the kinetic (second derivative) term itself which carries a 
dimensionful coupling. Of course one could assign to the metric a
mass dimension $2$, in which case Newton's constant would be dimensionless.
However one readily checks that then the wave function renormalization 
constant of a standard matter kinetic term acquires a mass dimension 
$d\!-\!2$ for bosons and $d\!-\!1$ for fermions, respectively. 
Assuming that the 
dimensionless parameter associated with them remains nonzero 
as $\mu \ra \infty$ one can repeat the above argument and finds 
now that {\it all} matter propagators have a $1/p^d$ type high momentum 
behavior, or a {\it logarithmic} short distance behavior. It is this 
universality which justifies to attribute the modification in the 
short distance behavior of the fields to a modification of the 
underlying (random) geometry. This may be viewed as a specific variant of 
the old expectation that gravity acts as a short distance regulator.

Let us stress that while the anomalous dimension always governs 
the UV behavior in the vicinity of a (UV) fixed point, it is 
in general {\it not} related to the geometry of field propagation, 
see~\cite{Kroeger} for a discussion in QCD. What is special about 
gravity is ultimately that the propagating field itself determines
distances. In the context of the above argument this is used in 
the reshuffling of the soft UV behavior to matter propagators. 
The propagators used here should be viewed as 
``test propagators'', not as physical ones. One 
transplants the information 
in $\eta_N$ derived from the gravitational functional integral into 
a conventional propagator on a (flat or curved) background spacetime. 
The reduced dimension two should be viewed as an ``interaction 
dimension'' specifying roughly the (normalized) number of 
independent degrees of freedom a randomly picked one interacts with.

The same conclusion ($1/p^d$ propagators or interaction dimension $2$) 
can be reached in a number of other ways as well, which are described 
in section 3. A more detailed understanding of the microstructure
of the random geometries occuring in an asymptotically safe 
functional integral remains to be found.

Accepting this dimensional reduction as a working hypothesis it is natural 
to ask whether there exists a two-dimensional field theory which provides 
a quantitatively accurate (`effective') description of this extreme UV
regime. Indeed, one can identify a number 
of characteristics such a field theory should have, using only the main ideas
of the scenario, see the end of Section 3.
The asymptotic safety of such a field theory would then strongly support 
the corresponding property of the full theory and the selfconsistency of the 
scenario. In summary, we have argued that 
the qualitative properties of the gravitational functional integral in 
the extreme ultraviolet follow directly from the previously highlighted 
principles: the existence of a nontrivial UV fixed point, asymptotic 
safety of the couplings, and antiscreening. Moreover these 
UV properties can be probed for selfconsistency.

\newsubsection{Coarse graining and dynamically adjusted background data}

Since renormalization implicitly (in perturbation theory) or 
explicitly (in the Kadanoff-Wilson framework) depends on the 
choice of a  coarse graining operation one is in a quantum gravity
context lead to address the question ``with respect to 
what'' field configurations are coarsely or finely grained. 
The piecemeal evaluation of the functional integral  
(decomposition of the presumed `critical' problem 
into `subcritical' ones) requires a physically motivated
notion of the slicing. For statistical field theories 
on a non-dynamical background the spectrum of a covariant 
differential operator (generalized momenta) can be used. In quantum 
gravity the determination of an averaged geometry is part of 
the dynamical problem, and one has to proceed differently. 
The retroactive dynamical adjustment of initially prescribed 
background data provides a natural generalization. 
The principle has already been outlined in the discussion 
around Eq.~(\ref{adjust}).

Here we describe the construction in somewhat more detail using the 
background effective action formalism, see \cite{Honerkamp1,Abbott,Boulware} 
for the 
latter. In this formalism the effective action $\Gamma$ of a scalar field 
theory becomes a (highly nonlocal) functional of two fields, $\Gamma = 
\Gamma[f, \bar{\chi}]$. The second field is the initially prescribed 
background field $\bar{\chi}$, the first can be interpreted as the 
source dependent average $\bra \chi - \bar{\chi} \ket_{J_*}$ of the 
quantum field $\chi$ shifted by $-\bar{\chi}$,
where the source $J_*$ is given by $J_*[f;\bar{\chi}] = 
\delta \Gamma[f;\bar{\chi}]/\delta f$. Switching off the 
source, $J_*[f; \bar{\chi}] =0$,  correlates both fields 
and one may assume that locally (in function space) one can be 
expressed as a functional of the other. We write the relation as 
$\bar{\chi} = F(f+\bar{\chi})$, so that $F = {\rm id}$ corresponds to 
$f=0$, and assume that it can be solved locally for $f$, 
i.e.~$f = - \bar{\chi} + F^{-1}(\bar{\chi})$. 
Then $\bar{\Gamma}[\phi] := \Gamma[\phi - \bar{\chi}; F(\phi)]$, 
obeys (\cite{MNnotes,Gravirept}) $\delta \bar{\Gamma}[\phi]/\delta \phi = 
\bra \delta S/\delta \bar{\chi} \ket_{J_*=0}\, \delta F/\delta \phi$, 
where the derivative of the action $S$ is taken with respect to the 
explicit background dependence, if any. From the viewpoint of the 
underlying functional integral a dynamically adjusted
background $\bar{\chi} = F(f + \bar{\chi})$ is optimal with 
regard to a small field expansion around it, where ``small'',
however,  now means ``selfconsistently small with respect 
to the full quantum dynamics''.

This construction can be transferred to gravity, where 
the dynamically adjusted background metric can in addition 
be used to define an intrinsic coarse graining scale.   
As remarked before the use of a background metric as opposed 
to other, more specific, background data is dispensible, 
for concreteness we use here the metric itself, both as a dynamical 
variable, $q_{\alpha\beta}$, in the functional integral and 
to specify the background data needed. This leads to 
an effective action $\Gamma[f;\bar{g}, \ldots]$ which is a reparameterization 
invariant functional of two symmetric second rank tensors. 
The second, $\bar{g}_{\alpha\beta}$, 
is an initially independently prescribed ``background metric''. 
The first, $f_{\alpha\beta}$, is interpreted as an initially source 
dependent average $\bra q_{\alpha\beta} - \gbab \ket_{J_*}$ of the 
dynamical metric $q_{\alpha\beta}$ shifted by $-\gbab$,
where the source $J_*$ is given by $J_*[f;\bar{g}] = 
\delta \Gamma[f;\bar{g}]/\delta f$. The dots in $\Gamma[f;\bar{g}, \ldots]$
indicate other fields, dual to ghost sources, which are inessential to 
the discussion. Switching off the source, $J_*[f, \bar{g}] =0$,  now 
correlates $f$ with $\bar{g}$ and as before we may assume that 
a functional relation $\bar{g} = F(f + \bar{g})$ with inverse
$f = - \bar{g} + F^{-1}(\bar{g})$ holds, at least locally 
in the space of metrics. Then $\bar{\Gamma}[g] := \Gamma[g - \bar{g}, F(g)]$, 
obeys \cite{MNnotes} 
\be 
\frac{\delta \bar{\Gamma}[g]}{\delta g_{\alpha\beta}} = 
\Big\langle \frac{\delta S}{\delta \bar{g}_{\gamma\delta}} 
\Big\rangle_{\!J_*=0}\; \frac{\delta F_{\gamma\delta}}{\delta g_{\alpha\beta}}\,,
\label{gadjust}
\end{equation}
where the derivative of the action $S$ is taken with respect to the 
explicit background dependence. Starting with a reparameterization 
invariant microscopic action $S_0[q]$, the gauge fixing and ghost
terms will introduce such an explicit dependence on $\gbab$; 
schematically $S[q;\bar{g}] = S_0[q] + S_{\rm g.f.}[q-\bar{g};\bar{g}]
+ S_{\rm ghost}[q-\bar{g};\bar{g}]$. The solutions of (\ref{gadjust}) 
contain information about: the source-free condition $J_*[f;\bar{g}] =0$, 
about the state vector underlying the functional integral through 
the choice of boundary terms, and about the choice of gauge-slice. 
We now comment on each of these dependencies successively:

The construction of the extremizing sources $J_*[f;\bar{g}]$ 
entering the definition of $\Gamma[f;\bar{g},\ldots]$ (as the 
Legendre transform of $W[J;\bar{g}, \ldots]$, the generating 
functional of connected correlation functions) is usually done 
within a formal power series ansatz. This gives a solution 
$J_*[f;\bar{g}](x) = \sum_{n \geq 1} j_n(x, x_1, \ldots , x_n) 
\, f(x_1) \ldots f(x_n)$, where the $j_n$'s can be expressed in 
terms of the $f$ moments $W^{(n)}$ of $W$, `amputated' with 
the exact $\Gamma^{(2)}[f;\bar{g}] = \delta \Gamma[f;\bar{g}]/
(\delta f \delta f)$. Clearly $J_* \equiv 0$ iff $f \equiv 0$ 
within a formal power series ansatz. This amounts to $F = {\rm id}$ 
and the dynamically adjusted background coincides with 
the prescribed one. Conversely, in order to get a genuine 
dynamical adjustment, one has to go beyond a formal power series 
ansatz. Assuming $f = - \bar{g} + F^{-1}(\bar{g})$ one gets 
$\bra q_{\alpha\beta} \ket_{J_* =0} = F^{-1}_{\alpha\beta}(\bar{g})$,
The functional $F^{-1}$ here contains the dynamical information 
inherited from the full $\Gamma$ via $\delta 
\Gamma[f;\bar{g},\ldots]/\delta f = 0$. The right hand side 
of (\ref{adjust}) can now be viewed as a new background 
$\bar{g}^*_{\alpha\beta}$ and the parameteric dependence 
of the state $\bra \;\;\ket_{J_*=0}$ can be relabelled 
to  $\bra \;\;\ket_{F(\bar{g})}$. The equation characterizing 
the class of dynamically adjusted backgrounds then becomes       
\be 
\bra q_{\alpha\beta} \ket_{F(\bar{g}^*)} = \bar{g}^*_{\alpha\beta}\,,
\label{gadjust2}
\end{equation} 
as anticipated in (\ref{adjust}).

The notion of a state is implicitly encoded in the effective 
action. Recall that the standard effective action for a scalar field theory, 
when evaluated on a given time-independent function $\varphi^i = 
\bra \chi^i \ket$, is proportional to the minimum value of the 
Hamiltonian $H$ in that part of the Hilbert space spanned 
by normalizable states $|\psi\ket$ satisfying $\bra \psi| \chi^i |\psi \ket 
= \varphi^i$. A similar interpretation holds formally for the various 
background effective actions~\cite{BurgKunst}. 
In a functional integral formulation the information about 
the state can be encoded in suitable (though often not explicitly known) 
boundary terms for the microscopic action. An alternative way to see 
that $\Gamma$ in principle also encodes the information about the 
underlying state vector, is via reconstruction. Let $\Gamma^{(n)}$, 
$n \geq 2$, be the vertex functions associated with $\Gamma$, i.e.
\be 
\Gamma^{(n)}(x_1,\ldots,x_n;g) := 
\frac{\delta}{\delta g(x_1)}\ldots 
\frac{\delta}{\delta g(x_n)} \Gamma[g - \bar{g}; \bar{g},
 \ldots]\Big|_{\bar{g} = F(g)}\,.
\label{consistentb3}
\end{equation}
In a flat space quantum field theory the Wightman or Osterwalder-Schrader 
reconstruction procedures would allow one to (re-)construct the 
state space and field operators from knowledge of the $\Gamma^{(n)}$. 
In a quantum gravity context little is known about the feasibility 
of such a reconstruction from e.g.~the vertex functions (\ref{consistentb3}).
The use of metric correlators (or quantities tentatively interpreted as 
such) may also not be ideal from the viewpoint of such a reconstruction. 
One would expect that correlators of (nonlocal) quantities 
closer to (Dirac) observables are better suited for a reconstruction.
Returning to $\Gamma$, one should think of it as a functional of both 
the selected state and of the fields. The selected state will indirectly 
(co-)determine the space of functionals on which the renormalization flow acts.
For example the type of nonlocalities which actually occur 
in $\Gamma$ should know about the fact that $\Gamma$ stems from 
a microscopic action suited for the appropriate notion 
of positivity and from a physically acceptable state.

The notion of a physically acceptable state is another unexplored 
issue in this context. In conventional flat space quantum field theories 
there is a clear-cut notion of a ground state and of the physical state space 
based on it. Already in quantum field theories on curved but non-dynamical 
spacetimes a preferred vacuum is typically absent and physically 
acceptable states have to be selected by suitable conditions 
(like, for example, the well-known Hadamard condition imposed on the 
short distance behavior of the two point function, which for 
free quantum field theories in curved spacetime selects 
states with desirable stability properties.) 
In quantum gravity the formulation of analogous selection criteria 
is an open problem. As a tentative example we mention the 
condition \cite{MNnotes}
\ba 
&& \bra P_q(T) \ket \sim T^{-d/2}\,, \;\sspace\; T \ra \infty\,,  
\nonum
&& P_q(T) := \int \! dx \sqrt{q} \exp( T \Delta_q)(x,x)\,.
\label{stateselect}
\end{eqnarray}
Here $\Delta_q$ is the Laplace-Beltrami operator of a 
(pseudo-) riemannian metric $q_{\alpha\beta}$, and 
$\exp( T \Delta_q)(x,y)$ is the associated heat kernel. 
When $q = \eta$ is flat $P_\eta(T)$ decays like 
$T^{-d/2}$ for $T \ra \infty$. The condition (\ref{stateselect}) 
therefore indirectly characterizes a class of states which favor 
geometries that are smooth and almost flat on large scales.

Finally we comment on the gauge dependence of $\Gamma$ or 
$\bar{\Gamma}$.  The right hand side of (\ref{gadjust}) renders the 
dependence of $\bar{\Gamma}$ on the choice of gauge slice manifest. 
Had a (technically more complicated) Vilkovisky-deWitt type effective 
action \cite{Vilkovisky,Rebhan,Pawlowski2} been used this dependence 
should be absent. 
As an approximative shortcut one can continue to work with the previous 
background effective action and consider solutions $\check{g}$ of 
$\delta \bar{\Gamma}[g]/\delta g_{\alpha \beta} =0$, which 
retroactively minimize the dependence on the choice of gauge. 
This condition will be used later on. Since $\bar{\Gamma}$ 
is highly nonlocal the identification of {\it physical} solutions 
of (\ref{gadjust}) or $\delta \bar{\Gamma}[g]/\delta g_{\alpha \beta} =0$, 
is a nontrivial problem. See e.g.~\cite{Balbinot} for examples 
based on an anomaly induced part of an effective action.  
The previous discussion suggests a partial  
characterization, namely those solutions of (\ref{gadjust}) should 
be regarded as physical which are associated with physically 
acceptable states.

The use of a dynamically adjusted background geometry has the 
additional advantage of allowing one to introduce an intrinsic 
coarse graining scale. Let $\bar{g}_{\alpha\beta}$ be again an 
initially prescribed background geometry. Let $\mu$ denote an 
(unphysical) scale parameter which schematically cuts off modes 
whose average extension with respect to $\bar{g}_{\alpha\beta}$ 
is larger than $\mu^{-1}$. Clearly there is a large degree of 
arbitrariness in defining such a mode cut-off and for each choice 
there is an effective action $\Gamma_\mu[f, \bar{g}]$ containing 
mostly the dynamical information about modes larger than $\mu$. 
In a two step procedure one can now replace $\mu$ with an intrinsic 
coarse graining scale. In a first step $\bar{g}_{\alpha\beta}$ 
is replaced with a dynamically adjusted background $\check{g}_\mu$ 
solving the counterpart of (\ref{gadjust}) for $\Gamma_\mu$.    
In a second step one considers the spectrum 
$\{\cE_{\om}(\check{g}_\mu)\,|\, \om \in S \subset \R\}$ 
of a covariant differential operator, say $\Delta_{\check{g}_\mu}$,  
built from $\check{g}_\mu$. The implicit equation 
\be 
\mu^2 = {\cal E}_{\om}(\check{g}_\mu)\,,
\end{equation} 
then determines $\mu = \mu(\om)$ or $\mu = \mu(\cE)$ and hence allows one 
to replace $\mu$ with the spectral scale $\cE$ intrinsic to 
the dynamically adjusted background.

\newsubsection{Evidence for asymptotic safety}

Presently the evidence for asymptotic safety in quantum gravity 
comes from the following very different computational settings: 
(a) the $2+ \eps$ expansion, (b) perturbation theory of higher derivative 
theories and a large N expansion in the number of matter fields,  
(c) the study of symmetry truncations, and (d) that of truncated
functional flow equations. Arguably none of the pieces of evidence is
individually compelling but taken together they make a strong case 
for asymptotic safety.

The results from the $2+\eps$ expansion were part of Weinberg's 
original motivation to propose the existence of a non-Gaussian fixed point. 
Since gravity in two and three dimensions is non-dynamical, however, 
the lessons for a genuine quantum 
gravitational dynamics are somewhat limited. Higher derivative 
derivative theories were known to be strictly renormalizable with 
a finite number of couplings, at the expense of having unphysical 
propagating modes, see \cite{Stelle1,FradkinHD,AvramidiHD,PeixetoHD}. 
With hindsight one can identify a non-Gaussian fixed point for Newton's 
constant already in this setting, see Section 2.2. 
The occurance of this 
non-Gaussian fixed point is closely related to the $1/p^4$-type propagator 
that is used. The same happens when (Einstein- or a higher derivative) gravity 
is coupled to a large number $N$ of matter fields and a $1/N$ expansion is 
performed. A nontrivial fixed point is found that goes hand in hand 
with a $1/p^4$-type progagator (modulo logs), which here arises from 
a resummation of matter selfenergy bubbles, however. 
 
As emphasized before the challenge of Quantum Gravidynamics is not 
so much to achieve (perturbative or nonperturbative) renormalizability 
but to reconcile  asymptotically safe couplings with the absence of 
unphysical propagating modes. Two recent developments provide complementary 
evidence that this might indeed be feasible. Both of these developments 
take into account the dynamics of infinitely many physical degrees of 
freedom of the four dimensional gravitational field. In order to be 
computationally feasible the `coarse graining' has to be constrained somehow. 
To do this the following two strategies have been pursued (which we 
label here (c) and (d) according to the subsection in which they will be 
discussed below):  

(c) The metric fluctuations are constrained 
by a symmetry requirement but the full (infinite dimensional) renormalization 
group dynamics is considered. This is the strategy via {\it symmetry reductions}. 

(d) All metric fluctuations are taken 
into account but the renormalization group dynamics is projected 
onto a low dimensional submanifold. Since this is done using truncations of 
functional renormalization group equations we shall refer to this as the 
strategy via {\it truncated functional flow equations}. 

Both strategies (truncation in the fluctuations but unconstrained flow
and unconstrained quantum fluctuations but constrained flow) are complementary.
Tentatively both results are related by the dimensional reduction 
phenomenon outlined earlier. 

For the remainder of this Section we now describe the 
pieces of evidence from the various computational settings (a) -- (d) 
mentioned. To emphasize its auxiliary role we shall write 
$q = (q_{\alpha\beta})_{1 \leq \alpha ,\beta \leq d}$ 
for the `quantum metric' playing the role of the integration 
variable in the functional integral. Averages thereof or macroscopic
metrics are denoted by $g_{\alpha\beta}$ and reference metrics 
by $\bar{g}_{\alpha\beta}$. Our curvature conventions are set by 
$(\nabla_{\alpha} \nabla_{\beta} - \nabla_{\beta} \nabla_{\alpha}) 
v^{\gamma} = R^{\gamma}_{\;\;\delta \alpha \beta} v^{\delta}$
and $R_{\alpha\beta} = R^{\gamma}_{\;\;\alpha \gamma \beta}$. 
For a $(-,+, \ldots,+)$ signature metric the Einstein-Hilbert and scalar 
field action read, $S_{\rm EH} = + (16 \pi G)^{-1} \int dx \sqrt{g} 
[R(g) - 2 \Lambda]$ and $S = - \frac{1}{2} \int dx \sqrt{g} g^{\alpha 
\beta} \dd_{\alpha} \phi \dd_{\beta} \phi$, respectively. 
Occasionally we shall switch to Euclidean signature metrics, 
$(+, +, \ldots, +)$, in which case $i S_{\rm EH} \mapsto - S^E_{\rm EH}$,
$i S \mapsto -S^E$, and the Euclidean signature Lagrangians are 
obtained by formally flipping the sign of the Lorentzian signature ones.


{\bf (a) Evidence from $2 + \eps$ expansions:} In the non-gravitational 
examples of perturbatively nonrenormalizable field theories with a non-Gaussian 
fixed point the non-Gaussian fixed point can be viewed as a `remnant' of an 
asymptotically free fixed point in a lower dimensional version of the theory. 
It is thus natural to ask how gravity behaves in this respect. 
In $d=2$ spacetime dimensions Newton's constant $\gn$ is dimensionless 
and formally the theory with the bare action $\gn^{-1} \int \! d^2 x \sqrt{q} R(q)$
is power counting renormalizable in perturbation theory. However, as 
the Einstein--Hilbert term is purely topological in two dimensions 
the inclusion of local dynamical degrees of freedom requires, 
at the very least, starting from $2 + \eps$ dimensions and then 
studying the behavior near $\eps \ra 0^+$. The resulting 
``$\eps$-expansion'' amounts to a double expansion in the number of 
`graviton' loops and in the dimensionality parameter $\eps$. Typically 
dimensional regularization is used, in which case the UV divergencies 
give rise to the usual poles in $1/\eps$. Specific for gravity are 
however two types of complications. The first one is due to the fact 
that $\int \! d^{2+\eps} x \sqrt{q} R(q)$ is topological at $\eps =0$, 
which gives rise to additional ``kinematical'' poles of order $1/\eps$ 
in the graviton propagator. The goal of the renormalization 
process is to remove both the ultraviolet and the kinematical 
poles in physical quantities. The second problem is that in pure 
gravity Newton's constant is an inessential parameter, i.e.~it 
can be changed at will by a field redefinition. Newton's constant $\gn$ 
can be promoted to a coupling proper by comparing its flow with that 
of the coefficient of some reference operator, which is fixed 
to be constant.

For the reference operator various 
choices have been adopted (we follow the discussion in 
Kawai et al~\cite{Kawai1, Kawai2, Kawai3, Kawai4} with the conventions 
of~\cite{Kawai3}): (i) a cosmological constant 
term $\int\! d^{2+\eps} x \sqrt{q}$, (ii) monomials from matter fields 
which are quantum mechanically 
non-scale invariant in $d=2$, (iii) monomials from matter fields which are 
quantum mechanically scale invariant in $d=2$, and (iv) the conformal 
mode of the metric itself in a background field expansion. 
All choices lead to flow equation of the form 
\be 
\mu \frac{d}{d\mu} \gn = \eps\, \gn - \gamma \, \gn^2\,,
\label{i0}
\end{equation}
but the coefficient $\gamma$ depends on the choice of the reference 
operator~\cite{Kawai1}. For all $\gamma>0$ there is a nontrivial fixed 
point $\gn^* = \eps/\gamma >0$ with a one-dimensional unstable 
manifold. In other words $\gn$ is an asymptotically safe coupling 
in $2+ \eps$ dimensions and the above rule of thumb suggests that 
this a remnant of a nontrivial fixed point in $d=4$ with respect 
to which $\gn$ is asymptotically safe (see Section 1.3 for the 
renormalization group terminology).

Technically the non-universality of $\gamma$ arises from 
the before-mentioned kinematical poles. In the early papers 
\cite{Gastmans, Christensen, Weinberg} the choice (i) 
was adopted giving $\gamma = 19/24 \pi$, or $\gamma= (19 -c)/24 \pi$ 
if free matter of central charge $c$ is minimally coupled. 
A typical choice for (ii) is a mass term of a Dirac fermion,
a typical choice for (iii) is the coupling of a four-fermion 
(Thirring) interaction. Then $\gamma$ comes out as 
$\gamma= (19 + 6 \Delta_0 -c)/24 \pi$, where $\Delta_0 = 1/2,1$,
respectively. Here $\Delta_0$ is the scaling dimension of the 
reference operator, and again free matter of central charge $c$ 
has been minimally coupled. It has been argued in~\cite{Kawai1} that 
the loop expansion in this context should be viewed as 
double expansion in powers of $\eps$ and $1/c$, and that 
reference operators with $\Delta_0 =1$ are optimal. 
The choice (iv) has been pursued systematically in a series 
of papers by Kawai et al~\cite{Kawai2, Kawai3, Kawai4}. It is based 
on a parameterization of the metric in terms of a background 
metric $\bar{g}_{\mu\nu}$, the conformal factor $e^{\sigma}$, 
and a part $f_{\mu\nu}$ which is traceless, $\bar{g}^{\mu\nu} f_{\mu\nu}=0$.
Specifically $q_{\mu\nu} = \bar{g}_{\mu\rho} (e^f)^{\rho}_{\;\;\nu} 
e^{\sigma}$ is inserted into the Einstein--Hilbert action; propagators 
are defined (after gauge fixing) by the terms quadratic in 
$\sigma$ and $f_{\mu\nu}$, vertices correspond to the higher order terms. 
This procedure turns out to have a number of advantages. First 
the conformal mode $\sigma$ is renormalized differently from 
the $f_{\mu\nu}$ modes and can be viewed as defining a reference 
operator in itself; in particular the coefficient 
$\gamma$ comes out as $\gamma= (25 -c)/24\pi$. Second, and related to the 
first point, the system has a well-defined $\eps$-expansion (absence of poles) 
to all loop orders. Finally this setting allows one to make contact 
to the exact (KPZ~\cite{Knizhnik}) solution of two-dimensional quantum gravity 
in the limit $\eps \ra 0$.

{\bf (b) Evidence from perturbation theory and large N:} Modifications of 
the Einstein-Hilbert action where fourth derivative terms are 
included are known to be perturbatively renormalizable \cite{Stelle1}. 
A convenient parameterization is 
\be 
S = -\!\int\! dx \sqrt{q} \Big[ \tilde{\Lambda} - \frac{1}{c_d G_N} R
+ \frac{1}{2s} C^2 - \frac{\om}{3 s} R^2 + 
\frac{\th}{s} E\Big]\,. 
\label{iHD}
\end{equation}
Here $d=4 + \eps$, $c_d$ is a constant such that $c_4 = 16 \pi$, $C^2$ is the 
square of the Weyl tensor and $E$ is the integrand of the Gauss-Bonnet 
term. The sign of the crucial $C^2$ coupling $s>0$ is 
fixed by the requirement that the Euclidean functional integral is 
damping. The one-loop beta functions for 
the (nonnegative) couplings, $s,\,\om,\,\th$, are known and on the 
basis of them these couplings are expected to be asymptotically safe. 
In particular $s$ is asymptotically free, $\lim_{\mu \ra 0} s(\mu) =0$.
The remaining couplings $\tilde{\Lambda}$ and $c_d G_N$ are made dimensionless
via ${\rm g}_N(\mu) = \mu^{d-2} c_d G_N$, $\lb(\mu) = \mu^{-d} \tilde{\Lambda} 
{\rm g}_N(\mu)/2$,
where $\mu$ is the renormalization scale. At $s=0$ these flow equations
are compatible with the existence of a non-trivial fixed point 
for Newton's constant, ${\rm g}^*_N \neq 0$, see Section 2.2. The value of 
${\rm g}^*_N$ is highly nonuniversal but it cannot naturally be 
made to vanish, i.e.~the nontrivial and the trivial fixed point, 
${\rm g}_N^* =0$, do not merge. The rationale for identifying a 
nontrivial fixed point by perturbative means is explained in Appendix A1. 
The benign renormalizability properties seen in this framework 
are due to the $1/p^4$ type propagator in combination with
diffeomorphism invariance, at the expense of unphysical 
propagating modes.

The action (\ref{iHD}) can be supplemented by a matter action, 
containing a large number, $O(N)$, of free matter fields. 
One can then keep the product $N \cdot c_d G_N$ fixed,
retain the usual normalization of the matter kinetic terms,
and expand in powers of $1/N$. Renormalizability of the resulting 
`large N expansion' then amounts to being able to remove the UV 
cutoff order by order in the formal series in $1/N$. This 
type of studies was initiated by Tomboulis 
where the gravity action was taken either the pure Ricci scalar 
\cite{Tomboulis1}, Ricci plus cosmological term \cite{Smolin}, or 
a higher derivative action \cite{Tomboulis2},
with free fermionic matter in all cases. More recently 
the technique was reconsidered \cite{PercacciN} with (\ref{iHD})  
as the gravity action and free matter consisting of $N n_S$ scalar fields, 
$N n_D$ Dirac fields, and $N n_M$ Maxwell fields. 
 
Starting from the Einstein-Hilbert action the high energy behavior 
of the usual $1/p^2$-type propagator gets modified. To leading order in 
$1/N$ the modified propagator can be viewed as the graviton 
propagator with an infinite number of matter 
selfenergy bubbles inserted and resummed. The resummation changes the 
high momentum behavior from $1/p^2$ to $1/(p^4 \ln p^2)$, in four dimensions. 
In $2 < d < 4$ dimensions the resulting $1/N$ expansion is believed 
to be renormalizable in the sense that the UV cutoff $\Lambda$ can strictly be 
removed order by order in $1/N$ without additional (counter) terms in the 
Lagrangian. In $d=4$ the same is presumed to hold provided an 
extra $C^2$ term is included in the bare Lagrangian, as in (\ref{iHD}).
After removal of the cutoff the beta functions 
of the dimensionless couplings can be analyzed in the usual way and 
already their leading $1/N$ term will decide about the flow pattern. 

The qualitative result (due to \cite{Tomboulis1,Smolin}) is that there 
exists a nontrivial fixed point for the dimensionless couplings 
${\rm g}_N, \, \lb$, and $s$. Its unstable manifold is three 
dimensional, i.e.\ all couplings are asymptotically safe. 
Repeating the computation in $2+\eps$ dimensions the fixed point 
still exists and (taking into account the different UV regularization) 
corresponds to the large $c$ (central charge) limit of the 
fixed point found the $2+ \eps$ expansion. 

These results have recently been confirmed and extended by Percacci 
\cite{PercacciN} using the heat kernel expansion. In the presence of 
$N n_S$ scalar fields, $N n_D$ Dirac fields, and $N n_M$ Maxwell
fields, the flow equations for ${\rm g}_N, \, \lb$ and $s$ come out 
to leading order in $1/N$ as 
\ba 
\label{ilargeN}
\nspace \mu\frac{d}{d \mu} {\rm g}_N \is \phantom{-}2 {\rm g}_N + 
\frac{1}{(4\pi)^2} \frac{1}{6}( n_S - 2 n_D - 4 n_M) {\rm g}_N^2 \,,
\nonum
\nspace \mu\frac{d}{d \mu} \lb \is -2 \lb + 
\frac{1}{(4\pi)^2} \Big[ \frac{1}{6}( n_S - 2 n_D - 4 n_M) \lb {\rm g}_N 
+ \frac{1}{4}( n_S - 4 n_D +2 n_M) {\rm g}_N\Big]\,.
\\
\nspace \mu\frac{d}{d \mu} s \is 
-\frac{1}{(4\pi)^2} \frac{1}{360}( 6n_S + 25 n_D +72 n_M)s^2 \,.
\nonumber
\end{eqnarray}
One sees that the $C^2$ coupling is always asymptotically free,
and that Newton's constant has a nontrivial fixed point, 
${\rm g}^*_N/(4\pi)^2 = 12/(-n_S + 2 n_D + 4 n_M)$, which is 
positive if the number of scalar matter fields is not too 
large.

As a caveat one should add that the $1/p^4$-type propagators  
occuring both in the perturbative and in the large $N$ framework
are bound to have an unphysical pole at some intermediate 
momentum scale. This pole corresponds to  unphysical propagating 
modes and it is the price to pay for (strict) perturbative  
renormalizability combined with asymptotically safe couplings.  
From this point of view, the main challenge of Quantum 
Gravidynamics lies in reconciling asymptotically safe 
couplings with the absence of unphysical propagating modes. 
This can be achieved in the context of the $2+2$ reduction.


{\bf (c) Evidence from symmetry reductions:} 
Here one considers the usual gravitational functional integral 
but restricts it from ``4-geometries modulo diffeomorphisms'' 
to ``4-geometries constant along a $2+2$ foliation modulo diffeomorphisms''. 
This means instead of the familiar $3+1$ foliation of geometries 
one considers a foliation in terms of two-dimensional 
hypersurfaces $\Sigma$ and performs the functional integral only over 
configurations that are constant as one moves along the stack of 
two-surfaces. Technically this constancy condition is 
formulated in terms of two commuting vectors fields $K_a = 
K_a^{\;\alpha} \dd_{\alpha}$, $a=1,2$, that are Killing vectors 
of the class of geometries $q$ considered, $\cL_{K_a} q_{\alpha\beta} =0$.  
For definiteness we consider here only the case where both 
Killing vectors are spacelike. From this pair of Killing vector fields 
one can form the symmetric $2\times 2$ matrix $M_{ab} := q_{\alpha\beta} 
K_a^{\;\alpha} K_b^{\;\beta}$. Then $\gamma_{\alpha \beta} 
:= q_{\alpha \beta} - M^{ab} K_{a\alpha} K_{b\beta}$ 
(with $M^{ab}$ the components of $M^{-1}$ and $K_{a\alpha} := 
q_{\alpha \beta} K_a^{\beta}$) defines a metric on the orbit space 
$\Sigma$ which obeys $\cL_{K_a} \gamma_{\alpha\beta} =0$ 
and $K_a^{\alpha} \gamma_{\alpha \beta} =0$. The functional 
integral is eventually performed over metrics of the form 
\be 
q_{\alpha\beta} = \gamma_{\alpha\beta} + M^{ab} K_{a\alpha} K_{b\beta}\,,
\label{i1} 
\end{equation}
where the $10$ components of a metric tensor are parameterized 
by the $3+3$ independent functions in $\gamma_{\alpha\beta}$ and $M_{ab}$.
Each of these functions is constant along the stack of two-surfaces
but may be arbitrarily rough within a two-surface.

In the context of the asymptotic safety scenario the restriction of the 
functional integral to metrics of the form (\ref{i1}) is a very fruitful one: 
(i) the restricted functional integral inherits the perturbative 
non-renormalizability 
(with finitely many relevant couplings) from the full theory. 
(ii) it takes into account the crucial `spin-2' aspect, that 
is, linear and nonlinear gravitational waves with two independent 
polarizations per spacetime point are included. 
(iii) it goes beyond the Eikonal approximation \cite{tHooft,WDitt} whose dynamics 
can be understood via a related $2+2$ decomposition \cite{Kabat,Fabbrichesi}.
(iv) based on heuristic arguments the dynamics of full Quantum Gravidynamics 
is expected to be effectively two-dimensional in the extreme ultraviolet
with qualitative properties resembling that of the $2+2$ truncation.
The renormalization of the $2+2$ truncation can thus serve as a prototype
study and its asymptotic safety probes the selfconsistency of the 
scenario. 
(v) for the restricted functional integral the full infinite dimensional 
renormalization group dynamics can be studied; it reveals both a Gaussian and 
a non-Gaussian fixed point, where the properties of the latter are compatible with 
the existence of a non-perturbative continuum limit. 
Two additional bonus features are: in this sector the explicit construction 
of Dirac observables is feasible (classically and presumably also in 
the quantum theory). Finally a large class of matter couplings is 
easily incorporated. 

As mentioned the effective dynamics looks two-dimensional. 
Concretely the classical action describing the dynamics 
of the 2-Killing vector subsector is that of a non-compact symmetric space 
sigma-model non-minimally coupled to 2D gravity via the ``area radius'' 
$\rho := \sqrt{\det(M_{ab})_{1 \leq a,b \leq 2}}$, 
of the two Killing vectors. To avoid a possible confusion let 
us stress, however, that the system is very different from 
most other models of quantum gravity (mini-superspace, 
2D quantum gravity or dilaton gravity, Liouville theory, topological 
theories) in that it has infinitely many local and selfinteracting dynamical 
degrees of freedom. Moreover these are literally (an infinite subset of) 
the degrees of freedom of the 4-dimensional gravitational field, not 
just analogues thereof. The corresponding classical solutions 
(for both signatures of the Killing vectors) have been widely 
studied in the General Relativity literature, 
c.f.~\cite{Griffiths,Belinski,Klein}. We refer to \cite{BGM88,BM00} for 
details on the reduction procedure and \cite{Torre2K} 
for a canonical formulation. The case with aligned polarizations 
(Beck-Einstein-Rosen waves) is much simpler and the essential aspects 
can be modelled by a massive free field on ${\rm AdS}_2$ \cite{ERwaves}.

For generic polarizations strongly selfinteracting systems arise
whose the renormalization \cite{PTernst} can be achieved  by borrowing 
covariant background field 
techniques from Riemannian sigma-models; see~\cite{Honerkamp1,Frie85, HPS88,
  Shore, CurciPaff, Tseytlin87, Osb87}. 
In the particular application here the sigma-model perturbation theory is 
partially nonperturbative from the viewpoint of a graviton loop expansion 
as not all of the metric degrees of freedom are Taylor expanded in the bare 
action, see \cite{Gravirept}.  
This together with the field reparameterization invariance 
blurs the distinction between a perturbative and a non-perturbative 
treatment of the gravitational modes. 
The renormalization can be done to all orders of sigma-model perturbation 
theory, which is `not-really-perturbative' for the gravitational modes.
It turns out that strict cutoff independence can be achieved only 
by allowing for infinitely many essential couplings. They are conveniently 
combined into a generating functional $h$, which is a positive function 
of one real variable. Schematically the renormalized action takes the 
form~\cite{PTernst} 
\be 
S[q] = S_{EH}\Big[ \frac{h(\rho)}{\rho} q\Big] + 
\mbox{other second derivative terms}\,.
\label{i2} 
\end{equation}
Here $q$ is a metric of the form (\ref{i1}), $S_{EH}[q]$ is the Einstein--Hilbert
action evaluated on it, and $h(\rho)$ is the generating coupling function evaluated 
on the renormalized area radius field $\rho$. Higher derivative terms
are not needed in this subsector for the absorption of counter terms;
the ``other second derivative terms'' needed are known explicitly.

This ``coupling functional'' is scale dependent 
and is subject to a flow equation of the form 
\be 
\mu\frac{d}{d\mu} h = \betabf_h(h)\,,
\label{i3}
\end{equation}
where $\mu$ is the renormalization scale and $\mu \mapsto 
h(\,\cdot\,,\mu)$ is the `running' generating functional. 
To preclude a misunderstanding let us stress that the 
function $h(\,\cdot\,,\mu)$ changes with $\mu$, 
irrespective of the name of the argument, not just its value 
on $\rho$, say. Interestingly a closed formula for the beta function (or functional) 
in (\ref{i3}) can be found~\cite{PTernst, Sernst}. The resulting flow equation 
is a nonlinear 
partial integro-differential equation and difficult to analyze. 
The fixed points however are easily found. Apart from the 
degenerate `Gaussian' one, $1/h \equiv 0$, there is a nontrivial 
fixed point $h^{\mathrm{beta}}$. For the Gaussian fixed point 
a linearized stability analysis is empty, the structure of the 
quadratic perturbation equation suggests that it has both attractive 
and repulsive directions in the space of functions $h$. For the 
non-Gaussian fixed point $h^{\mathrm{beta}}$ a linearized stability  
analysis is non-empty and leads to a system of linear integro-differential 
equations. Since the fixed point $h^{\rm beta}$ has the form of a 
powerseries in the loop counting parameter $\lb$, the 
proper concept of a ``linearized perturbation'' has the form 
\ba
h(\rho,\lb,\mu) \is 
h^{\mathrm{beta}}(\rho,\lb) + \delta h(\rho,\lb,\mu)\,,
\nonum
\delta h(\rho,\lb,\mu) \is 
\frac{\lb}{2\pi} {\mathsf s}_1(\rho,t)
+ \Big(\frac{\lb}{2\pi}\Big)^2 {\mathsf s}_2(\rho,t) + \ldots. 
\label{2Kflow5}
\end{eqnarray}
where the ${\mathsf s}_l(\rho,t)$ are functions of $\rho$ 
and $t = \frac{1}{2\pi} \ln \mu/\mu_0$. Note that the perturbation 
involves infinitely many functions of two variables. 
Inserting the ansatz (\ref{2Kflow5}) into the flow equation 
$\mu \frac{d}{d\mu} h = \betabf_h(h/\lb)$ and linearizing 
in $\delta h(\rho,\lb,\mu)$ gives a recursive system of 
inhomogeneous integro-differential equations for the ${\mathsf s}_l$, 
The boundary conditions are fixed such that the full $h$ flow is 
driven by the counterterms only, which amounts to the requirement 
that all the ${\mathsf s}_l(\rho,t)$ vanish for $\rho \ra \infty$ uniformly in $t$. 
Subject to these boundary conditions the recursive system of 
integro-differential equations can be shown to have a unique solution
for arbitary smooth initial data. The solution for ${\mathsf s}_1$ reads 
\be 
{\mathsf s}_1(\rho,t) = \rho \int_{\rho}^{\infty} 
\frac{du}{u} r_1(u -\zeta_1 t)\;, 
\label{2Kflow7}
\end{equation}
where $r_1$ is an arbitrary smooth function of one variable satisfying 
$u\, r_1(u) \ra 0$ for $u \ra 0$. This function can essentially be 
identified with the initial datum at some renormalization time $t=0$,
as $r_1(\rho) = - \rho \dd_{\rho}[{\mathsf s}_1(\rho, t=0)/
\rho]$. Evidently ${\mathsf s}_1(\rho, t) \ra 0$ for 
$t \ra \infty$, if $\zeta_1 <0$. 

This condition is indeed satisfied by all the symmetry reduced gravity 
theories considered in \cite{Sernst}, precisely because the coset space $G/H$ is 
noncompact. If sigma-model scalars and abelian gauge fields are present 
in the 4D action one has the simple formula
\be
\zeta_1 = - \frac{k+2}{2} \;,\sspace k = \# \mbox{abelian vector fields}\,. 
\label{2Kflow8}
\end{equation}
Equation~(\ref{2Kflow7}) shows that the lowest order perturbation 
${\mathsf s}_1$ will always die out for $t \ra \infty$, for arbitrary smooth initial data
prescribed at $t=0$. It can be shown that this continues to
hold for {\it all} higher order ${\mathsf s}_l$ irrespective of the 
signs of the coefficients $\zeta_l,\,l \geq 2$. 
The situation is illustrated in the Figure below. The proof of this 
result is somewhat technical and can be found in~\cite{Sernst}. 

\bigskip

\hspace{4cm}
\epsfig{file=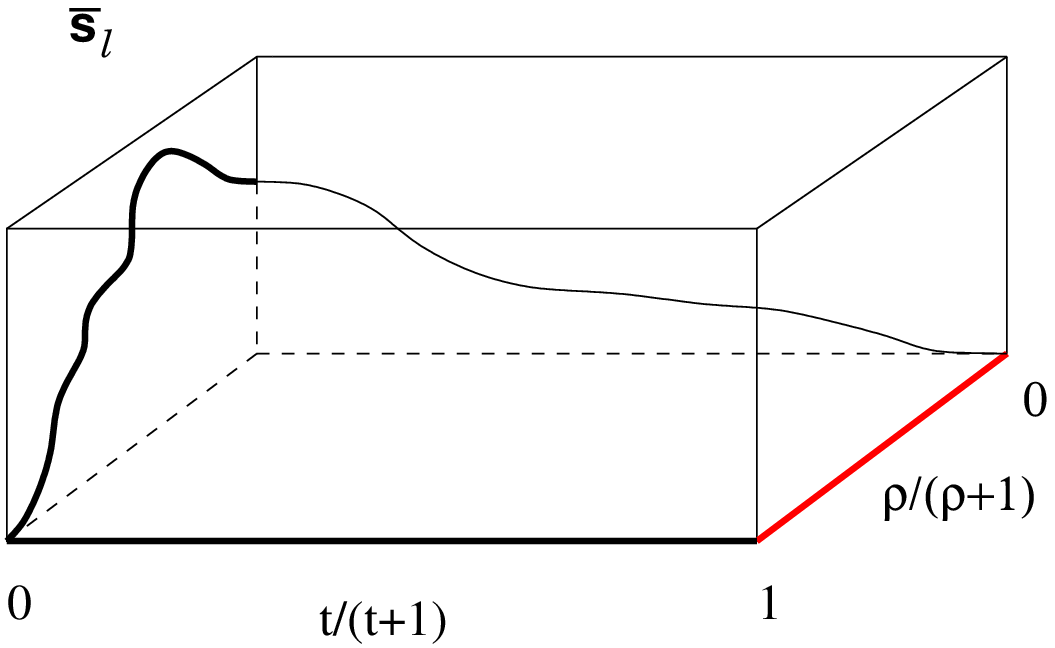,width=65mm,height=50mm}

\bigskip

Thus {\it all} linearized perturbations decay for $\mu \ra \infty$, 
which is precisely what Weinberg's criterion for asymptotic safety asks for. 
Moreover the basic propagator used is free from unphysical poles. This 
suggests that a genuine continuum limit exist 
for the $2+2$ reduced Quantum Gravidynamics beyond approximations 
(like the sigma-model perturbation theory/partially 
nonperturbative graviton expansion used to compute (\ref{i3})). 
See~\cite{Qernst,Korotkin} for a proposed `exact' bootstrap construction,
whose relation to a $2+2$ truncated functional integral however remains to 
be understood.

In summary, in the context of the $2\!+\!2$ reduction an asymptotically 
safe coupling flow can be reconciled with the absence of unphysical 
propagating modes. In contrast to the technique on which evidence (d) 
below is based the existence of an infinite cutoff limit here can be 
{\it shown} and does not have to be stipulated as a hypothesis subsequently 
probed for selfconsistency. Since the properties of the 
$2\!+\!2$ truncation qualitatively are the ones one would expect from 
an `effective' field theory describing the extreme UV aspects 
of Quantum Gravidynamics (see the end of Section 3), its asymptotic 
safety is a strong argument for the selfconsistency of the scenario.


{\bf (d) Evidence from truncated flows of the effective average action:} 
The {\it effective average action} $\Gamma_{\Lambda,k}$ for a scalar 
field theory is a generating functional generalizing the usual effective action,
to which it reduces for $k=0$. Here $\Gamma_{\Lambda,k}$ depends on the 
UV cutoff $\Lambda$ and an additional scale $k$, indicating 
that in the defining functional integral roughly the field modes 
with momenta $p$ in the range $k \leq p \leq \Lambda$ have been integrated 
out. Correspondingly $\Gamma_{\Lambda,\Lambda}$ gives back the bare action 
and $\Gamma_{\Lambda,0} = \Gamma_{\Lambda}$ is the usual quantum effective action, 
in the presence of the UV cutoff $\Lambda$. The modes in the momentum range 
$k \leq p \leq \Lambda$ are omitted or suppressed by a mode cutoff 
`action' $C_{\Lambda,k}$, and one can think of $\Gamma_{\Lambda,k}$ 
as being the conventional effective action $\Gamma_{\Lambda}$
but computed with a bare action that differs from the original one 
by the addition of $C_{\Lambda,k}$; specifically
\be 
\Gamma_{\Lambda,k} = - C_{\Lambda,k} + \Gamma_{\Lambda}\Big|_{S \mapsto 
S + C_{\Lambda,k}}\,.
\label{iGamma}
\end{equation}   
From the regularized functional integral defining $\Gamma_{\Lambda,k}$ an 
(`exact') functional renormalization group equation (FRGE) can be derived. 
Schematically it has the form $k \frac{d}{d k} \Gamma_{\Lambda,k} = 
{\mathrm{rhs}}$, where the ``right hand side'' involves the Hessian of 
$\Gamma_{\Lambda,k}$ with respect to the dynamical fields. 
The FRGE itself (that is, its rhs) carries no {\it explicit} 
dependence on the UV cutoff, or one which can trivially be removed. 
However the removal of the UV regulator $\Lambda$ implicit in the definition 
of $\Gamma_{\Lambda,k}$ is nontrivial and is related to the 
traditional UV renormalization problem. 
Whenever massless degrees of freedom are involved also the 
existence of the $k \ra 0$ limit of $\Gamma_{\Lambda,k}$ 
is nontrivial and requires identification of the proper 
infrared degrees of freedom. In the present context we take 
this for granted and focus on the UV aspects.

The effective average action has been generalized to gravity by Reuter~\cite{Reuter}.
The substitution (\ref{iGamma}) is now applied to the (highly nonlocal) 
background effective 
action $\Gamma[g, \bar{g}]$ which in addition to the average $g$ of the 
`quantum' metric $q$ depends on a background metric $\bar{g}$.  
The mode cutoff functional $C_{\Lambda,k}[q,\bar{g}]$ depends 
covariantly on $\bar{g}$ and the bare action $S_{\Lambda}[q, \bar{g}] = 
\Gamma_{\Lambda,\Lambda}[q, \bar{g}]$ is not specified from 
the outset. In fact, conceptually  it is largely determined by the requirement 
that a continuum limit exists, see the criterion in Appendix A.2.
$\Gamma_{\Lambda,\Lambda}$ can be expected to have a well-defined derivative 
expansion with the leading terms roughly of the form (\ref{iHD}). Also the 
gravitational effective average action $\Gamma_{\Lambda,k}$ obeys an `exact' FRGE, 
which is a new computational tool in quantum gravity not limited
to perturbation theory. In practice $\Gamma_{\Lambda,k}$ is replaced
in this equation with a $\Lambda$ independent functional interpreted 
as $\Gamma_{\infty,k}$. The assumption that the `continuum limit' 
$\Gamma_{\infty,k}$ for the gravitational effective average action exists 
is of course what is at stake here. 
The strategy in the FRGE approach is 
to show that this assumption, although without a-priori justification, 
is consistent with the solutions of the flow equation 
$k \frac{d}{d k} \Gamma_{\infty,k} = {\mathrm{rhs}}$ (where right hand 
side now also refers to the Hessian of $\Gamma_{\infty,k}$). 
The structure of the solutions $\Gamma_k$ of this cut-off independent 
FRGE should be such that they can plausibly be identified with 
$\Gamma_{\infty,k}$. Presupposing the `infrared safety' in the 
above sense, a necessary condition for this is that the 
limits $\lim_{k \ra \infty} \Gamma_k$ and $\lim_{k \ra 0} \Gamma_k$ 
exist. Since $k \leq \Lambda$ the first limit probes whether $\Lambda$ 
can be made large; the second condition is needed to have all modes 
integrated out. In other words one asks for {\it global existence} of the 
$\Gamma_k$ flow obtained by solving the cut-off independent FRGE. 
Being a functional differential equation the 
cutoff independent FRGE requires an initial condition, 
i.e.\ the specification of a functional $\Gamma_{\mathrm{initial}}$ 
which coincides with $\Gamma_k$ at some scale $k=k_{\mathrm{initial}}$. 
The point is that only for very special `fine tuned' 
initial functionals $\Gamma_{\mathrm{initial}}$ will the associated 
solution of the cutoff independent FRGE exist globally. 
The existence of the $k \ra \infty$ limit in this sense can be viewed 
as the counterpart of the UV renormalization problem, 
namely the determination of the unstable manifold associated with 
the fixed point $\lim_{k \ra \infty} \Gamma_k$. We refer to Appendix A.2
for a more detailed discussion of this issue.

The full nonlinear functional differential equation is 
of course intractable. To make the FRGE computationally useful 
the space of functionals is truncated typically to a finite dimensional 
one of the form
\be 
\Gamma_k[g,\bar{g}] = \sum_{i=0}^N {\mathrm{g}}_i(k) k^{d_i} I_i[g] 
+ \mbox{gauge fixing term}\,,  
\label{i4}
\end{equation}
where the $I_i$ are `well-chosen' -- local and nonlocal -- 
functionals of $g$, and $\bar{g}$ is identified with $g$ after 
functional differentiation. The ${\mathrm{g}}_i(k)$ are numerical 
parameters that carry the scale dependence. For $I_i$'s obeying 
a non-redundancy condition, the ${\mathrm{g}}_i$ play the role of essential 
couplings which have been normalized to have vanishing mass dimension 
by taking out a power $k^{d_i}$. Beyond perturbation theory 
unfortunately little is known about the type of nonlocal terms to 
expect in $\Gamma_k[g,\bar{g}]$, leaving the choice of such $I_i$ 
somewhat arbitrary. Conceptually the truncation implicitly 
replaces the full gravitational dynamics by one whose functional 
renormalization flow is confined to the subspace (\ref{i4}), 
similar to what happens in a hierarchical approximation.

The original FRGE then can be converted into a system of nonlinear 
ordinary differential equations for the couplings ${\rm g}_i$. In 
the case of gravity the following ansatz has been made by Lauscher 
and Reuter~\cite{LR1, LR2} (with Euclidean signature) 
\be 
I_0[g] = \int \! dx \,\sqrt{g} \,,\quad 
I_1[g] = -\!\int \! dx \, \sqrt{g} R(g)\,,\quad 
I_2[g] = \int \! dx \sqrt{g} R(g)^2\,,
\label{i5}
\end{equation}
where $g=(g_{\alpha \beta})_{1 \leq \alpha,\beta \leq 4}$ is the 
metric and $R(g)$ is the associated curvature scalar. 
The flow pattern $k \mapsto ({\mathrm{g}}_0(k), {\mathrm{g}}_1(k), 
{\mathrm{g}}_2(k))$ displays a number of remarkable properties. Most 
importantly a non-Gaussian fixed point exists
(first found in~\cite{Souma99} based on~\cite{Reuter} and 
corroborated in~\cite{Souma00, LR1, LR2, LitimFP, BonannoR}). 
Within the truncation (\ref{i5}) a {\it three}-dimensional subset of 
initial data is attracted to the fixed point under the reversed flow 
\be 
\lim_{k\ra \infty} ({\mathrm{g}}_0(k), {\mathrm{g}}_1(k), {\mathrm{g}}_2(k)) = 
({\mathrm{g}}^*_{0}, {\mathrm{g}}^*_{1}, {\mathrm{g}}^*_{2})\,,
\end{equation}
where the fixed point couplings ${\mathrm{g}}^*_{i},\,i=0,1,2$, are finite and 
positive and no blow-up occurs in the flow for large $k$. 
Again this adheres precisely to the asymptotic safety criterion. 

The flow equations for the ${\mathrm{g}}_i(k)$'s  
depend on the choice of the mode cutoff function and on the choice of 
gauge fixing. In general they do not assume a transparent 
analytical form. An exception is when all but $I_0$ and $I_1$ are 
omitted from the truncation (so that only the Einstein-Hilbert terms remain) 
and an optimized mode cutoff is used in combination with a limiting 
version of the gauge fixing term \cite{LitimFP}. In terms of the 
parameterization ${\rm g}_0 = 2 \lb/{\rm g}_N$ and ${\rm g}_1 = 1/{\rm g}_N$ 
used later on the flow equations then take the form
\ba 
k \frac{d}{dk} {\rm g}_N \is 
2 {\rm g}_N + \frac{6 {\rm g}_N^2}{{\rm g}_N - 6 (4\pi)^2 (1-2 \lb)^2}\,,
\nonum
k \frac{d}{dk} \lb \is
- 2\lb - \frac{{\rm g}_N}{2 (4\pi)^2} 
\bigg( 1 + 2 \frac{3 {\rm g}_N + 12 (4\pi)^2 (1- 3 \lb)}%
{\;2 {\rm g}_N - 12 (4\pi)^2 (1-2\lb)^2} \bigg) \,.
\label{litimflow}
\end{eqnarray} 
The above properties can then be verified analytically.  
 
Some of the trajectories with initial data
in the unstable manifold cannot be extended to $k \ra 0$
due to (infrared) singularities. 
This problem is familiar from nongravitational theories 
and is presumably an artifact of the truncation. 
In the vicinity of the fixed point, on the other hand, all 
trajectories show remarkable robustness properties against 
modifications of the mode cutoff scheme which 
provide good reasons to believe that the structural aspects of the 
above results are not an artifact of the truncation used. The upshot 
is that there is a clear signal for 
asymptotic safety in the subsector (\ref{i4}), obtained via 
truncated functional renormalization flow equations.

The impact of matter has been studied by Percacci et al
\cite{Dou, PercacciP1, PercacciP2}. 
Minimally coupling free fields (bosons, fermions, or abelian 
gauge fields) one finds that the non-Gaussian fixed point 
is robust, but the positivity of the fixed point couplings 
${\mathrm{g}}_0^* >0,\,{\mathrm{g}}_1^* >0$ puts certain constraints on the 
allowed number of copies. When a selfinteracting scalar $\chi$ is 
coupled nonminmally via $-\sqrt{g} [(\kappa_0 + \kappa_2 \chi^2 + 
\kappa_4 \chi^4 + \ldots) R(g) + \lb_0 + \lb_2 \chi^2 + \lb_4 
\chi^4 + \ldots + \dd \chi \dd \chi]$, one finds a fixed point $\kappa_0^* >0,\,
\lb_0^* >0$ (whose values are with matched normalizations the 
same as ${\mathrm{g}}^*_{1}, {\mathrm{g}}^*_{0}$ in the pure gravity computation)
while all selfcouplings vanish, $\kappa_2^* = \kappa_4^* = \ldots =0$, 
$\lb_2^* = \lb_4^* = \ldots =0$. In the vicinity of the fixed point 
a linearized stability analysis can be performed; the admixture 
with $\lb_0$ and $\kappa_0$ then lifts the marginality of $\lb_4$,
which becomes marginally irrelevant ~\cite{PercacciP1,PercacciP2}. 
The running of $\kappa_0$ and $\lb_0$ is qualitatively
unchanged as compared to pure gravity, indicating that 
the asymptotic safety property is robust also with respect 
to the inclusion of selfinteracting scalars.

This concludes our survey of the evidence for asymptotic safety. 
More details on the results (c) and (d) can be found in 
the review \cite{Gravirept}. The perturbative identification of 
the non-Gaussian fixed point is detailed in section 2.2. The 
results (c) and (d) are genuinely surprising. With hindsight, 
the most natural explanation is to view them as manifestations of 
the asymptotic safety of the {\it full} dynamics with 
respect to a nontrivial fixed point. Tentatively (c) reflects 
a property of the full dynamics in the extreme ultraviolet via 
the dimensional reduction of the residual interactions. Since 
$\Gamma_{\Lambda,\Lambda} = S_{\Lambda}$ the origin of (d) 
could be the match to the perturbatively visible non-Gaussian 
fixed point.

\subsection{Some working definitions}

Here we attempt working definitions for some of the key terms used 
before.

{\bf Quantum Gravidynamics:} The term is coined in analogy to 
``Quantum Chromodynamics'' indicating, first, 
that the theory is supposed to be defined not only as an effective field 
theory and, second, that the selfinteraction of the quantized gravitational 
field is predominantly antiscreening in the ultraviolet.     
 
In contrast to ``Quantum General Relativity'' the microscopic action is 
allowed to be different from the Einstein-Hilbert action or a 
discretization thereof. 
Plausibly it should be still quasilocal, i.e.\ have a well-defined 
derivative expansion, and based on perturbatively renormalizable higher 
derivative theories one would expect it to contain at least quartic 
derivative terms. This means that also the number of physical propagating 
degrees of freedom (with respect to a background) may be different from the 
number entailed by the Einstein--Hilbert action. 
As with ``Quantum General Relativity'' we take the 
term ``Gravidynamics'' in a broad sense, allowing for 
any set of field variables (e.g.\ vielbein and spin connection,
Sen-Ashtekar variables, Plebanski and BF type formulations, teleparallel etc.) 
that can be used to recast general relativity (see 
e.g.\ the review~\cite{Peldan}). It is of course {\it not} assumed 
from the outset that the quantum 
gravidynamics based on the various set of field variables are necessarily 
equivalent.

{\bf Gaussian fixed point:} A fixed point is called 
{\it Gaussian} if there exists a choice of field variables for which the 
fixed point action is quadratic in the fields and the 
functional measure is Gaussian. This includes the local case but also 
allows for nonlocal quadratic actions. The drawback of this definition 
is that the proper choice of field variables 
in which the measure reveals its Gaussian nature may be hard to find. 
(For example in the correlation functions of the spin field 
in the two-dimensional Ising model the underlying free fermionic 
theory is not visible.) 

A non-Gaussian fixed point is simply one where 
no choice of fields can be found in which the measure becomes 
Gaussian. Unfortunately this, too, is not a very operational criterion.

{\bf Unstable manifold:} The unstable manifold of a 
fixed point with respect to a coarse graining operation is the set of 
all points that can be reached along flow lines emanating from the 
fixed point, the so-called {\it renormalized trajectories}. Points 
on such a flow line correspond to {\it perfect actions}. 
The {\it stable} manifold is the set of points attracted to the fixed 
point in the direction of coarse graining.

{\bf Strict (weak) renormalizability:} We call a field theory 
strictly (weakly) renormalizable with respect to a fixed point and 
a coarse graining operation if the dimension of its unstable 
manifold is finite (infinite). It is implied that if a field theory 
has this property with respect to one coarse graining operation it 
will have it with respect to many others (``universality''). Strict or weak 
renormalizability is believed to be a sufficient condition 
for the existence of a genuine continuum limit for observables.

{\bf Relevant coupling:} Given an expansion ``sum over couplings 
times interaction monomials'', a coarse graining operation, and a
fixed point of it, a coupling is called {\it relevant} ({\it irrelevant}) 
if it is driven away from (towards) the value the corresponding 
coordinate has at the fixed point, under a sufficient number of 
coarse graining steps. Note that this distinction makes sense 
even for trajectories not connected to the fixed point (because they 
terminate). It is however an explicitly `coordinate 
dependent' notion. The same terms are used for the interaction 
monomials associated with the couplings. The dimension of the 
unstable manifold equals the maximal number of independent relevant 
interaction monomials `connected' to the fixed point. 
All points on the unstable manifold are thus parameterized by 
relevant couplings but not vice versa. 

Couplings which are relevant or irrelevant in a linearized analysis 
are called linearly relevant or linearly 
irrelevant, respectively. A coupling which is neither linearly 
relevant nor linearly irrelevant, is called (linearly) marginal.

{\bf Continuum limit:} By a genuine continuum limit we mean here a 
limit in which physical quantities become: (C1) strictly independent of the 
UV cutoff, (C2) independent of the choice of the coarse graining 
operation (within a certain class), and (C3) independent of the 
choice of gauge slice and invariant under point 
transformations of the fields. 
Usually one stipulates properties (C1) and (C2) for the functional measure 
after which (C3) should be a provable property of physical quantities like the 
S-matrix. The requirement of having also (C1) and (C2) only for 
observables is somewhat weaker and in the spirit of the asymptotic 
safety scenario. For the issue of gauge-independence see 
\cite{Kummer, Pawlowski2}.  

Typically the properties (C1-C3) 
cannot be rigorously established, but there are useful criteria which 
render the existence of a genuine continuum limit plausible in 
different computational frameworks. In Appendices A1 and A2 we
discuss in some detail such criteria for the perturbative
and for the FRGE approach, respectively. For convenience we 
summarize the main points here. 

In {\it renormalized perturbation} theory the criterion involves two parts:
(PTC1) Existence of a formal continuum limit. This means, the removal of the 
UV cutoff is possible and the renormalized physical quantities 
are independent of the scheme and of the choice of interpolating fields
--  all termwise in a formal power series in the loop counting parameter. 
The perturbative beta functions always have a have a trivial (Gaussian) 
fixed-point but may also have a nontrivial (non-Gaussian) fixed point. 
The second part of the criterion is: (PTC2) The dimension of the unstable 
manifold of the (Gaussian or non-Gaussian) fixed point as computed from the 
perturbative beta functions equals the number of independent essential 
couplings. For example $\phi^4_4$ and QED meet (PTC1) but not (PTC2) 
while QCD satisfies both (PTC1) and (PTC2).

In the framework of the {\it functional renormalization group equations} 
(FRGE) similar criteria for the existence of a genuine 
continuum limit can be formulated. Specifically for the 
FRGE of the effective average action one has: (FRGC1) The solution of 
the FRG equation admits (for fine tuned initial data 
$\Gamma_{\mathrm{initial}}$ at some $k=k_{\mathrm{initial}}$) a global solution 
$\Gamma_k$, i.e.\ one that can be extended both to $k \ra \infty$ and to $k \ra 0$
(where the latter limit is not part of the UV problem in itself).
(FRGC2) The functional derivatives of $\lim_{k\ra 0} \Gamma_k$ 
(vertex functions) meet certain requirements which ensure 
stability/positivity/unitarity. 

In (FRGE1) the existence of the $k \ra 0$ limit in 
theories with massless degrees of freedom is nontrivial and 
the problem of gaining computational control over the infrared physics 
should be separated from the UV aspects of the continuum limit as 
much as possible. However the $k \ra 0$ limit is essential to probe 
stability/positivity/unitarity. For example, to obtain a (massive) Euclidean 
quantum field theory the Schwinger functions constructed from the vertex 
functions have to obey nonlinear relations which ensure that the Hilbert space 
reconstructed via the Osterwalder-Schrader procedure has a
positive definite inner product.

{\bf Perturbative (weak) renormalizability:} We call a theory 
perturbatively (weakly) renormalizable if (PTC1) can be achieved 
with finitely (infinitely) many essential couplings. A theory 
were neither can be achieved is called perturbatively nonrenormalizable. 
Perturbative (weak) renormalizability is neither necessary nor 
sufficient for (weak or strict) renormalizability in the above 
nonperturbative sense. It is only in combination with (PTC2) that 
perturbative results are indicative for the existence of a 
genuine continuum limit.

{\bf Asymptotically free coupling:} A non-constant coupling 
in the unstable manifold of a Gaussian fixed point. 

The ``non-constant'' proviso is needed to exclude cases like a 
trivial $\phi_4^4$ coupling. In a nonperturbative lattice construction
of $\phi_4^4$ theory only a Gaussian fixed point with a one-dimensional
unstable manifold (parameterized by the renormalized mass) is thought 
to exist, along which the renormalized $\phi_4^4$ coupling is constant 
and identically zero. The Gaussian nature of the fixed-point, on 
the other hand, is not crucial and we define:

{\bf Asymptotically safe coupling:} A non-constant coupling 
in the unstable manifold of a fixed point.

{\bf Asymptoticaly safe functional measure}: The functional measure of a 
statistical field theory is said to be asymptotically safe 
if it is perturbatively weakly renormalizable or non-renormalizable, but it 
possesses a fixed point with respect to which it is strictly 
renormalizable. Subject to the regularity assumption that the 
space of actions can in the vicinity of the fixed point 
be decomposed into a stable and an unstable manifold this 
is equivalent to the following requirement: all relevant couplings 
are asymptotically safe and there is only a finite number of them. 
Note that unitarity or other desirable properties that would 
manifest itself on the level of observables are not part of this 
definition.

In a non-gravitational context the functional measure of the 3D 
Gross--Neveu model is presently the best candidate to be asymptotically 
safe in the above sense (see~\cite{GNKogut, GNMagnen, GNW1, GNW2}
and references therein). Also 5D Yang--Mills theories 
(see~\cite{giesYM, MorrisYM} and references therein) are believed to 
provide examples. In a gravitational context, however, 
there are good reasons to modify this definition. 

First the choice of couplings has to be physically motivated,
which requires to make contact to observables. 
In the above nongravitational examples with a single coupling 
the `meaning' of the coupling is obvious; in particular it is clear 
that it must be finite and positive at the non-Gaussian fixed point.  
In general however one does not know whether ill behaved couplings 
are perverse redefinitions of better behaved ones. To avoid this 
problem the couplings should be defined as coefficients
in a power series expansion of the observables themselves 
(Weinberg's ``reaction rates'', see the discussion in section 1.1).
Of course painfully little is known about (generic) quantum gravity   
observables, but as a matter of principle this is how couplings  
should be defined. In particular this will pin down the physically 
aedequate notion of positivity or unitarity.

Second, there may be good reasons to work initially with infinitely 
many essential couplings. Recall that the  
number of essential couplings entering the initial construction of 
the functional measure is not necessarily equal to the number 
eventually indispensable. 
In a secondary step a reduction of couplings might be feasible. 
That is, relations among the couplings might exist which are compatible 
with the renormalization flow. If these relations are sufficiently 
complicated, it might be better to impose them retroactively 
than to try to switch to a more adapted basis of interaction 
monomials from the beginning.

Specifically in the context of quantum gravity microscopic actions with 
infinitely many essential couplings occur naturally in several ways: 
when starting from the Gomis and Weinberg picture \cite{GomisWeinb} of 
perturbative quantum gravity and in the $2\!+\!2$ reduction \cite{PTernst}, 
where a coupling function is needed for  a dimensionless scalar.  
Further, the (Wilsonian) effective actions induced by the conformal anomaly 
can be rewritten in terms of dimensionless scalars \cite{Balbinot,AMM}. 
Their functional form is only partially constrained by the requirement 
to reproduce the anomaly and the fate of the associated  couplings 
or coupling functions in the ultraviolet is in principle a matter of 
dynamics.

Third, the dimension of the unstable manifold is of secondary importance 
in this context. Recall that the dimension of the unstable manifold 
is the {\it maximal} number of independent relevant interaction 
monomials `connected' to the fixed point. This maximal number 
may be difficult to determine in Quantum Gravidynamics for the above 
reasons. Moreover the identification of {\it all} renormalized trajectories
emanating from the fixed point may be more than what is needed physicswise;
the successful construction of a subset of renormalized trajectories for physically 
motivated couplings may already be enough to obtain predictions/explanations 
for some observables. What matters is not so much the total number of 
relevant couplings but the way how observables depend on them. 
We remark that even in conventional perturbation theory based on the 
Einstein-Hilbert action the divergencies in the S-matrix seem to be less 
severe than those in the effective action \cite{BernS}. Generally,     
since generic observables (in
the sense used in Section 1.1) are likely to be nonlinearly and
nonlocally related to the metric or to the usual basis of interaction
monomials (scalars built from polynomials in the curvature tensors, for 
instance) the condition that the theory should allow 
for predictions in terms of observables is only 
indirectly related to the total number of relevant couplings. 

In summary, the interplay between the microscopic action,
its parameterization through essential or relevant couplings, 
and observables is considerably more subtle than in the 
presumed non-gravitational examples of asymptotically safe 
theories with a single coupling. The existence of an 
asymptotically safe functional measure in the above sense seems 
to be neither necessary nor sufficient for a physically viable theory 
of Quantum Gravidynamics. This leads to our final working 
definition.      

{\bf Asymptotically safe Quantum Gravidynamics}: A quantum theory 
of gravity based on a notion of an asymptotically safe functional 
integral measure which incorporates the interplay between couplings 
and observables described above. In brief: (i) the choice of 
couplings has to be based on observables; this will pin 
down the physically relevant notion of positivity/unitarity. 
(ii) the number of essential or relevant couplings is not 
a-priori finite. (iii) what matters is not so much the dimension 
of the unstable manifold than how observables depend on the 
relevant couplings.

\subsection{Discussion of possible objections} 
 
Here we discuss some of the possible objections to a 
physically viable theory of Quantum Gravidynamics. 

\begin{itemize}
\item[{\bf Q1}] Since the microscopic action is likely to contain 
higher derivative terms don't the problems with non-unitarity 
notorious in higher derivative gravity theories reappear? 
\item[{\bf A1}] 
In brief, the unitarity issue has not much been 
investigated so far, but the presumed answer is No. 

First, the problems with perturbatively strictly renormalizable 
higher derivative theories stem mostly from the $1/p^4$-type 
propagator used. The alternative perturbative framework
already mentioned, namely to use a $1/p^2$-type propagator 
at the expense of infinitely many essential (potentially `unsafe')
couplings avoids this problem \cite{GomisWeinb, anselmi1}.
The example of the $2\!+\!2$ reduction shows that the reconcilation 
of safe couplings with the absence of unphysical propagating modes 
can be achieved in principle. Also the superrenormalizable gravity 
theories with unitary propagators proposed in~\cite{Tomboulis3} are 
intriguing in this respect. 

Second, even for higher derivative theories on flat space a well-defined 
Euclidean functional integral can exist, free of negative norm states or 
negative probabilities \cite{Lghosts1}. Physical unitarity is then 
thought to be restored at low energies, in which case one could 
`live with' higher derivative ghosts. The same would presumably hold 
for higher derivative theories on a fixed curved background. 

Third, when the background effective action is used as the central object 
to define the quantum theory, the `background' is {\it not} a solution of 
the classical field equations. Rather it is adjusted selfconsistenly 
by a condition involving the full quantum effective action. 
If the background effective action is computed nonperturbatively (by 
whatever technique) the intrinsic notion of unitarity will 
not be related to the `propagator unitarity' around a solution 
of the classical field equations in any simple way. 

One aspect of this intrinsic positivity is the convexity 
of the background effective action. In the flow equation 
for the effective average action one can see, for example, that 
the wrong-sign of the propagator is not an issue: if $\Gamma_k$ 
is of the $R + R^2$ type, the running 
inverse propagator $\Gamma_k^{(2)}$ when expanded around 
flat space has ghosts similar to those in perturbation theory.
For the $\Gamma_k$ flow, however, this is irrelevant
since in the derivation of the beta functions no background needs to 
be specified explicitly. All one needs is that the RG trajectories
are well defined down to $k=0$. This requires that 
$\Gamma_k^{(2)} + \cR_k$ is a positive operator for all $k$.
In the untruncated functional flow this is believed to be the case. 
A rather encouraging first result in this direction 
comes from the $R^2$ truncation~\cite{LR2}.

More generally, the reservations towards higher derivative theories 
came from a loop expansion around flat space and near the perturbative 
Gaussian fixed point. In contrast in Quantum Gravidynamics one aims at 
constructing the continuum limit nonperturbatively at a different 
fixed point and with respect to a dynamically adjusted background. 
The status of unitarity and causality have then not even been 
explored in toy models.  

In the previous discussion we implicitly assumed that 
generic physical quantities are related in a rather simple way 
to the interaction monomials entering the microscopic action. 
For Dirac observables however this is clearly not the case.
Assuming that the physically correct notion of unitarity 
concerns such observables it is clear that the final word 
on unitarity issues can only be spoken once actual observables 
are understood. 
\end{itemize}

\begin{itemize}
\item[{\bf Q2}] Doesn't the very notion of renormalizability 
presuppose a length or momentum scale referring to a prescribed 
background spacetime? 
\item[{\bf A2}] In some sense a coarse graining procedure is 
a `background structure', not intrinsic to generic physical 
quantities, which is as inevitable as it is innocuous.   
However a fixed background spacetime is not needed in principle.
As sketched in the intoduction it is part of the physics premise of 
a functional integral based 
approach that there is a physically relevant distinction between 
coarse grained and fine grained geometries. Geometrically motivated 
and tested proposals for coarse graining operations are presently 
not available, but there is certainly no obstruction of principle. 
Since the background field formalism is well-tested one 
may for the time being define the coarse graining with 
respect to a dynamically adjusted background metric, as 
described in section 1.2. 
\end{itemize}

\begin{itemize}
\item[{\bf Q3}] Doesn't such a non-perturbative renormalizability 
scenario require a hidden enhanced symmetry? 
\item[{\bf A3}] Improved renormalizability properties around a given 
fixed point are indeed often rooted in symmetries. 
A good example is QCD in a lightfront formulation 
where gauge invariance is an `emergent phenomenon' occuring 
only after an infinite reduction of couplings~\cite{PerryWilson}. 
In the case of Quantum Gravidynamics, the symmetry in question 
would be one that becomes visible only around the non-Gaussian 
fixed point. If it  exists, its identification would constitute a 
breakthrough. From the Kadanoff--Wilson view of renormalization 
it is however the fixed point which is fundamental -- the 
enhanced symmetry properties are a consequence (see 
the notion of generalized symmetries in~\cite{Zimmermann,
  OehmeZimmermann}. 
\end{itemize}

\begin{itemize}
\item[{\bf Q4}] Shouldn't the proposed anti-screening be seen in 
perturbation theory? 
\item[{\bf A4}] Maybe maybe not. Presently no good criterion for 
antiscreening in this context is known. For the reasons explained
in section 1.1 it should not merely be identified with the sign of the 
dominant contribution to some beta function.  The answer will thus 
depend somewhat on the identification of 
the proper degrees of freedom and the quantity considered.  

In the literature quantum gravity corrections to the Newton potential have 
been considered in some detail. The result is always of the form 
$$
V(r) = - \frac{G m_1m_2}{r} \bigg[ 
1+ 3 \frac{G(m_1\! + \!m_2)}{r} \,+\,\zeta 
\frac{\hbar G}{r^2} \bigg]\,.
$$
Interpreted as a modification of Newton's constant 
$G(r) = G(1 + \zeta \hbar G/r^2)$, one sees that 
$\zeta >0$ roughly corresponds to screening and 
$\zeta <0$ to anti-screening behavior. 
The value of $\zeta$ is unambigously defined in 1-loop 
perturbation theory and is a genuine prediction of quantum 
gravity viewed as an effective field theory (as stressed by 
Donoghue). However $\zeta$ will depend on the precise 
definition of the nonrelativistic potential and there 
are various options for it. 

One is via the 
$2 \rightarrow 2$ scattering amplitude. The coefficient
$\zeta_{\mathrm{scatt}}$ was computed initially by Donoghue and later by 
Khriplovich--Kirilin; the result considered definite 
in~\cite{Bjerrum-BohrD} is $\zeta_{\mathrm{scatt}} = \frac{41}{10\pi}$. 
It decomposes into a negative
vertex and triangle contributions $\zeta_v = -\frac{105}{3\pi}$,
and a just slightly larger positive remainder $\zeta_{\mathrm{scatt}} - 
\zeta_v = \frac{117.3}{3\pi}$ coming from box, seagull, and vacuum 
polarization diagrams. 

Another option is to consider corrections to the Schwarzschild 
metric. Different sets of diagrams have been used for the 
definition~\cite{KhriKirilin,BDHreply} and affect the 
reparameterization (in-)dependence and other properties 
of the corrections. Both choices advocated lead to 
$\zeta_{\mathrm{metric}} < 0$, which amounts to anti-screening.

Let us also mention alternative definitions 
of an effective Newton potential via Wilson lines 
in Regge calculus~\cite{HamberN} or by resummation 
of scalar matter loops~\cite{Ward}. The latter gives rise to
an ``antiscreening'' Yukawa type correction of the form 
$V(r) = - \frac{G}{r}(1 - e^{-r/\sqrt{\zeta G}})$, with $\zeta >0$. 
Via $V(r) = \int \frac{d^3k}{(2\pi)^3} e^{i \vec{k} \cdot 
\vec{x}} G(k)/\vec{k}^2$ it can be interpreted as 
a running Newton constant $G(k) = G/(1 + \zeta \vec{k}^2)$. 
Finally, in higher derivative theories of the form 
(\ref{iHD}) Yukawa type corrections already occurs 
at tree level, e.g.~$V(r) = -\frac{G}{r}(1 - \frac{4}{3} 
e^{-r\sqrt{s/(16 \pi G)}})$, for $\om =\th =0$  \cite{StelleCHD}. 
\end{itemize}

\begin{itemize}
\item[{\bf Q5}] There are several thought experiments suggesting 
a fundamental limit to giving an operational meaning to spacetime 
resolutions, for example via generalized uncertainty 
relations of the form (\cite{Garay, Padmanabhan, Mead} and references 
therein)
$$ 
\Delta x \approx \frac{\hbar}{\Delta p} + \frac{G \hbar}{c^3}\,
\frac{\Delta p}{\hbar}\,. 
$$ 
These relations are sometimes taken as hinting at a 
``fundamental discretum''. If so, doesn't this contradict the asymptotic 
safety scenario, where in the fixed point regime the microscopic spacetimes 
become selfsimilar?

\item[{\bf A5}] No, the arguments assume that Newton's constant 
$G$ is constant. (We momentarily write $G$ for $G_N$ in $(16 \pi G_N)^{-1}
\int dx \sqrt{g} R(g)$.) 
If $G$ is treated as a running coupling the 
derivations of the uncertainty relations break down. 
As an example consider a photon-electron scattering 
process as in~\cite{Mead, Padmanabhan}: $G$ refers to gravity in the 
(`photon' $k$ -- `electron' $\Delta p$) interaction
region with a pointlike `electron'. If viewed as running 
one expects $G(k) \approx G_*/k^2$ in the fixed point 
regime. Hence in the above relation one should replace $G$ by 
$\frac{G_* \,c^3}{(\Delta p)^2}$. This gives 
$$ 
\Delta x \Delta p \approx \hbar (1 + G_*)\,,
$$ 
and there is no limit on the spatial resolution. 
One can of course decide to choose units in which $G$ is 
constant by definition (see~\cite{PercacciPN}) in which case 
the derivations go through. 
Our conclusion is that the perceived dichotomy between 
a fundamentally `discrete' versus `continuum' 
geometry may itself not be fundamental. 
\end{itemize}

Each of the issues raised clearly deserves much further investigation. 
For the time being we conclude however that the asymptotic 
safety scenario is conceptually selfconsistent.

\newpage
\newsection{The role of perturbation theory} 

The application of renormalization group ideas to quantum 
gravity has a long history. In the following we focus 
on the role of perturbation theory for two reasons. First, 
because arguably the best non-perturbative implementation 
of a gravitational renormalization group still remains to found.
Second, because nonperturbative techniques are typically 
ill-adapted to make definite statements about the UV regime --
the regime under scrutiny here. In contrast, 
UV renormalized perturbation theory (PT), provided it is indicative 
for the behavior of an underlying `exact theory'  at all, is 
tailor-made for the investigation of the UV regime. 

Technically the key advantage of PT is that the 
UV cutoff can strictly be removed, termwise to 
all loop orders in a perturbatively (strictly or weakly) 
renormalizable field theory. The price to pay for 
this advantage is that the series is at best asymptotic 
to the (unknown) exact answer supposed to exist. 
Indeed, whether or not an underlying `exact theory' exists 
can then only be assessed in  terms of a plausibility criterion 
involving the perturbative beta functions of the theory; 
see Appendix A for a discussion.   

Despite the perturbative nonrenormalizability of the 
Einstein-Hilbert action there are two instructive ways 
in which strict or weak renormalizability can be achieved 
perturbatively in higher derivative theories. 
By higher derivative theories we mean here gravitational theories 
whose bare action contains, in addition to the Einstein-Hilbert term, 
scalars built from powers of the Riemann tensor and its covariant 
derivatives. In overview there are two distinct perturbative 
treatments of such theories, which will be surveyed in sections
2.2 and 2.3, respectively.  

The first one, initiated by Stelle \cite{Stelle1}, uses $1/p^4$ type 
propagators (in four dimensions) in which case a higher derivative 
action containing all (three)  quartic derivative terms can be expected 
to be power counting renormalizable. In this case strict 
renormalizability with only $4$ (or $5$, if Newton's constant is 
included) couplings can be achieved \cite{Stelle1}. However the 
$1/p^4$ type propagators are problematic from the point of view of unitarity. 

An alternative perturbative treatment of higher derivative theories
was first advocated by Gomis and Weinberg \cite{GomisWeinb}. 
The idea is try to maintain a $1/p^2$ type propagator and 
include all (infinitely many) counterterms generated in the bare action. 
Consistency requires that quadratic counterterms (those which 
contribute to the propagator) can be absorbed by field 
redefinitions. As verified in \cite{anselmi1} this is the case 
either in the absence of a cosmological constant term or when the 
background spacetime admits a metric with constant curvature.


\newsubsection{Does Newton's constant run?}

Before turning to 
renormalization aspects proper, let us describe the special role of 
Newton's constant in a diffeomorphism invariant theory with a dynamical 
metric. Let $S[g,{\mathrm{matter}}]$ be any local action,
where $g = (g_{\alpha\beta})_{1 \leq \alpha ,\beta \leq d}$ is the metric
and the ``matter'' fields are not scaled when the metric is. 
Scale changes in the metric then give rise to a variation of 
the Lagrangian which vanishes on shell:
\be 
\frac{d}{d \om^2} S[\om^2 g,{\mathrm{matter}}]\Big|_{\om =1} = 
\int\! dx \sqrt{g} \, g_{\alpha\beta}
\frac{\delta S[g,{\mathrm{matter}}]}{\delta g_{\alpha\beta}}\,.
\label{n1}
\end{equation}
As a consequence one of the coupling parameters which in the absence 
of gravity would be {\it essential} (i.e.\ a genuine coupling) becomes
{\it inessential} (i.e.\ can be changed at will by a redefinition of the 
fields). The running of this parameter, like that of a wave function 
renormalization constant, has no direct significance. 
If the pure gravity part contains the usual 
Ricci scalar term $Z_N\sqrt{g} R(g)$ the parameter that becomes 
inessential may be taken as its prefactor, i.e.\ may be identified 
with the inverse of Newton's constant, via 
\be 
Z_N^{-1}  = 2 \frac{d\!-\!2}{d\!-\!3} {\rm Vol}(S^{d-2}) \,
G_{\rm Newton} =: c_d \,G_{\rm Newton}\,.
\label{nnorm}
\end{equation}
The normalization factor $c_d$, $d \geq 4$ \cite{Robinson}, is chosen such that 
the coefficient in the nonrelativistic force law, as computed from 
$Z_N\sqrt{g} R(g) + L_{\rm matter}$, equals $G_{\rm Newton} {\rm Vol}(S^{d-2})$. 
For $d=2,3$ a different normalization has to be adopted.

The physics interpretation of the inessential parameter $\omega$ is that 
it sets the absolute momentum or spectral scale. To see this we can 
think of $g_{\alpha\beta}$ as a reference metric in the background 
field formalism. For example for the spectral values $\nu(g)$ of 
the covariant Laplacian $\Delta_g$ associated with $g_{\alpha\beta}$
one has 
\be 
\nu(\om^2 g) = \om^{-2} \nu(g)\,,
\label{n2}
\end{equation}
since $\Delta_{\om^2 g} = \om^{-2} \Delta_g$. The spectral 
values play the role of a covariant momentum squared. Indeed, 
if the metric is taken dimensionless $\nu(g)$ carries dimension  
$2$ (since $\Delta_g$ does) and for a flat metric $g_{\alpha\beta} 
=\eta_{\alpha\beta}$ they reduce to $-\nu(\eta) = k^2$, for 
plane waves labeled by $k$. From (\ref{n2}) one sees that   
rescaling of the metric and rescaling of the spectral values 
amout to the same thing. Since the former parameter is inessential 
the latter is too. Hence in a theory with a dynamical metric 
the three (conceptually distinct) inessential parameters: overall 
scale of the metric $\om$, the inverse of Newton's constant $Z_N^{-1} 
= c_d G_{\rm Newton}$, and the overall normalization of the 
spectral/momentum values are in one-to-one correspondence. 
For definiteness we take Newton's constant as the variant under 
consideration.   

Being inessential the quantum field theoretical running  of 
$G_{\mathrm{Newton}}$ has significance only relative to the 
running coefficient of some reference operator. The most commonly 
used choice is a cosmological constant term $\tilde{\Lambda} 
\int\! dx \sqrt{g}$. Indeed 
\be 
G_{\mathrm{Newton}} \tilde{\Lambda}^{\frac{d-2}{d}} = 
{\rm const}\, \tau(\mu)^{2/d}\,,
\label{n3}
\end{equation}  
is dimensionless and invariant under constant rescalings of 
the metric \cite{Kawai2}. One usually switches to dimensionless 
parameters via 
\be 
c_d  G_{\mathrm{Newton}} = \mu^{2-d} {\rm g}_N(\mu) \,,\sspace 
\tilde{\Lambda} = 2 \mu^d \frac{\lb(\mu)}{{\rm g}_N(\mu)}\,,
\label{n4}
\end{equation}  
where $\mu$ is some dimension one parameter which will be taken as 
`renormalization group time'. The Einstein-Hilbert action then reads 
\be 
\frac{\mu^{d-2}}{{\rm g}_N(\mu)} \int\! dx \sqrt{g} [R(g) - 
2 \mu^2 \lb(\mu) ]\,.
\label{n5}
\end{equation}
Being dimensionless one expects the running of 
${\rm g}_N(\mu)$ and $\lb(\mu)$ to be governed by flow equations without
explicit $\mu$ dependence  
\be 
\mu \frac{\dd}{\dd \mu} {\rm g}_N = \gamma_{\rm g}({\rm g}_N,\lb)\,,
\sspace 
\mu \frac{\dd}{\dd \mu} {\rm \lb} = \beta_{\rm \lb}({\rm g}_N,\lb)\,.
\label{n6}
\end{equation}
For the essential parameter $\tau(\mu) = {\rm g}_N(\mu) \lb(\mu)^{(d-2)/2}$ 
obtained from (\ref{n3}) this gives 
\be 
\mu \frac{\dd}{\dd \mu} \tau = \tau \Big[ \frac{\gamma_{\rm g}}{\rm g_N}
+ \frac{d-2}{2} \frac{\beta_{\lb}}{\lb} \Big]\,.
\label{n7}
\end{equation}   
In the present context we assume $\tau$ to be asymptotically safe, i.e.
\be 
\sup_{\mu_0 \leq \mu \leq \infty} \tau(\mu) < \infty\,,\sspace 
\lim_{\mu \ra \infty} \tau(\mu) = \tau_* < \infty\,,
\label{n8}
\end{equation}   
where here $0 < \tau_* < \infty$. Given (\ref{n8}) there are two possibilities. 
First, the various scheme choices are such that the parameters ${\rm g}_N(\mu)$ 
and $\lb(\mu)$ are both nonsingular and approach finite values ${\rm g}_*$ and 
$\lb_*$ for $\mu \ra \infty$. Second, the scheme choices are such that one of them 
becomes singular and the other vanishes for  $\mu \ra \infty$.  
Usually the first possibility is chosen; then the ${\rm g}_N(\mu)$ 
flow defined by the first equation in (\ref{n6}) has all the 
properties required for an essential asymptotically safe 
coupling. This `nonsingular parametric representation' of the 
$\tau(\mu)$ coupling flow is advantageous for most purposes. 
   
The second possibility is realized when inserting a singular 
solution of the equation for ${\rm g}_N(\mu)$ into the equation 
for $\lb(\mu)$. This naturally occurs when working in Planck 
units. One makes use of the fact that an inessential parameter 
can be frozen at a prescribed value. Specifically fixing 
\be 
[G_{\mathrm{Newton}}]^{\frac{1}{d-2}} = M_{\mathrm{Pl}} \approx 1.4 
\times 10^{19}\; {\rm GeV}\,, 
\label{n9}
\end{equation}
amounts to working with Planck units \cite{PercacciPN}.  From (\ref{n4}) it then follows 
that 
\be 
{\rm g}_N(\mu) = c_d \Big( \frac{\mu}{M_{\rm Pl}} \Big)^{d-2}\,,\sspace  
(d-2)  c_d \Big( \frac{\mu}{M_{\rm Pl}} \Big)^{d-2} = 
\gamma_g({\rm g},\lb)\,.
\label{n10}
\end{equation}
We may assume that the second equation has a local solution 
${\rm g}_N(\mu) = f(\lb, \mu/M_{\rm Pl})$. Reinserted into the $\lb$ 
equation gives a flow equation 
\be 
\mu \frac{\dd}{\dd k} \lb(\mu) = \widetilde{\beta}_{\lb}(\lb, \mu/M_{\rm Pl})\,,
\label{n11}
\end{equation}
which now explicitly depends on $\mu$. Writing similarly $\tilde{\tau}_* :=
\tau_*(f(\lb, \mu/M_{\rm Pl}),\lb)$ the condition defining the $\tau(\mu)$ 
fixed point becomes 
\be 
\widetilde{\beta}_{\lb}\Big|_{\tilde{\tau}_*} = -2 \lb\,.
\label{n12}
\end{equation}
Both formulations are mathematically equivalent to the extent 
the inversion formula ${\rm g}_N(\mu) = f(\lb, \mu/M_{\rm Pl})$
is globally defined. For definiteness we considered here the 
cosmological constant term as a reference operator, but 
the principle clearly generalizes.  

\newsubsection{Is a non-Gaussian fixed point visible in PT?}

Here we present the perturbative treatment of higher derivative theories 
which renders them strictly renormalizable. Our proposed answer to 
the question raised is: Yes, in a setting which can be taken as 
indicative for the genuine renormalization flow.  
We write again $q_{\alpha\beta}$ for the metric entering the 
functional integral and use $(-,+ ,\ldots, +)$ signature metrics here.  

Initially in $d=4+ \eps$ dimensions we consider the general action 
containing up to four derivatives of the metric
\ba 
S \is  -\!\int\! dx \sqrt{q} \Big[ \tilde{\Lambda} - \frac{1}{c_d G_N} R
+ \frac{1}{2s} C^2 -\frac{\om}{3 s} R^2 + 
\frac{\th}{s} E\Big] 
\nonum
\is  -\!\int\! dx \sqrt{q} \Big[ \frac{1}{c_dG_N}(2\Lambda - R) 
+ z R^2 + y R_{\alpha\beta} R^{\alpha\beta} + x 
R_{\alpha\beta\gamma\delta} R^{\alpha\beta\gamma\delta} \Big]\,. 
\label{HD1}
\end{eqnarray}
Here $C^2$ is the square of the Weyl tensor, $E$ is the 
integrand of the Gauss-Bonnet term, and a total derivative term, 
$\nabla^2 R$, has been omitted. The parameterization of the coefficients 
by couplings $s, \om, \th$ is chosen for later convenience. 
The parameters in the second line are related to those in the first by 
$\Lambda = c_d  G_{\!N} \tilde{\Lambda}/2$ and 
\be
s\, x = \frac{1}{2} + \th\,,\quad 
s\, y = - \frac{2}{d-2} - 4\th\,,\quad 
s\, z = -\frac{\om}{3} + \th + \frac{1}{(d-1)(d-2)}\,.
\label{HD2}
\end{equation} 
In $d=4$ the Gauss-Bonnet term is negligible; however if
dimensional regularization is used, $d \neq 4$, it must be kept. 
For $d=3$ both $E$ and $C^2$ vanish. 

Expanding around flat space, $q_{\alpha\beta} = \eta_{\alpha\beta} 
+ f_{\alpha\beta}$, the quadratic part of the action reads 
\ba 
S_{\tilde{\Lambda}=0}^{(2)} &=& \int \! d^4 x \Big\{ 
\frac{1}{4} \dd^2 f^{\alpha\beta} \Big( \frac{1}{c_4 G_N} - 
\frac{1}{s} \dd^2 \Big) [P^{(2)} f]_{\alpha\beta} 
\nonum
&& + \,
\dd^2 f^{\alpha\beta} \Big(\!\! -\!\frac{1}{2c_4 G_N} + 
\frac{\om}{s} \dd^2 \Big) [P^{(0)} f]_{\alpha\beta} \Big\}\,. 
\label{HD3}
\end{eqnarray}
For simplicity we took $d=4$ here and omitted the cosmological 
constant term. Further $P^{(0)}$, $P^{(2)}$ are the projectors 
onto the spin $0,2$ parts of $f_{\alpha\beta}$, i.e.~
\be
P^{(0)}_{\alpha\beta,\gamma\delta} = \frac{1}{3} P_{\alpha\beta}
P_{\gamma\delta}\,,
\quad
P^{(2)}_{\alpha\beta,\gamma\delta} = 
\frac{1}{2}[P_{\alpha\gamma} P_{\beta\delta} +P_{\alpha\delta} P_{\beta\gamma}] 
- \frac{1}{3} P_{\alpha\beta} P_{\gamma\delta}\,,
\label{HD4}
\end{equation}
with $P_{\alpha\beta} = \eta_{\alpha\beta} - \dd^{-2} \dd_{\alpha}\dd_{\beta}$.
One sees that the signs in (\ref{HD3}) are such that for $s >0$ the spin 
two part gives rise to a positive definite Euclidean action. The corresponding 
part of the Euclidean propagator can be written as 
\be 
2 P_{\alpha \beta,\gamma\delta} \Big( 
\frac{1}{p^2} - \frac{1}{p^2 + s/(c_4 G_N)} \Big)\,,
\label{HD5}
\end{equation}
and thus for fixed (bare) $s/(c_4 G_N)$ displays the characteristic 
$+1/p^4$ behavior. The spin zero part does not necessarily enter (\ref{HD3}) 
with the `good' sign, but it will be affected by gauge fixing and measure
terms anyhow. See \cite{Stelle1,Buchetal} for a discussion.  

The perturbative quantization of (\ref{HD1}) proceeds as 
usual. Gauge fixing and ghost terms are added and the total 
action is expanded in powers of $f_{\alpha\beta} = q_{\alpha\beta} - 
\eta_{\alpha\beta}$. The one loop counter term (minus the divergent part of
the effective action) has been computed by a number of authors  
\cite{FradkinHD,AvramidiHD,avramidi,PeixetoHD,PercHD}, using different 
regularizations. 
The result of Avramidi and Barvinsky \cite{AvramidiHD,avramidi} 
in dimensional regularization has been confirmed in \cite{PeixetoHD} 
and in the present conventions reads 
\ba 
\Delta S^{(1)}\! &\!=\!&\! -\Gamma_{\rm div}^{(1)} = 
\frac{\mu^{d-4}}{(4\pi)^2 (d-4)} \int \! d^dx \sqrt{q} 
\bigg\{ 2 s (u(\om) - \gamma(\om))\,\tilde{\Lambda} + 
\Big( \frac{s}{c_d G_N} \Big)^2 \,\frac{1 + 20 \om^2}{8 \om^2} 
\nonumber\\[3mm]
&& + \gamma(\om) \frac{s}{c_d G_N} R 
+ \frac{133}{20} C^2 + \frac{5}{36}(1 + 12 \om + 8 \om^2) R^2 
+ \frac{196}{45} E
\bigg\}\,. 
\label{HDcounter}
\end{eqnarray}
Here $u(\om) = (-1 + 30 \om + 40 \om^2)/(12 \om)$   
is independent of the gauge fixing parameters, while 
$\gamma(\om)$ is gauge dependent. E.g.~in harmonic gauge one has 
$\gamma(\om) = (-1 + 10 \om^2)/(3\om)$, while in the 
Vilkovisky-deWitt effective action $\gamma(\om) = 
(-13 + 18 \om + 40\om^2)/(12 \om)$ enters. 

To absorb (\ref{HDcounter}) singular field redefinitions should be    
taken into account. By inspection of the equations of 
motion operator  for (\ref{HD1}) one sees that only 
field redefinitions proportional to $q_{\alpha\beta}$   
are of immediate use. We thus take 
\be 
q^{\rm B}_{\alpha\beta} = q_{\alpha\beta} + 
\frac{1}{(4\pi)^2 (d-4)} \xi\, q_{\alpha\beta} \,,
\label{HD6}
\end{equation}  
where $\xi$ can be a function of 
$s/{\rm g}_N, \lb, \om, \th$. The 1-loop flow equations for 
$s,\om,\th$ obtained from (\ref{HDcounter}) are universal 
\ba 
(4\pi)^2 \mu \frac{d}{d\mu} s  \is - \frac{133}{10} s^2 \,, 
\nonum
(4\pi)^2 \mu \frac{d}{d\mu} \om  \is - 
\frac{25 + 1098 \om + 200 \om^2}{60}s \,, 
\nonum
(4\pi)^2 \mu \frac{d}{d\mu} \th  \is \frac{7(56 - 171 \th)}{90}s\,. 
\label{HD7}
\end{eqnarray}  
These equations have a trivial fixed point $s_* =0, 
\,\om_* = {\rm const},\, \th_* = {\rm const}$, and a nontrivial 
fixed point $s_* = 0,\, \om_* = -(549 \pm 7 \sqrt{6049})/200,\,
\th_* = 56/171$. Importantly the $C^2$ coupling $s$ is asymptotically free.

To describe the flow of the Newton and cosmological constants 
one switches to the dimensionless parameters ${\rm g}_N$ and $\lb$ 
as in Section 2.1. From (\ref{HDcounter}) and (\ref{HD6}) one finds 
\ba 
\mu \frac{d}{d\mu} {\rm g}_N \is 2 {\rm g}_N + 
\frac{1}{(4\pi)^2} s {\rm g}_N [\gamma(\om) + \dd_s \xi]\,,
\nonum
\mu \frac{d}{d\mu} \lb \is -2 \lb + 
\frac{1}{(4\pi)^2} s \lb \Big[2 u(\om) - \gamma(\om) - \dd_s \xi 
+ \frac{s}{{\rm g}_N \lb} \frac{1 + 20 \om^2}{8 \om^2} \Big]\,.
\label{HD8}
\end{eqnarray}   
The $({\rm g}_N,\lb)$  flow is highly non-universal but the flow of the scale 
invariant combination $\tau = {\rm g}_N \lb$ 
\be 
(4\pi)^2 \mu \frac{d}{d\mu} \tau = 2 s \tau\, u(\om)  + 
\frac{s^2}{8 \om^2} (1 + 20 \om^2)\,,
\label{HD9}
\end{equation} 
is independent of the gauge fixing parameters and $\xi$.  
Note that in principle one is free to freeze the ${\rm g}_N$ 
evolution by taking $\xi = -s \gamma(\om)$ and work with Planck 
units. As discussed in section 2.1 this amounts to using a potentially 
singular parametric representation of the $\tau$ flow and we prefer 
to let both ${\rm g}_N$ and $\lb$ evolve, e.g.~with $\xi =0$. 

The flow equations (\ref{HD8}), (\ref{HD9}) are also scheme 
dependent. The scheme dependence turns out to be crucial in 
the present context, so we discuss it in slightly more detail 
following \cite{MNnotes}. We momentarily display the loop counting 
parameter $\hbar$ (not to be identified with Planck's constant here) 
and make use of the fact that changes of scheme correspond to finite 
redefinitions of the renormalized couplings compatible with the 
$\hbar$ grading. One readily sees that $\hbar$ and the dimensionless 
couplings should enter only in the combinations $\hbar s, \hbar {\rm g}_N,      
\lb, \om ,\th$. For simplicity we assume that $\om$ and $\th$ 
couplings are set to their values $\om_*, \th_*$ at their 
nontrivial fixed point. Taking $s$ as the reference coupling the 
the most general $O(\hbar)$ redefinitions compatible with the 
above grading then are 
\ba 
\tilde{\rm g}_N \is {\rm g}_N + \hbar(c_1 {\rm g}_N^2 + c_2 {\rm g}_N \,s
+ c_3 s^2) \,,
\nonum
\tilde{\lb} \is \lb + \hbar (d_1 {\rm g}_N + d_2 s)\,, 
\label{HD10}
\end{eqnarray} 
where $c_1,c_2,c_3,d_1,d_2$ are functions (powerseries) of $s/{\rm g}_N$, 
$\lb$ and $\om_*,\th_*$. The flow equations for 
the new couplings are readily worked out, after which one can restore 
$\hbar =1$. At $s = s_* =0$ the new flow equations simplify to 
\ba 
\mu \frac{d}{d\mu} \tilde{\rm g}_N \is 2 \tilde{\rm g}_N + 
2 \tilde{\rm g}_N^2 c_1\,,
\nonum
\mu \frac{d}{d\mu} \tilde{\lb} \is -2 \tilde{\lb} + 
2 \tilde{\rm g}_N \Big(  2 - \tilde{\lb} \frac{\dd}{\dd \tilde{\lb}}
\Big) 
d_1\,.
\label{HD11}
\end{eqnarray}   
The point at issue now is that these equations have a nontrivial 
fixed point for $\tilde{\rm g}_N$ whenever $c_1 \neq 0$ and generically 
also one in $\tilde{\lb}$: 
\be 
\tilde{\rm g}_N^* = -\frac{1}{c_1} \,,\sspace
\tilde{\lb}^* =0 \quad \mbox{or} \quad \tilde{\lb}^2 
\frac{\dd}{\dd \tilde{\lb}} (\tilde{\lb}^{-2} d_1) 
\Big|_{\tilde{\lb} = \tilde{\lb}^*} =
c_1 \neq 0\,.
\label{HD12}
\end{equation}
The rationale for identifying a non-Gaussian fixed point in perturbation 
theory is explained in Appendix A1. 

There are several reasons why in the specific context here the 
flow equations (\ref{HD11}) rather than (\ref{HD8}) should be considered as 
indicative for the genuine $({\rm g}_N,\lb)$ flow: (i) Dimensional 
regularization suppresses powerlike divergencies in the UV cutoff,
in particular singular contributions from the measure are not 
taken into account. On the other hand the proper treatment of measure 
terms is  essential for the perturbative cure of the 
conformal factor instability \cite{Mazur}. 
(ii) Dimensional regularization has no nonperturbative counterpart, 
so the matching to a nonperturbatively defined $({\rm g}_N,\lb)$ flow 
should be done for the flow (\ref{HD11}) with nontrivial $c_1, \,d_1$. 
(iii) A recent reevaluation of $\Gamma_{\rm div}^{(1)}$ using tabulated 
heat kernels asymptotics \cite{PercHD} confirmed the presence of 
powerlike singularities 
leading to additional terms in the flow equations (\ref{HD8}). These 
can be checked \cite{MNnotes} to correspond to a change of scheme 
(\ref{HD10}) with suitable $c_1 = c_1(\om),\, d_1 = d_{10}(\om) + 
d_{11}(\om) \tilde{\lb}$, $d_2 = d_2(\om), c_2 = c_3 =0$.  
(iv) Upon expanding a $({\rm g}_N,\lb)$ flow defined through 
the truncated effective average action in powers of ${\rm g}_N$ 
a match to the perturbative 1-loop flow should be found. 
This is indeed the case for (\ref{HD11}) with nontrivial $c_1, d_1$.

Let us briefly elaborate on point (iv). In the so-called Einstein-Hilbert 
truncation using an optimed cutoff and a limiting version of the 
gauge-fixing parameter, the `beta' functions $\gamma_g,\,\beta_{\lb}$ 
reduce to ratios of polynomials in ${\rm g}_N,\,\lb$,
see Eq.~(\ref{litimflow}). Upon expansion to one finds 
\ba 
\mu \frac{d}{d \mu} {\rm g}_N \is 
2 {\rm g}_N - \frac{1}{(4\pi)^2} {\rm g}_N^2 + O({\rm g}_N^2 \lb)\,,
\nonum
\mu \frac{d}{d \mu} \lb \is 
-2 \lb  + \frac{1}{(4\pi)^2} \frac{{\rm g}_N}{2}( 1 + 2 \lb) + O({\rm g}_N^2)\,.
\label{litimlin}
\end{eqnarray}
This is of the form (\ref{HD11}) with $(4\pi)^2 c_1 = -1/2,\, 
(4\pi)^2 d_1 = 1/8 + \tilde{\lb}/2$, with the ensued nontrivial 
fixed point (\ref{HD12}).

The flow equations  (\ref{n7}), (\ref{HD5}) of course also admit the 
Gaussian fixed point ${\rm g}_N^* = 0 = \lb_*$, and one may be tempted 
to identify the `realm' of perturbation theory (PT) with the `expansion' 
around a Gaussian fixed point. As explained in Appendix A1, however, in a 
theory with several couplings the 
conceptual status of PT referring to a non-Gaussian fixed point is not 
significantly different from that referring to a Gaussian fixed point. 
In other words there is no reason to take the perturbative non-Gaussian 
fixed point (\ref{HD12}) any less serious than the perturbative Gaussian one. 
This point is  also relevant in the framework of the $2+2$ reduction, 
where a non-Gaussian fixed point is also identified by perturbative means.

The fact that a non-Gaussian fixed point can already be identified in PT 
is important for several reasons. First, it provides an important 
consistency check on the scenario. It is only in PT that the ultraviolet 
cutoff can strictly be removed; granting the usual assumption that 
it is asymptotic to the exact result, the putative non-Gaussian 
fixed point should be visible already in first order PT -- and it is!    
Second, although the value of ${\rm g}_N^*$ 
in (\ref{HD6}) is non-universal, the anomalous dimension 
$\eta_N = \gamma_g/{\rm g}_N -2$ is exactly $-2$ at the fixed point 
(\ref{HD6}). The general argument for the dimensional reduction  
of the residual interactions outlined after Eq.~(\ref{N2}) can thus 
already be based on PT alone! Together the result (\ref{HD6}) 
suggests that the interplay between the perturbative and the nonperturbative 
dynamics might be similar to that of non-abelian gauge theories, 
where the nonperturbative dynamics is qualitatively and quantitatively 
important mostly in the infrared.

In summary there is good evidence for the existence of 
a non-Gaussian fixed point in higher derivative gravity theories. 
Here we limited the discussion to the one loop level, the 
qualitative aspects should be unaffected by loop corrections,
however. Indeed, the theories (\ref{HD1}) are thought to be 
strictly renormalizable to all loop orders \cite{Stelle1}. 
This is not trivially a consequence of the $1/p^4$ propagator 
(see \cite{Lghosts2} for counter examples) but arises through the 
interplay with diffeomorphism invariance.  By much of the same rationale 
that underlies the belief that Yang-Mills theories exist nonperturbatively one is 
then lead to the following conjecture (which, although not verbatim 
contained in \cite{Weinberg} is much along the same lines):  

Higher derivative gravity theories are renormalizable in $d=4$ 
in the Kadanoff-Wilson sense with finitely many asymptotically 
safe couplings based on a non-Gaussian fixed point.

\newsubsection{Can unitarity and renormalizability be reconciled?} 

The main drawback of the above renormalizable gravity 
theories is that the status of unitarity in them  is uncertain. 
In a perturbative formulation based on the free propagator 
obtained by linearizing (\ref{HD1}) one encounters unphysical 
propagating modes, see section 1.5 for a discussion of this issue.  
As already mentioned these problem are absent in an alternative 
perturbative formulation where a $1/p^2$ type propagator 
is used throughout \cite{GomisWeinb}. We now describe this construction in 
slightly more detail. See also \cite{Kreimer} for a recent alternative 
setting.

Starting from the $d=4$ Lagrangian $\frac{1}{c_d G_N} \sqrt{q} R(q)$ without 
cosmological constant the one-loop divergencies come out in dimensional 
regularization as \cite{tHVelt74}
\be 
\frac{\mu^{d-4}}{(4\pi)^2(d\!-\!4)} \sqrt{q} 
\Big( \frac{1}{60} R^2 + \frac{7}{10} R_{\alpha\beta} R^{\alpha\beta} \Big)\,. 
\label{HD13}
\end{equation}   
They can be removed in two different ways. One is by adding new couplings 
so that a higher derivative action of the form (\ref{HD1}) 
arises with parameters $\Lambda =0$, $x =0$, and 
\be 
z_B = \mu^{d-4} \Big( z + \frac{1}{(4 \pi)^2 (d\!-\!4)} \frac{1}{60} \Big) \,,\quad 
y_B = \mu^{d-4} \Big( y + \frac{1}{(4\pi)^2 (d\!-\!4)} \frac{7}{10} \Big) \,. 
\label{HD14}
\end{equation}
The renormalizability of the resulting theory is mostly due to the modified 
propagator which can be viewed as a resummed graviton propagator in a
power series in $z,y$. The unphysical singularities are of order $1/z,\,1/y$. 
The second option to remove (\ref{HD13}) is  by a singular field redefinition 
\be 
q_{\alpha\beta} \mapsto q_{\alpha\beta} + 
\frac{c_4 G_N}{ (4 \pi)^2 (d\!-\!4)} \frac{1}{10} \Big( - 7 R_{\alpha\beta} 
+ \frac{11}{3} g_{\alpha\beta} R\Big) \,.
\label{HD15}
\end{equation} 
This restores the original $\sqrt{q} R(q)$ Lagrangian up to two- and 
higher loop contributions. However this feature is specific to one loop.
As shown in \cite{Goroff,vandeVen} at two loops there is a divergence 
proportional to $R_{\alpha\beta}^{\;\;\;\gamma\delta} 
R^{\alpha\beta}_{\;\;\;\rho \sigma} 
R^{\rho \sigma}_{\;\;\;\gamma\delta}$, which cannot be absorbed by a 
field redefinition. A counterterm proportional to it must thus be added 
to $\sqrt{q} R(q)$. Importantly, when re-expanded in powers of 
$f_{\alpha\beta} = q_{\alpha\beta} - \delta_{\alpha\beta}$, this counterterm,
however, produces only terms quadratic in $f$ that are proportional to the Ricci 
tensor or the Ricci scalar. These can be removed by a covariant field redefinition,
so that the initial $1/p^2$ type propagator does not receive corrections. 
A simple argument \cite{anselmi1} shows that this property also holds for 
all higher order counterterms that can be expected to occur. Explicitly, 
consider a Lagrangian of the form 
\be 
L = \frac{1}{c_d G_N} \sqrt{q} R(q) + \sum_{i \geq 1} 
(c_d G_N)^{\frac{d_i}{2-d} -d} \;{\rm  g}_i\, L_i(q)\,,
\label{HD16}
\end{equation}
where $L_i(q)$ are local curvature invariants of mass dimension $-d_i$,
the ${\rm g}_i$ are dimensionless couplings and the power 
of $c_d G_N$ (with $c_d$ e.g.~as in (\ref{n3})) gives each term in the sum mass 
dimension $-d$. 

Let us briefly recap the 
power counting and scaling dimensions of local curvature 
invariants. These are integrals $P_i[g] = \int \! d^dx L_i(g)$ over 
densities $L_i(g)$ which are products of factors of the form 
$\nabla_{\alpha_1} \ldots \nabla_{\alpha_{l-4}} 
R_{\alpha_{l-3} \ldots \alpha_l}$, suitably contracted to get a scalar 
and then multiplied by $\sqrt{g}$. 
One easily checks $L_i(\om^2 g) = \om^{s_i} L_i(g)$, $\om >0$, with 
$s_i = d -2 r -s$, where $r$ is the total power of the Riemann tensor 
and $s$ is the (necessarily even) total number of covariant derivatives. 
This scaling dimension matches minus the mass dimension of $P_i(g)$ 
if $g$ is taken dimensionless. For the mass dimension $d_i$ of the 
associated coupling $u_i$ in a product $u_i P_i[g]$ one thus gets 
$d_i = s_i = d - 2 r - s$. For example, the three local invariants 
in (\ref{i5}) have mass dimensions $-d_0 = -d$, $-d_1 =-(d\!-\!2)$, 
$-d_2 = -(d\!-\!4)$, respectively. There are three other local invariants
with mass dimension $-(d\!-\!4)$, namely the ones with integrands 
$C^2 = R^{\alpha\beta\gamma\delta} R_{\alpha\beta\gamma\delta} - 
2 R^{\alpha\beta} R_{\alpha\beta} + R^2/3$ (the square of the Weyl tensor),
$E = R^{\alpha\beta\gamma\delta} R_{\alpha\beta\gamma\delta} -  
4 R^{\alpha\beta} R_{\alpha\beta} + R^2$ (the generalized Euler density), 
and $\nabla^2 R$. 
Then there is a set of dimension $-(d\!-\!6)$ local invariants, and so on. 
Note that in $d=4$ the integrands of the last two of the dimensionless 
invariants are total divergencies so that in $d=4$ there are only $4$ local 
invariants with non-positive mass dimension; see (\ref{HD1}).  

A generic term in $P_i$ will be symbolically of the form $\nabla^s R^r$,
where all possible contractions of the $4r + s$ indices may occur. Since 
the Ricci tensor is schematically of the form $R = \nabla^2 f + O(f^2)$,
the piece in $P_i$ quadratic in $f$ is of the form $\nabla^{s+4} R^{r-2} f^2$. 
The coefficient of $f^2$ is a tensor with 4 free indices and one 
can verify by inspection that the possible index contractions are 
such that the Ricci tensor or Ricci scalar either occurs directly,  
or after using the contracted Bianchi identity.  
In summary, one may restrict the sum in (\ref{HD10}) to terms with 
$-d_i = -d + 2r +s$,  $r \geq 3$, and the propagator derived from it 
will remain of the $1/p^2$ type to all loop orders. This suggests 
that (\ref{HD10}) will give rise to a renormalizable Lagrangian.
A proof requires to show that after gauge fixing and ghost terms 
have been included all counter terms can be chosen local and covariant
and has been given in \cite{GomisWeinb}.

In the terminology of section 1.4 the above results then show 
the existence of a ``weakly renormalizable''  but ``propagator unitary'' 
Quantum Gravidynamics based on a perturbative Gaussian fixed point. 
The beta functions for this infinite set of 
couplings are presently unknown. If they were known, expectations are that
at least a subset of the couplings would blow up at some finite 
momentum scale $\mu = \mu_{\mathrm{term}}$ and would be unphysical for 
$\mu > \mu_{\mathrm{term}}$. In this case the computed results for 
physical quantities are likely to blow up likewise at some (high) energy 
scale $\mu = \mu_{\mathrm{term}}$. In other words the couplings 
in (\ref{HD10}) are presumably not all asymptotically safe.

Let us add a brief comment on the relevant-irrelevant distinction 
in this context, if only to point out that it is no longer useful. 
Recall from Section 1.3 that the notion of a relevant 
or irrelevant coupling applies even to flow lines {\it not} connected 
to a fixed point. This is the situation here. All but a few of the 
interaction monomials in (\ref{HD10}) are powercounting irrelevant 
with respect to the $1/p^2$ propagator. Equivalently   
all but a few couplings $u_i(\mu) = \mu^{d_i} {\rm g}_i(\mu)$ have 
non-negative mass dimensions $d_i \geq 0$. These are the only ones not 
irrelevant with respect to the stability matrix computed at the 
perturbative Gaussian fixed point. However in (\ref{HD10}) these 
power counting irrelevant couplings with $d_i < 0$ are crucial for 
the absorption of infinities and thus are converted into practically 
relevant ones. In the context of (\ref{HD10}) we shall therefore 
discontinue to use the terms relevant/irrelevant. 
\medskip

Comparing the two perturbative constructions in section 
2.2 and 2.3 one arrives at the conclusion anticipated in 
section 1.1: the challenge of Quantum Gravidynamics lies not so 
much in achieving renormalizability, but to reconcile asymptotically 
safe couplings with the absence of unphysical propagating modes. 
As surveyed in section 1.3c this program is realized for the $2+2$ 
reduction \cite{PTernst,Sernst}. Within the framework of truncated 
flow equations (see section 1.3d and A2 in section 1.5 here) the 
results for the $R + R^2$ type truncation likewise are compatible 
with the absence of unphysical propagating modes.

\newsection{Dimensional reduction of residual interactions in UV} 

In order to realize this program without reductions or truncations 
a mathematically controllable nonperturbative definition of 
Quantum Gravidynamics is needed. Within a functional integral formulation 
this involves the following main steps: definition of a kinematical 
measure, setting up a coarse graining flow for the dynamical measures 
and then probing its asymptotic safety. This is probably best done 
in a discretized setting. However an important qualitative 
feature of an asymptotically safe functional integral can be inferred 
without actually evaluating it, namely that in the extreme 
ultraviolet the residual interactions appear two-dimensional.  
There are a number of interconnected heuristic arguments 
for this phenomenon which we present here.

{\bf (a) Scaling of fixed point action:} Consider a candidate for a 
quasilocal microscopic action 
\be 
S_k[q] = \sum_i u_i(k) P_i[q]\,,
\label{gren1}\
\end{equation}
where the $u_i(k)$ are running couplings of mass dimension $d_i$ and 
$P_i[q]$ are local invariants of mass dimension $-d_i$. 
By quasilocal we mean here that the sum may be infinite 
and off hand arbitrarily high derivative terms may occur. 
For example such an action arises in the perturbative framework 
described in section 2.3. When viewed
as a renormalized action perturbatively defined in the above 
sense (with the UV cutoff strictly removed) the running of the 
$u^{\mathrm{PT}}_i(k)$ is unknown but expectations are that 
${\mathrm{g}}^{\mathrm{PT}}_i(k) = k^{-d_i} u^{\mathrm{PT}}_i(k)$ 
are not uniformly bounded functions in $k$, the dimensionless couplings 
then are not asymptotically safe but blow up at various ($i$-dependent) 
intermediate scales. The situation is drastically different if 
all the couplings are assumed to be asymptotically safe. Then $u_i(k) = 
{\mathrm{g}}_i(k) k^{d_i} \sim {\mathrm{g}}_i^* k^{d_i}$ as $k \ra \infty$ 
and if one uses the fact that $s_i = d_i$ (see the discussion after 
Eq.~(\ref{HD10}) for all local invariants one gets 
\be 
S_k[q] \sim \sum_i {\mathrm{g}}_i^* P_i[k^2 q] 
= S_*[k^2 q]\,,
\label{gren2}
\end{equation}
for $k \ra \infty$, with $S_*[q] = \sum_i {\mathrm{g}}_i^* S_i[q]$ the 
candidate fixed point action. The overall scale of the metric is 
an inessential parameter, see section 2.1, and a fixed point action always 
refers to an equivalence class modulo possibly running inessential 
parameters. 

One sees that in the fixed point regime ${\mathrm{g}}_i(k) \sim 
{\mathrm{g}}_i^*$ the $k$-dependence enters only through the combination 
$k^2 g_{\alpha\beta}$, a kind of selfsimilarity. This simple 
but momentous fact eventually underlies all the subsequent arguments.   
It is `as if' in the fixed point regime only a rescaled metric 
$\tilde{q}_{\alpha\beta} = k^2 q_{\alpha\beta}$ entered which carries 
dimension two. 
This has consequences for the `effective dimensionality' of Newton's 
constant: recall that conventionally the Ricci scalar term, 
$\int \! dx\, \sqrt{q} R(q)$, has dimension $2-d$ in $d$ dimensions. 
Upon substitution $q_{\alpha \beta} \mapsto \tilde{q}_{\alpha\beta}$ one 
quickly verifies that $\int \! dx \sqrt{\tilde{q}} R(\tilde{q})$ 
is dimensionless. Its prefactor, i.e.~the inverse of Newton's constant, 
then can be taken dimensionless -- as it is in two dimensions. 
Compared to the infrared regime it looks `as if' Newton's 
constant changed its effective dimensionality from $d-2$ to zero,
i.e.~at the fixed point there must be a large anomalous dimension 
$\eta_N = 2-d$. 

Formally what is special about the Einstein--Hilbert term is that 
the kinetic (second derivative) term itself carries a 
dimensionful coupling. To avoid the above conclusion one might 
try to assign the metric a mass dimension $2$ from the beginning 
(i.e.~not just in the asymptotic regime). However this would merely 
shift the effect from the gravity to the matter sector, as we 
wish to argue now. 

In addition to the dimensionful metric $\tilde{q}_{\alpha\beta} := 
k^2 q_{\alpha\beta}$, we introduce a dimensionful vielbein by 
$\tilde{E}_{\alpha}^{\;m} := k E_{\alpha}^m$, if $q_{\alpha\beta} = 
E_{\alpha}^{\;\;m} E_{\beta}^{\;\;n} \eta_{mn}$ is the 
dimensionless metric. With respect to a dimensionless metric 
$\int \! dx\, \sqrt{q} R(q)$ has mass dimension $2-d$ in $d$ dimensions, 
while the mass dimensions $d_{\chi}$ of a Bose field $\chi$ and that 
$d_{\psi}$ of a Fermi field $\psi$ are set such that their 
kinetic terms are dimensionless, i.e.\ $d_{\chi} = (d\!-\!2)/2$ 
and $d_{\psi} = (d\!-\!1)/2$. Upon substitution 
$q_{\alpha \beta} \mapsto \tilde{q}_{\alpha\beta}$ the 
gravity part $\int \! dx \sqrt{\tilde{q}} R(\tilde{q})$ 
becomes dimensionless, while the kinetic terms of a Bose 
and Fermi field pick up a mass dimension of $d\!-\!2$ and $d\!-\!1$, 
respectively. This means their wave function renormalization 
constants $Z_{\chi}(k)$ and $Z_{\psi}(k)$ are now dimensionful 
and should be written in terms of dimensionless parameters 
as $Z_{\chi}(k) = k^{d-2}/{\mathrm{g}}_{\chi}(k)$ and 
$Z_{\psi}(k) = k^{d-1}/{\mathrm{g}}_{\psi}(k)$, say. 
For the dimensionless parameters one expects finite 
limit values $\lim_{k \ra \infty} {\mathrm{g}}_{\chi}(k) = {\mathrm{g}}_{\chi}^*>0$ 
and $\lim_{k \ra \infty} {\mathrm{g}}_{\psi}(k) = {\mathrm{g}}_{\psi}^*>0$,
since otherwise the corresponding (free) field would 
simply decouple. Defining the anomalous dimension as 
usual $\eta_{\chi} = - k \dd_k \ln Z_{\chi}$ 
and $\eta_{\psi} = - k \dd_k \ln Z_{\psi}$, the argument 
presented after Equation~(\ref{N2}) can be repeated and gives
that $\eta_{\chi}^* = 2\!-\!d$, $\eta_{\psi}^* = 1\!-\!d$ for the 
fixed point values, respectively. The original large momentum 
behavior $1/p^2$ for bosons and $1/p$ for fermions is thus modified to 
a $1/p^d$ behavior in the fixed point regime, in both cases. 

This translates into a logarithmic short distance behavior 
which is universal for all (free) matter.  Initially the propagators 
used here should be viewed as ``test propagators'', in the sense that 
one transplants the information in the $\eta$'s derived from the gravitational 
functional integral into a conventional propagator on a (flat or curved) 
background spacetime. Since the short distance asymptotics is 
the same on any (flat or curved) reference spacetime, one can plausibly 
convert this into a prediction for a genuine quantum gravity 
correlator: 

Consider as in \cite{Smit} a geodesic two-point correlator of a scalar field 
\be 
G(R) = \int\! \cD q \cD \phi \, e^{i S[q,\phi]} \int \!
dx dy \sqrt{q(x)} \sqrt{q(y)}\, \phi(x) \phi(y) \,\delta(\Sigma_q(x,y) -R)\,,
\label{geocorr}
\end{equation}
where $\Sigma_q(x,y)$ is the minimal geodesic distance between the 
points $x$ and $y$. The first integral is the heuristic geometry and 
matter functional integral, all configurations are taken into 
account which produce the given geodesic distance $R$. 
The previous considerations then lead to the prediction that 
if (\ref{geocorr}) is based on an asymptotically safe functional measure
a {\it logarithmic} (powers of $\log R$ and $\log(-\log R)$) behavior 
for $R \ra 0$ is expected.

On the other hand the universality of the logarithmic short distance 
behavior in the matter propagators also justifies to attribute the 
phenomenon to a modification in the underlying random geometry,
a kind of ``quantum equivalence principle''.


{\bf (b) Anomalous dimension at non-Gaussian fixed point:} 
The ``anomalous dimension argument'' has already been sketched 
in the introduction, see also \cite{LR1}. Here we present a few more 
details and relate it to (a). 

Suppose again that the unkown microscopic action of Quantum Gravidynamics 
is quasilocal and reparameterization invariant. The only term containing 
second derivatives then is the familiar Einstein--Hilbert term, 
$Z_N \int \! dx \sqrt{q} R(q)$, of mass dimension $2\!-\!d$ in $d$ dimensions, 
if the metric is taken 
dimensionless. As explained in section 2.1 the dimensionful running prefactor 
multiplying it, $Z_N(k)$, ($N$ for ``Newton'') can be treated either as a wave 
function renormalization or as a quasi-essential dimensionless coupling 
${\rm g}_N$, where 
\be 
c_d G_{\rm N} = Z_N(k)^{-1} = {\mathrm{g}}_N(k) k^{2-d}\,.
\label{Dred1}
\end{equation}
Here we treat ${\rm g}_N$ as running in which case its running may also be 
affected by all the other couplings (gravitational and non-gravitational, made 
dimensionless by taking out a suitable power of $k$). The short distance behavior of the 
propagator will now be governed by the ``anomalous dimension''
$\eta_N = - k \dd_k \ln Z_N(k)$, by general field theoretical 
arguments. On the other hand the flow equation for ${\mathrm{g}}_N$ 
can be expressed in terms of $\eta_N$ as 
$k \dd_k {\mathrm{g}}_N = [d-2 + \eta({\mathrm{g}}_N,{\mathrm{other}})]\, 
{\mathrm{g}}_N$, where we schematically indicated the dependence on the other 
dimensionless couplings. {\it If} this flow equation now has a nontrivial fixed point 
$\infty > {\mathrm{g}}_N^* >0$, the only other way how the right hand side can 
vanish is for 
\be 
\eta_N({\mathrm{g}}_N^*, {\mathrm{other}}) = 2-d, 
\label{Dred2}
\end{equation}
irrespective of 
the detailed behavior of the other couplings as long as no blow-up occurs. 
This is a huge anomalous dimension. We can now transplant this anomalous 
dimension into a ``test graviton propagator'' on a flat background. 
The characteristic property of $\eta_N$ then is that it gives rise to a a high 
momentum behavior of the form $(p^2)^{-1 + \eta_N/2}$ modulo logarithms, 
or a short distance behavior of the form $(\sqrt{x^2})^{2-d -\eta_N}$ modulo 
logarithms. This follows from general field theoretical principles: 
a Callan-Symanzik equation for the effective action, the vanishing of 
the beta function at the fixed point and the decoupling of the low 
momentum modes. Keeping only the leading part the vanishing power at $\eta_N =
2-d$ translates into a logarithmic behavior, $\ln x^2$, formally 
the same as for a massless scalar propagator in a two-dimensional 
field theory.

The fact that a large anomalous dimension occurs at a non-Gaussian 
fixed point was initially observed in the context of the $2+\eps$ expansion 
\cite{Kawai2, Kawai3}
and later in computations based on the effective average action 
\cite{LR1, LR2}. 
The above argument shows that no specific computational information 
enters. 

Let us emphasize that in general an anomalous dimension is {\it not} 
related to the geometry of field propagation and in a conventional
field theory one cannot sensibly define a fractal dimension by 
looking at the high momentum behavior of a two-point function~\cite{Kroeger}. 
What is special about gravity is ultimately that the 
propagating field itself defines distances. One aspect thereof 
is the universal way matter is affected, as seen in (a). 
In contrast to an anomalous dimension in conventional field 
theories, this allows one to attribute a geometric significance to the 
modified short distance behavior of the test propagators, see (d). 


{\bf (c) Strict renormalizability and $1/p^4$ propagators:}
With hindsight the above patterns are already implicit 
in earlier work on strictly renormalizable gravity theories.  
As emphasized repeatedly the benign renormalizability 
properties of higher derivative theories are mostly 
due to the use of $1/p^4$ type propagator (in $d=4$ dimensions). 
As seen in section 2.2 this $1/p^4$ type behavior 
goes hand in hand with asymptotically safe couplings.
Specifically for the dimensionless Newton's constant ${\rm g}_N$
it is compatible with the existence of a nontrivial fixed point,
see (\ref{HD6}). This in turn enforces an anomalous dimension
$\eta_N =-2$ at the fixed point which links back to the $1/p^4$ 
type propagator.   
 
Similarly in the $1/N$ expansion \cite{Tomboulis1, Tomboulis2, Smolin} 
a nontrivial fixed point goes hand in hand with a propagator whose high 
momentum behavior is of the form $1/(p^4 \ln p^2)$, in four 
dimensions, and formally $1/p^d$ in $d$ dimensions. In position space 
this amounts to a $\ln x^2$ behavior, once again.  


{\bf (d) Spectral dimension and scaling of fixed point action:}
The scaling (\ref{gren2}) of the fixed point action also allows one 
to  estimate the behavior of the spectral dimension in the 
ultraviolet. This leads to a model-independent variant \cite{MNnotes} 
of an argument used in~\cite{LRfractal1, LRfractal2}.

If one wants to probe the functional measure over geometries only 
an interesting operator insertion is the trace of the heat kernel 
\cite{Kawaifrac1, Kawai1, AmbjornDT} 
\be 
G(T) = \int\! \cD q \,e^{i S[q]} \int \!dx \sqrt{q(x)} 
\exp(T \Delta_q)(x,x)\,. 
\label{heatcorr}
\end{equation}
Here $\Delta_q := q^{\alpha\beta} \nabla_{\alpha} \nabla_{\beta} = 
\sqrt{q}^{-1} \dd_{\alpha}(\sqrt{q} q^{\alpha\beta} \dd_{\beta})$ 
is the Laplace-Beltrami operator, and the heat kernel 
$\exp(T \Delta_q)(x,x')$ associated with it is the symmetric (in $x,x'$) 
bi-solution of the heat equation $\dd_T K = \Delta_q K$ with initial condition 
$\lim_{T \ra 0} \exp(T \Delta_q)(x,x') = \delta(x,x')$. 
The $T \ra \infty$ limit will then probe the large scale 
structure of the typical geometries in the measure and 
the $T \ra 0$ limit will probe the micro aspects. 
Again the expressions (\ref{heatcorr}) are here only
heuristic, in particular normalization factors have been omitted 
and the functional measure over geometries would have to be 
defined as previously outlined.

Let us briefly recapitulate the definition 
of the heat kernel and some basic properties. For a smooth 
Riemannian metric $g$ on a compact closed $d$-manifold let $\Delta_g 
:= g^{\alpha\beta} \nabla_{\alpha} \nabla_{\beta} = 
\sqrt{g}^{-1} \dd_{\alpha}(\sqrt{g} g^{\alpha\beta} \dd_{\beta})$ 
be the Laplace-Beltrami operator. The heat kernel 
$\exp(T \Delta_g)(x,x')$ associated with it is the symmetric 
(in $x,x'$) bi-solution 
of the heat equation $\dd_T K = \Delta_g K$ with initial condition 
$\lim_{T \ra 0} \exp(T \Delta_g)(x,x') = \delta(x,x')$. 
Since $(\cM,g)$ is compact $\Delta_g$ has purely discrete spectrum 
with finite multiplicities. We write $-\Delta_g \phi_n(g) = \cE_n(g) 
\phi_n(g)$, $n \geq 0$, for the spectral problem and assume that the 
eigenfunctions $\phi_n$ are normalized and the eigenvalues 
monotonically ordered $\cE_n(g) \leq \cE_{n+1}(g)$. 
We write $V(g) = \int \!dx \sqrt{g}$ for the volume of $(\cM,g)$ 
and
\be 
P_g(T) = \frac{1}{V(g)} \int\! dx \sqrt{g}\, \exp(T \Delta_g)(x,x)
= \frac{1}{V(g)} \sum_n e^{-\cE_n(g) T} \,,
\label{Dred3} 
\end{equation}
for the trace of the heat kernel. In the random walk picture 
$P_g(T)$ can be interpreted as the probability of a test particle
diffusing away from a point $x \in \cM$ and to return to it after 
the fictitious diffusion time $T$ has elapsed. In flat 
Euclidean space $(\cM,g) = (\R^d,\eta)$ for example $P_\eta(T) = 
(4\pi T)^{-d/2}$ for all $T$. For a generic manifold the trace 
of the heat kernel cannot be evaluated exactly. However the 
short time and the long time asymptotics can to some extent be 
described in closed form. Clearly the $T \ra \infty$ limit probes the 
large scale structure of a Riemannian manifold (small eigenvalues $\cE_n(g)$) 
while the $T \ra 0$ limit probes the small scales (large eigenvalues
$\cE_n(g)$).

For $T \ra 0$ one has an asymptotic 
expansion $P_g(T) \sim (4\pi T)^{-d/2} \sum_{n\geq 0} T^n \int \! dx \sqrt{g} 
a_n(x)$, where the $a_n$ are the Seeley-deWitt coefficients. 
These are local curvature invariants, $a_0 = 1,\; 
a_1 = \frac{1}{6} R(g)$, etc. The series can be rearranged 
so as to collect terms with a fixed power in the curvature 
or with a fixed number of derivatives~\cite{avramidi}. 
Both produces nonlocal curvature invariants. The second 
rearrangement is relevant when the curvatures are small 
but rapidly varying (so that the derivatives of the curvatures are 
more important then their powers). The leading derivative terms then are 
given by $P_g(T) \sim (4\pi T)^{-d/2}[V(g) + T \int \! dx \sqrt{g} a_1
+ T^2 N_2(T) + \ldots]$, where $N_2(T)$ is a known nonlocal 
quadratic expression in the curvature tensors. 
The $T \ra \infty$ behavior is more subtle as also global information 
on the manifold enters. For compact manifolds a typical behavior is 
$P_g(T) \sim (4\pi T)^{-d/2}[1 + O(\exp(- c T))]$, where the rate 
of decay $c$ of the subleading term is governed by the smallest non-zero 
eigenvalue.

Returning now to the quantum gravity average $G(T) \sim \bra P_q(T) \ket$,
one sees that on any state on which all local curvature polynomials
vanish the leading short distance behavior of $\bra P_q(T) \ket$ 
will always be $\sim T^{-d/2}$, as on a fixed manifold. 
The same will hold if the nonlocal invariants occurring 
in the derivative expansion all have vanishing averages in the 
state considered. A leading short distance behavior of the 
form 
\be 
\bra P_q(T) \ket \sim T^{-d_s/2} \,,\quad T \ra 0\,,
\label{Dred4}
\end{equation}
with $d_s \neq d$ will thus indicate that either the operations
``taking the average'' and ``performing the asymptotic expansion
for $T \ra 0$'' no longer commute, or that the microscopic 
geometry is very rough so that the termwise averages no longer vanish,
or both. Whenever well-defined the quantity $d_s(T) := -2 d\ln \bra P_q(T) \ket/
d\ln T$ is known as the spectral dimension (of the micro-aspects 
of the random geometries probed by the state $\cO \mapsto \bra \cO\ket$). 
See~\cite{Kawaifrac1, Kawai1, AmbjornDT} for earlier uses in random 
geometry, and~\cite{Havlin} for an evaluation of the 
spectral dimension for diffusion on the Sierpinski gasket based on 
a principle similar to (\ref{Dred7}) below.

We assume now that the states considered are such that the $T \ra \infty$ 
behavior of $\bra P_g(T) \ket$ is like that in flat space, i.e.\ $\bra P_q(T) 
\ket \sim T^{-d/2}$ for $T \ra \infty$, see Eq.~(\ref{stateselect}). 
Since $(4 \pi T)^{-d/2} = \int \! \frac{d^dp}{(2\pi)^d} 
\exp(-p^2 T)$ one can give the stipulated $T \ra \infty$ 
asymptotics an interpretation in terms of the spectrum 
$\{ p^2,\; p \in \R^d\}$ of the Laplacian of a `typical' reference 
metric $\bar{g}_{\alpha\beta}$ which is smooth and almost flat 
at large scales. The spectrum of $\Delta_q$ must be such that 
the small spectral values can be well approximated by 
$\{ p^2 < C,\; p \in \R^d\}$ for some constant $C>0$. 
Its unknown large eigenvalues will then determine the short 
distance behavior of $\bra P_q(T)\ket$. We can incorporate this 
modification of the spectrum by introducing a function $F_{\bar{g}}(p^2)$ 
which tends to $1$ for $p^2 \ra 0$, and whose large $p^2$ behavior 
remains to be determined. Thus 
\be 
\bra P_q(T) \ket \approx \int\! \frac{d^dp}{(2\pi)^d} 
\exp\{ - p^2 F_{\bar{g}}(p^2) T \}\,.
\label{Dred5}
\end{equation}
The following argument now suggests that within the asymptotic safety 
scenario $F_{\bar{g}}(p^2) \sim p^2$ for $p^2 \ra \infty$. Before turning to the 
argument let us note that this property of $F_{\bar{g}}(p^2)$
entails 
\be 
\bra P_q(T) \ket \sim T^{-d/4} \quad \mbox{for} \quad T \ra 0\,,
\quad {\mathrm{i}.e.} \;\; d_s= d/2\,.
\label{Dred6}
\end{equation}
The ``microscopic'' spectral dimension equals half the ``macroscopic''
$d$. Notably this equals $2$, as suggested by the ``anomalous 
dimension argument'' precisely in $d=4$ dimensions.

The argument for $F_{\bar{g}}(p^2) \sim p^2$ for $p^2 \ra \infty$ 
goes as follows: We return to discrete description
$P_q(T) = \sum_n e^{-\cE_n(q) T}$ for $(\cM,q)$ compact, and consider the 
average of one term in the sum $\bra e^{- \cE_n(q) T}\ket$, with 
$\cE_n(q)$ large. The computation of this average is a single scale 
problem in the terminology of Appendix A. As such it should allow 
for a good description via an effective field theory at scale $k$.
Here only the fact is needed that the average $\bra e^{- \cE_n(q) T}\ket$ 
can approximately be evaluated as~\cite{LRfractal1, LRfractal2} 
\be 
\bra e^{-\cE_n(q) T} \ket \approx e^{- \cE_n(\check{g}_k) T} \;, 
\label{Dred7}
\end{equation}
where $(\check{g}_k)_{\alpha\beta}$ is a saddle point configuration 
of the effective field theory at scale $k$, defined 
e.g.~as a stationary point of the effective action $\bar{\Gamma}_k[g]$ 
at scale $k$, see the discussion at the end of Section 1.2
Since the only scale available is $\cE_n$ itself the relevant 
scale $k$ is for given $n$ determined by the implicit equation $k^2 
= \cE_n(\check{g}_k)$. Next we consider how these spectral values 
scale in the fixed point regime where the dimensionless couplings are 
approximately constant, ${\mathrm{g}}_i(k) \approx {\mathrm{g}}_i$. 
Recall from (\ref{gren2}) the limiting behavior $\bar{\Gamma}_k[g] \ra 
S_*[k^2 g]$ as $k \ra \infty$. 
Two stationary points $(\check{g}_k)_{\alpha\beta}$ for $\bar{\Gamma}_k$ 
and $(\check{g}_{k_0})_{\alpha\beta}$ for $\bar{\Gamma}_{k_0}$ 
will thus in the fixed point regime be simply related by 
$k^2 \check{g}_k = k_0^2 \check{g}_{k_0}$. Since $k^2 \Delta_{k^2 g} = 
\Delta_g$ this means for the spectral values $k^2 \,\cE_n(\check{g}_k) 
= k_0^2\, \cE_n(\check{g}_{k_0})$. In order to make contact to 
the continuum parameterization in (\ref{Dred5}) we now identify for 
given $p$ the $n$'s such that for the typical metric $\bar{g}_{\alpha\beta}$ 
entering (\ref{Dred5}) one has $\cE_n(\bar{g}) \sim p^2$ for large $n$. 
After this reparameterization $\cE_n = \cE_p$, $p = \sqrt{p^2}$, 
one can identify the $F_{\bar{g}}(p^2)$ in (\ref{Dred5})
with $F_{\bar{g}}(p^2) = \cE_p(\check{g}_{k = p})/
\cE_{p}(\check{g}_{k_0})$. This scales for $p \ra \infty$ like 
$p^2$, which completes the argument.

In summary, the asymptotic safety scenario leads to the specific 
(theoretical) prediction that the residual interactions in the 
exteme ultraviolet are effectively two-dimensional. One manifestation 
is that the (normally powerlike) shortdistance singularities of {\it all} 
free matter propagators are softened to logarithmic ones. In quantum 
gravity averages like $G(R)$ in (\ref{geocorr}) this leads to the 
expectation that they should scale like logarithmically (powers of 
$\log R$ and $\log(-\log R)$) for $R \ra 0$. On the other hand 
this universality allows one to shuffle the effect from matter to 
gravity propagators. This justifies to attribute the effect to a modification 
in the underlying random geometry. The spectral dimension $G(T)$ of the random 
geometries probed by a certain class of ``macroscopic'' states comes out $d/2$, 
which (notably!) equals $2$ precisely in $d=4$ dimensions. 
Technically all aspects (a) -- (d) have their origin in the 
scaling relation (\ref{gren2}).

Accepting this dimensional reduction in the extreme ultraviolet as 
a working hypothesis one is lead to the following conjecture: 
The functional averages of an asymptotically safe theory 
of quantum gravity can in the extreme UV be approximately
(but more and more accurately as one approaches the fixed point) 
reproduced by a two-dimensional 
statistical field theory with the following properties: 
(i) It is two-dimensional and self-interacting; the latter 
because of the non-Gaussian nature of the original fixed point. 
(ii) It is not a conformal field theory in the usual 
sense, as the extreme UV regime in the original theory is reached 
from outside the critical surface (``massive continuum limit''). 
(iii) It is asymptotically safe itself (accounting for the 
antiscreening behavior presumed to be responsible for 
the stabilization of the UV properties). 

Note that in principle the identification of such a UV field 
theory is a well-posed problem. Presupposing that the functional 
integral has been made well-defined and through suitable operator 
insertions data for its extreme UV properties have been obtained,
for any proposed field theory with the properties (i) --(iii) one 
can test whether or not these data are reproduced.

\section{Conclusions} 

The asymptotic safety scenario delineates conditions under which 
a functional integral based quantum theory of gravity can be 
viable beyond the level of an effective field theory. It 
combines the lessons drawn from an advanced quantum field 
theory perspective on the problem of quantum gravity in an 
apparently  selfconsistent way. The moral drawn from renormalization 
is that the main challenge lies not so much in achieving renormalizability 
but to reconcile asymptotically safe couplings with the absence of 
unphysical propagating modes. The asymptotic safety property 
should lead to a dynamical reduction of the interacting 
degrees of freedom in the extreme ultraviolet. Finally this 
exteme UV regime is conjectured to be described by an effectively
two-dimensional quantum field theory.  

The goal of this review would be reached if a reasonably
convincing case for this `heretically orthodox' scenario has been made. 
Future work will have to focus on four areas: 
(i) Consolidating the existence of a non-Gaussian fixed point 
and the asymptotic safety property of the couplings. This may be 
done in various formalisms, field variables, and approximations. 
(ii) Clarifying the microstructure of the geometries and identification of
the antiscreening degrees of freedom. 
(iii) Understanding of the physically adequate notion of 
unitarity and its interplay with (i) and (ii). 
(iv) Characterization of generic observables and working out 
sound consequences for the macrophysics.
\vspace{5mm}

{\bf Acknowledgements:} I wish to thank M. Reuter for the 
collaboration on the Living Review article with the same title. 
I am grateful to E.~Seiler for a critical reading of the manuscript 
and for suggesting various improvements. In addition I wish to thank 
A.~Ashtekar, C.~Bervillier, R.~Loll, E.~Mottola, and A.~Niemi for 
comments and correspondence.


\newpage

\renewcommand{\theequation}{\Alph{section}.\arabic{equation}}
\setcounter{equation}{0}

\setcounter{section}{0}
\appendix
\newappendix{Renormalizing the nonrenormalizable}
\setcounter{equation}{0}

The modern view of renormalization has been shaped by 
Kadanoff and Wilson. See~\cite{Kadanoff} and~\cite{Wilson,
Wilson1,Wilson2} for first hand accounts and a guide to
the original articles. In the present context the relevance of a Kadanoff-Wilson 
view on renormalization is two-fold: first it allows one 
to formulate the notion of renormalizability without reference 
to perturbation theory, and second it allows one to treat at least 
in principle renormalizable and nonrenormalizable theories on the 
same footing. For convenience we briefly summarize the 
main principles of the Kadanoff-Wilson approach to renormalization here:

{\bf Kadanoff--Wilson view on renormalization -- main principles:}
\vspace{-3mm}

\begin{itemize}
\item[(i)] A theory is not defined in terms of a given action, but in 
terms of a field content and the steps (ii)-(v) below. 
\item[(ii)] The functional integral is performed in piecemeal, integrating out 
fast modes, retaining slow modes, while keeping the values of observables
fixed. This ``coarse graining'' process results in a flow in the 
space of actions which depends on the chosen coarse graining operation. 
\item[(iii)] Starting from a retroactively justified initial action 
ideally {\it all} interaction monomials generated 
by the flow are included in a typical action; in any case many more than just the 
power-counting renormalizable ones. Then one classifies the coefficients of the 
monomials into essential (couplings) and inessential (field redefinitions). 
\item[(iv)] A fixed point (FP) in the flow of couplings is searched for. 
The position of the FP depends on the chosen coarse graining operation, 
but the rates of approach to it typically do not (``universality''). 
\item[(v)] The flow itself decides which monomials are relevant in 
the vicinity of a FP and hence defines the dynamics. The scaling dimensions 
with respect to a non-Gaussian FP may be different from (corrected) 
power-counting dimensions referring to the Gaussian FP.
\item[(vi)] The dimension of the unstable manifold and hence the ``degree'' 
of renormalizability depends on the FP! 
\end{itemize}
We add some remarks: The more familar perturbative notion of renormalizability 
is neither sufficient (e.g.~$\Phi_4$ theory in $d=4$) nor necessary 
(e.g.~Gross-Neveu model in $d=3$) for renormalizability in the 
above sense. The title of this Section is borrowed from a paper 
by Gawedzki and Kupiainen~\cite{GawKup2}.

As summarized here, these principles describe the construction of a so-called 
massive continuum limit of a  statistical field theory initially 
formulated on a lattice, say. 
A brief reminder: in a lattice field theory there is typically a dynamically 
generated scale, the correlation length $\xi$, which allows one to convert 
lattice distances into a physical length scale, such that say, $\xi$ 
lattice spacings equal 1 fm. The lattice points $n = (n_1,\ldots, n_d) \in \Z^d$ 
are then traded for dimensionful distances $x_i = (n_i/\xi)$ fm. Taking the 
lattice spacing to zero amounts to sending $\xi$ to infinity while 
keeping $x_i$ fixed. If the correlation functions of some lattice 
fields are rescaled accordingly (including a `wave function' renormalization 
factor) and the limit exists this defines a massive continuum limit
of the lattice theory. 

The rationale for the 
piecemeal performance of the functional integral is that in statistical 
mechanics language a critical problem is decomposed into a 
sequence of subcritical ones. Here a critical problem is one where 
fluctuations of the dynamical variables over 
vastly different length scales have to be taken into account; 
for a subcritical problem the opposite is true. In more detail, 
let $\cO$ be a function of the fields $\chi$ whose functional 
average is meant to be a macroscopic observable, but whose 
statistical average 
is sensitive to fluctuations of the microscopic fields 
$\chi$ on very different length scales. The replacement 
by a sequence of subcritical problems is done by specifying an 
asymmetric blocking kernel $K: {\mathrm{Configurations}} \times 
{\mathrm{Configurations}} \ra \R$, $K(\chi', \chi) = 
K(\{\chi'_p\},(\{\chi_p\})$, such that 
\begin{itemize}
\item[(a)] $K(\chi', \chi)= K_{l,\delta l}(\chi',\chi)$ has support 
mostly on configurations $\{\chi_p\}$ with $l - \delta l \leq p \leq l$. 
\item[(b)] $\int \prod_{p \leq \Lambda} d\chi'_p \, K(\chi', \chi) =1$. 
\end{itemize}
Then 
\be 
\bra \cO \ket = \int \prod_{p \leq \Lambda} d\chi_p 
\,\cO(\chi) \, e^{-S[\chi]} = 
\int \prod_{p \leq \Lambda - \delta l} \! d\chi'_p\, 
\,\cO'(\chi') \, e^{-S'[\chi']}\,, 
\label{flowscheme1}
\end{equation}
with 
\be 
\cO'(\chi') e^{-S'[\chi']} = 
\int \! \prod_{p \leq \Lambda} d\chi_p \, K_{\Lambda,\delta l}(\chi', \chi) 
\cO(\chi) e^{-S[\chi]}\,.
\label{flowscheme2}
\end{equation}
Taking $\cO =\1$ defines the coarse grained action functional $S'$, 
after which~(\ref{flowscheme2}) can be used to define the coarse 
grained observables $\cO'$. Property (a) entails that only 
field configurations with a similar `degree of roughness' have to 
be considered in evaluating the functional integral in 
(\ref{flowscheme2}). It should thus be much more amenable to 
(numerical or analytical) approximation techniques than 
the original functional integral (\ref{flowscheme1}). 

Once~(\ref{flowscheme2}) has been evaluated one can 
iterate the procedure. The formulas (\ref{flowscheme1}), (\ref{flowscheme2}) 
remain valid with the basic kernel $K$ replaced by its $n$-fold 
convolution product, for which we write $K_{\Lambda, n \delta l}(\chi',\chi)$. 
For most choices a kernel $K_{l,\delta l}$ will not be reproducing,
i.e.\ $\int \! \prod_{p \leq \Lambda} d\chi'_p K_{l-\delta l,\delta l}(\chi'',
\chi') K_{l,\delta l}(\chi',\chi) =: K_{l - 2 \delta l,2 \delta l}(\chi'',\chi)$ 
will not (despite the suggestive notation) coincide with the original 
kernel $K_{l,\delta l}$, just with modified parameters. 
Technically it is thus easier to specify the iterated kernel directly,
which is of course still normalized. The $n$-fold 
iterated kernel will have support mostly on configurations with 
$e^{-t} := \frac{l}{\Lambda} 
\leq \frac{p}{\Lambda} \leq 1$, if $l = n \delta l$, and $e^{-t}$ is the fraction 
of the momentum modes over which the functional integral has been performed
after $n$ iterations. In the above terminology the critical 
problem~(\ref{flowscheme1}) has been replaced by the sequence of 
subcritical problems~(\ref{flowscheme2}). In each iteration, referred 
to as a {\it coarse graining} step defined by the kernel $K$, 
only a small fraction of the degrees of freedom 
is integrated out. The action $S = S_{\Lambda}$ 
at the cutoff scale $p= \Lambda$ is called the {\it microsocopic 
(or bare) action}, the $S' = S_l$ reached after integrating out 
the `fast' modes in the range $l/\Lambda \leq p/\Lambda \leq 1$ 
is called the {\it coarse grained action} at scale $l$, and similarly 
for the fields $\chi' = \chi_l$. Note that the action $S_l[\,\cdot\,]$ 
as a functional is defined for all field configurations though for 
the evaluation of~(\ref{flowscheme1}) only $S_l[\chi_l]$ is needed. 

Throughout we shall follow the sloppy field theory convention 
that the coarse graining operates on the action. Of course what 
really gets updated is the functional measure 
\be 
d\mu_l[\chi]= \prod_{p} 
d\chi_p \,e^{-S_l[\chi]}\,.
\label{flowscheme3} 
\end{equation}
In the (lattice) regularized theory the decomposition of the measure into 
a flat reference measure $\prod_p d\chi_p$ and a Boltzmann factor
parameterized by the action is unproblematic. The flow in the measures
can thus be traded for a flow in the actions (as long as the 
Jacobian is taken into account that comes from the reference 
measure upon a change of field variables $\chi \mapsto \chi'(\chi)$).
The Wilsonian ``space of actions'' refers to a cone of 
positive measures~(\ref{flowscheme3}) which 
is preserved under the coarse graining operation considered.

In a gravitational context the very concept of 
renormalizabilty is less clear cut, and one should presumably go 
back to the even more fundamental property for which renormalizability 
is believed to be instrumental, namely the existence 
of a genuine continuum limit, roughly in the sense outlined in Section 1.3.
Since rigorous results based on controlled approximations 
are unlikely to be obtained in the near future, we describe in 
the following criteria for the plausible existence of a genuine continuum 
limit based on two uncontrolled approximations: renormalized 
perturbation theory and the functional renormalization group 
approach. Such criteria are `implicit wisdom' and are hardly ever 
spelled out. In the context of Quantum Gravidynamics, 
however, the absence of an obvious counterpart of the correlation length 
and the lack of perturbative renormalizability makes things more subtle. 
In the two appendices below we therefore try to make the implicit explicit 
and to formulate critera for the existence of a genuine continuum limit
which are applicable to Quantum Gravidynamics as well.

\subsection{Perturbation theory and continuum limit} 

Perturbatively renormalizable field theories are a degenerate special 
case of the Wilson--Kadanoff framework. The main advantage of 
perturbation theory is that the UV cutoff $\Lambda$ can be removed 
exactly and independently of the properties of the coupling flow.
The existence of a $\Lambda \ra \infty$ limit with the required 
properties (PTC1) can often be rigorously proven, in contrast 
to most nonperturbative techniques where this can only be established 
approximately by assembling evidence. With (PTC1) satisfied, 
the coupling flow then can be studied in a second step and used 
to probe whether or not the criterion (PTC2) for the existence of 
a genuine continuum limit as anticipated in Section 1.4 is also satisfied. 
The main disadvantage of perturbation theory is that everything is 
initially defined as a formal power series in the loop counting parameter. 
Even if one trades the latter for a running coupling the series in this 
coupling remains a formal one, typically non-convergent and not Borel-summable. 
It is generally believed, however, that provided (PTC2) is satisfied 
for a perturbative Gaussian fixed point, the series is asymptotic 
to the (usually unknown) exact result. In this case the perturbative 
analysis should indicate the existence of a genuine continuum limit
based on an underlying Gaussian fixed point proper. Our main reason 
for going through this in some detail below is to point out 
that in a situation with several couplings the very same rationale 
applies if the perturbative fixed point is a non-Gaussian rather 
than a Gaussian one.

As mentioned, in perturbation theory one initially only aims at 
defining the expectations (\ref{flowscheme1}) as a formal powerseries 
in the loop counting parameter $\lb$. In the regularized functional 
integral $\lb$ enters via $\exp\{ - \frac{1}{\lb} S_{\Lambda}[\chi]\}$
and the (bare) perturbative expansion is a saddle point expansion 
around $\lb =0$. The fluctuations around a saddle point configuration 
are rescaled by a factor $\lb^{1/2}$ after which the sum of all 
$\ell$-loop contributions to a quantity occurs with a factor $\lb^\ell$. 
The quadratic part of the expansion defines the propagators of 
a set of free fields; for definiteness we consider here the 
case where these are formally massless. For the reasons explained 
in~\cite{HolsteinD} the loop expansion then does not necessarily 
coincide with an expansion in powers of Planck's constant 
$\hbar$. For example 1-loop diagrams can contribute to the classical 
limit $O(\hbar^0)$. Typically the field are given a reference 
mass $\mu$ and we write $S_{*,\mu}[\chi]$ for the (quadratic) 
action of this set of free fields. 
The interaction is described by a set of monomials $P_i[\chi]$,
$i \in E_{\mathrm{p}.c.}$, which are ``powercounting renormalizable''.
The latter means that their mass dimension $-d_i$ is such that $d_i \geq 0$.
It is also assumed that the $P_i$ are functionally independent, so 
that the corresponding couplings are essential. The so-called ``bare'' 
action functional then is $S_{\Lambda} = S_{*,\mu} + \sum_{i \in 
E_{\mathrm{p}.c.}} u_i(\Lambda) P_i$, where $u_i(\Lambda)$ 
are the essential ``bare'' couplings (including masses) corresponding to the 
interaction monomials $P_i[\chi]$. Inessential parameters are generated 
by subjecting $S_{\Lambda}$ to a suitable class of field redefinitions. 
In more detail one writes 
\ba 
u_i(\Lambda) \is u_i(\mu)\, V_{i,0}(\mu) + 
\sum_{\ell \geq 1} \lb^{\ell} \,V_{i,\ell}(u(\mu),\Lambda,\mu) \,,
\nonum
\chi_{\Lambda} \is \chi_{\mu} + \sum_{\ell \geq 1} \lb^{\ell}\, 
\Xi_{\ell}(\chi_{\mu};\, u(\mu),\Lambda,\mu)\,.
\label{PTscheme1}
\end{eqnarray}
Here $u_i(\mu)$ are the renormalized couplings which are $\Lambda$ 
independent and the $V_i(u(\mu),\Lambda,\mu)$ are counterterms which 
diverge in the limit $\Lambda \ra \infty$. This divergence is 
enforced by very general properties of QFTs. Similarly the $\chi_{\mu}$ 
are called renormalized fields and the
$\Xi_{\ell}(\chi_{\mu};u(\mu),\Lambda,\mu)$ are local functionals of 
the $\chi_{\mu}$ with coefficients depending on $u(\mu),\Lambda,\mu$;
the coefficients again diverge in the limit $\Lambda \ra \infty$. 
Often one aims at ``multiplicative renormalizability'', 
which means the ansatz for the $\Xi_\ell$ is taken to be linear 
in the fields $\Xi_{\ell}(\chi_{\mu};u(\mu),\Lambda,\mu) = 
Z_{\ell}(u(\mu),\Lambda,\mu) \chi_{\mu}$ and $Z_{\ell}$ is the 
$\ell$-loop ``wave function renormalization'' constant. 
One should emphasize, however, that multiplicative 
renormalizability can often not be achieved, and even in field theories
where it can be achieved, it evidently will work only with a particular 
choice of field coordinates; see~\cite{Bonneau} for a discussion. 

The normalizations in (\ref{PTscheme1}) can be chosen 
such that $u_i(\mu = \Lambda) = 
u_i(\Lambda)$ and $\chi_{\mu = \Lambda} = \chi_{\Lambda}$, but one is 
really interested in the regime where $\mu \ll \Lambda$. Inserting these 
parameterizations into $S_{\Lambda}[\chi_{\Lambda}]$ gives an 
expression of the form 
\be 
S_{\Lambda}[\chi_{\Lambda}] = S_{*,\mu}[\chi_{\mu}] + 
\sum_{\alpha} \Big( \sum_{\ell \geq 0} \lb^{\ell} \,u_{\alpha,\ell}(u(\mu),
\Lambda,\mu) \Big) P_{\alpha}[\chi_{\mu}]\,,
\label{PTscheme2}
\end{equation}
where the sum over $\alpha$ includes terms of a form which can be absorbed 
by a nonlinear field redefinition, see e.g.~Appendix A of \cite{Gravirept}.  
Often the $\mu$-dependence in the fields can 
be traded for one carried by (inessential) parameters $z_i(\mu),\,i \in I$.
Then (\ref{PTscheme2}) takes the form $S_{\Lambda}[\chi_{\Lambda}] = 
\sum_{\alpha'} u_{\alpha'}({\mathrm{g}}(\mu),z(\mu),\Lambda,\mu)
P_{\alpha'}[\chi]$, with some $\mu$-independent fields, $\chi = \chi_{\mu_0}$, 
say. The rhs clearly resembles the Wilsonian form 
$\sum_{\alpha} u_{\alpha} P_{\alpha}$, with the difference that modulo 
field redefinitions only powercounting renormalizable interaction 
monomials occur.

So far the counterterms in (\ref{PTscheme1}) have been left unspecified.
The point of introducing them is of course as a means to absorb
the cut-off dependence generated by the regularized functional integral
in (\ref{flowscheme1}). Specifically, one replaces the Boltzmann factor by its 
power series expansion in $\lb$, i.e.\ $\exp\{-S_{\Lambda}[\chi_{\Lambda}]\}
= \exp\{-S_{*,\mu}[\chi_{\mu}]\}(1 + \sum_{\ell \geq 1} \lb^{\ell} 
Q_{\ell}[\chi_{\mu}])$, and aims at an evaluation of multipoint 
functions $\bra \chi_{\Lambda}(x_1) \ldots \chi_{\Lambda}(x_n) 
\ket_{S_{\Lambda}}$ as formal power series in $\lb$. 
After inserting (\ref{PTscheme1}) and the expansion of 
$e^{-S_{\Lambda}[\chi_{\Lambda}]}$ this  reduces the problem to an 
evaluation of the free multipoint functions 
$\bra \chi_{\mu}(x_1) \ldots \chi_{\mu}(x_n)\, Q_l[\chi_{\mu}] \ket_{S_{*,\mu}}$ 
computed with the quadratic action $S_{*,\mu}$ on the field space with 
cutoff $\Lambda$. The free multipoint functions will contain contributions 
which diverge in the limit $\Lambda \ra \infty$. On the other hand via the 
parameterization (\ref{PTscheme1}),(\ref{PTscheme2}) the coefficients carry 
an adjustable $\Lambda$ dependence. In a renormalizable QFT the $\Lambda$ 
dependence in the coefficients can be chosen so as to cancel (for $\mu \ll \Lambda$) 
that generated by the multipoint functions 
$\bra \chi_{\mu}(x_1) \ldots \chi_{\mu}(x_n) Q_l[\chi_{\mu}]\ket_{S_{*,\mu}}$. 
With this adjustment the limits 
\be 
\sum_{\ell \geq 0} \lb^{\ell} \lim_{\Lambda \ra \infty} 
\bra \chi_{\Lambda}(x_1) \ldots \chi_{\Lambda}(x_n) 
\ket_{S_{\Lambda},\ell} =: \sum_{\ell \geq 0} \lb^{\ell} 
\bra \chi_{\mu}(x_1) \ldots
\chi_{\mu}(x_n) \ket_{S_{\mu},\ell}\,,
\label{PTscheme3}
\end{equation}
exist and define the renormalized multipoint functions. As indicated 
they can be interpreted as referring to the renormalized action 
$\lim_{\Lambda \ra \infty} S_{\Lambda}[\chi_{\Lambda}] = S_{\mu}[\chi_{\mu}]$. 
Eq.~(\ref{PTscheme3}) highlights the main advantage of renormalized 
perturbation theory: the existence of the infinite cutoff limit (\ref{PTscheme3}) 
is often a provable property of the system, while this is not the case for most 
nonperturbative techniques. In the terminology introduced in Section
1.3 the criterion (PTC1) is then satisfied. In order for this 
to be indicative for the existence of a genuine continuum 
limit, however, the additional condition (PTC2) must be satisfied, whose 
rationale we proceed to discuss now.

Since the renormalization scale $\mu$ is arbitrary, changing its value must 
not affect the values of observables. The impact of a change in $\mu$ can 
most readily be determined from (\ref{PTscheme1}). The left hand sides 
are $\mu$ independent, so by differentiating these relations with respect 
to $\mu$ and extracting the coefficients in a powerseries in 
(say) $\Lambda$ and/or $\log \Lambda$ consistency conditions arise for 
the derivatives 
$\mu \frac{d}{d\mu}u_i$ and $\mu \frac{d}{d \mu} \chi_{\mu}$. 
The ones obtained from the leading order are the 
most interesting relations. For the couplings one obtains a system 
of ordinary differential equations which define their 
renormalization flow under a change of $\mu$. As usual it is convenient 
to work with dimensionless couplings ${\mathrm{g}}_i := u_i \mu^{-d_i}$, 
where $d_i$ is the mass dimension of $u_i$. The flow equations then 
take the form 
\be 
\mu \frac{d}{d\mu} {\mathrm{g}}_i = \beta_i({\mathrm{g}}(\mu))\,,
\label{PTscheme4}
\end{equation}
where the $\beta_i$ are the perturbative beta functions. 
The flow equations for the renormalized fields are familiar 
only in the case of multiplicatively renormalizable fields, 
where one can work with scale independent fields and have the scale 
dependence carried by the wave function renormalization constant.  
In general however the fields are scale dependent. 
For example this ensures that the renormalized action evaluated 
on the renormalized fields is scale independent: $\mu \frac{d}{d\mu} 
S_{\mu}[\chi_{\mu}] =0$.

By construction the perturbative beta functions have a fixed point 
at ${\mathrm{g}}^*_i =0$, which is called the {\it perturbative 
Gaussian fixed point}. Nothing prevents them from having other 
fixed points, but the Gaussian one is built into the construction. 
This is because a free theory has vanishing beta functions and 
the couplings ${\mathrm{g}}_i = u_i \mu^{-d_i}$ have been introduced 
to parameterize the deviations from the free theory with action 
$S_{*,\mu}$. Not surprisingly the stability matrix $\Theta_{ij} = 
\dd \beta_i/\dd {\mathrm{g}}_j|_{{\mathrm{g}}^* =0}$ of the perturbative 
Gaussian fixed point just reproduces the information which has 
been put in. The eigenvalues come out to be $-d_i$ modulo corrections 
in the loop coupling parameter, where $-d_i$ are the mass dimensions
of the corresponding interaction monomials. 
For the eigenvectors one finds a one-to-one correspondence to 
the unit vectors in the `coupling direction' ${\mathrm{g}}_i$, again 
with power corrections in the loop counting parameter. 
One sees that the couplings $u_i$ not irrelevant with respect to the 
stability matrix $\Theta$ computed at the perturbative Gaussian 
fixed point are the ones with mass dimensions $d_i \geq 0$, i.e.\ just the 
power counting renormalizable ones.

The attribute ``perturbative Gaussian'' indicates 
that whenever in a nonperturbative construction of the renormalization 
flow in the same `basis' of interaction monomials ${\mathrm{g}}_i^*=0$ is 
also a fixed point (called the {\it Gaussian fixed point}) the 
perturbatively defined expectations are {\it believed} to 
provide an asymptotic (nonconvergent) expansion to the expectations 
defined nonperturbatively based on the Gaussian fixed point,
schematically
\be 
\bra \cO \ket_{\mathrm{Gaussian FP}} \sim \sum_{\ell \geq 0} 
\lb^\ell \bra \cO \ket_{\ell}\,.
\label{GaussvsPT}
\end{equation}
Here $\bra \cO \ket_{\ell}$ is the perturbatively computed $\ell$-loop contribution 
after a so-called renormalization group improvement. Roughly speaking the latter 
amounts to the following procedure: one assigns to the loop counting parameter $\lb$ 
a numerical value (ultimately related to the value of Planck's constant 
in the chosen 
units; see however~\cite{HolsteinD}) and solves $\mu\frac{d {\mathrm{g}}_i}{d \mu} = 
\beta_i({\mathrm{g}}(\mu))$ as an ordinary differential equation. One of the 
functions obtained, say ${\mathrm{g}}_1(\mu)$, is used to eliminate $\lb$ in favor 
of $\mu$ and an integration constant $\Lambda_{\mathrm{beta}}$ (not to be confused 
with the cutoff, which is gone for good). So $\bra \cO \ket_L := \sum_{\ell \leq L} 
{\bar \lb}(\mu,\Lambda_{\mathrm{beta}})^{\ell} \bra \cO \ket_{\ell,\mu}$ at this 
point carries a two-fold $\mu$-dependence, the one which comes out of 
the renormalization procedure (\ref{PTscheme3}) and the one carried now 
by $\bar{\lb}(\mu,\Lambda_{\mathrm{beta}})$. For an observable quantity $\cO$ both 
dependencies cancel out, modulo terms of higher order, leaving behind a 
dependence on the integration constant $\Lambda_{\mathrm{beta}}$.
We write $\bra \cO \ket_L(\Lambda_{\mathrm{beta}})$ to indicate this dependence. 
One then uses the expectation of one, suitably chosen, 
observable $\cO_0$ to match its value $\bra \cO_0 \ket$ (measured or 
otherwise known) with that of $\bra \cO_0\ket_{L}(\Lambda_{\mathrm{beta}})$ to a 
given small loop order $L$ (typically 
not larger than 2). For a well chosen $\cO_0$ this allows 
one to replace $\Lambda_{\mathrm{beta}}$ by a physical mass scale 
$m_{\mathrm{phys}}$. Eliminating $\Lambda_{\mathrm{beta}}$ in favor
of $m_{\mathrm{phys}}$ gives the perturbative predictions for all other 
observables. Apart 
from residual scheme dependences (which are believed to be numerically small) 
this defines the right hand side of (\ref{GaussvsPT}) unambigously as a 
functional over the observables.

Nevertheless, except for some special cases, 
it is difficult to give a mathematically precise meaning to the `$\sim$' in 
Equation~(\ref{GaussvsPT}). Ideally one would be able to prove that 
perturbation theory is asymptotic to the (usually unknown) exact 
answer for the same quantity. For lattice theories on a finite 
lattice this is often possible, the problems start when taking the  
limit of infinite lattice size, see \cite{NNW} for a discussion. 
In the continuum limit a proof that perturbation theory is 
asymptotic has been achieved in a number of low dimensional quantum field 
theories: the superrenormalizable $P_2(\phi)$ and $\phi_3^4$
theories~\cite{dimock,BovierFelder} and the two-dimensional
Gross--Neveu model, where the correlation functions are the Borel sum
of their renormalized perturbation expansion~\cite{GawKup1, GawKup3}. Strong 
evidence for the asymptotic correctness of perturbation theory 
has also been obtained in the O(3) nonlinear sigma-model 
via the form factor bootstrap~\cite{BalogNie}. 
In four or higher dimensional theories unfortunately no such 
results are available. 
It is still believed that whenever the above ${\mathrm{g}}_1$ is asymptotically 
free in perturbation theory, that the corresponding series is 
asymptotic to the unknown exact answer. 
On the other hand, to the best of our knowledge, a serious attempt to
establish the asymptotic nature of the expansion has never been made,
nor are plausible strategies available. 
The pragmatic attitude usually adopted is to refrain from the 
attempt to theoretically understand the domain of applicability 
of perturbation theory. Instead one interprets 
the `$\sim$' in (\ref{GaussvsPT}) as approximate numerical 
equality, to a suitable loop order $L$ and in a benign scheme, 
as long as it works, and attributes larger 
discrepancies to the `onset of nonperturbative physics'.
This is clearly unsatisfactory, but often the best one can do. Note
also that some of the predictive power of the QFT considered is wasted by
this procedure and that it amounts to a partial immunization of perturbative 
predictions against (experimental or theoretical) refutation.

So far the discussion was independent of the nature of the running
of $\bar{\lb}(\mu,\Lambda_{\mathrm{beta}})$ (which was traded for ${\mathrm{g}}_1$).
The chances that the vage approximate relation `$\sim$' in (\ref{GaussvsPT}) 
can be promoted to the status of an asymptotic expansion are of course 
way better if $\bar{\lb}(\mu,\Lambda_{\mathrm{beta}})$ is
driven towards $\bar{\lb}=0$ by the perturbative flow. 
Only then is it reasonable to expect that an asymptotic
relation of the form (\ref{GaussvsPT}) holds, linking the 
perturbative Gaussian fixed point to a genuine Gaussian 
fixed point defined by nonperturbative means. The perturbatively 
and the nonperturbatively defined coupling ${\mathrm{g}}_1$ can then be 
identified asymptotically and lie in the unstable manifold of the 
fixed point ${\mathrm{g}}_1^*=0$. On the other hand the existence of a 
Gaussian fixed point with a nontrivial unstable manifold is thought 
to entail the existence of a genuine continuum limit, in the sense 
discussed before. In summary, if ${\mathrm{g}}_1$ is traded for a running 
$\bar{\lb}(\mu,\Lambda_{\mathrm{beta}})$, a perturbative criterion for the 
existence of a genuine continuum limit is that the perturbative flow 
of ${\mathrm{g}}_1$ is regular with $\lim_{\mu \ra \infty}{\mathrm{g}}_1(\mu) =0$. 
Since the beta functions of the other couplings are formal power 
series in $\lb$ without constant coefficients the other couplings 
will vanish likewise as ${\mathrm{g}}_1 \ra 0$, and one recovers the 
local quadratic action $S_{*,\mu}[\chi]$ at the fixed point. 
The upshot is that the coupling with respect to which the
perturbative expansion is performed should be asymptotically free 
in perturbation theory in order to render the existence of 
a nonperturbative continuum limit plausible.

The reason for going through this discussion is to highlight that 
is applies just as well to a {\it perturbative non-Gaussian fixed point}.
This sounds like a contradiction in terms, but it isn't.
Suppose that in a situation with several couplings ${\mathrm{g}}_1,\ldots ,{\mathrm{g}}_n$ 
the perturbative beta functions (which are formal power series in $\lb$ 
without constant coefficients) admit a nontrivial zero, 
${\mathrm{g}}^*_1(\lb),\ldots ,{\mathrm{g}}^*_n(\lb)$. Suppose in addition 
that {\it all} the couplings lie in the unstable manifold of that 
zero, i.e.\ the flows ${\mathrm{g}}_i(\mu)$ are regular and $\lim_{\mu \ra \infty} 
{\mathrm{g}}_i(\mu) = {\mathrm{g}}_i^*$. We shall call a coupling with this 
property {\it asymptotically safe}, so that the additional assumption is 
that all couplings are asymptotically safe. As before one must assign 
$\lb$ a numerical
value in order to define the flow. Since the series in $\lb$ anyhow has zero 
radius of convergence, the `smallness' of $\lb$ is not off-hand a measure 
for the reliability of the perturbative result (the latter intuition 
in fact precisely presupposes (\ref{GaussvsPT})). Any one of the 
deviations $\delta {\mathrm{g}}_i = {\mathrm{g}}_i - {\mathrm{g}}_i^*$, which 
is of order $\lb$ at some $\mu$ can be used as well to parameterize
the original loop expansion. By a relabeling or reparameterization 
of the couplings we may assume that this is the case for $\delta {\mathrm{g}}_1$.
The original loop expansion can then be rearranged to read 
$\sum_{\ell \geq 0} (\delta {\mathrm{g}}_1)^{\ell} \bra \cO\ket_{\ell}$. 
However, if there is an underlying nonperturbative structure at all,
it is reasonable to assume that it refers to a non-Gaussian fixed point, 
\be 
\bra \cO \ket_{\mathrm{Non-Gaussian FP}} \sim \sum_{\ell \geq 0} 
(\delta {\mathrm{g}}_1)^\ell \bra \cO \ket_{\ell}\,.
\label{NonGaussvsPT}
\end{equation}
The rationale for (\ref{NonGaussvsPT}) is exactly the same 
as for (\ref{GaussvsPT}). What matters is not the value of the 
couplings at a perturbative fixed point, but their flow pattern.
For a nontrivial fixed point the couplings ${\mathrm{g}}_i^*$ in the above 
basis of interaction monomials are nonzero, but any one of the 
deviations $\delta {\mathrm{g}}_i = {\mathrm{g}}_i - {\mathrm{g}}_i^*$ can be 
made arbitrarily small as $\mu \ra \infty$. The relation 
`$\sim$' in (\ref{NonGaussvsPT}) then again plausibly 
amounts to an asymptotic expansion for the unknown exact answer, 
where the latter this time is based on a non-Gaussian fixed point.

Summarizing: in perturbation theory the termwise removal of the 
UV cutoff can be achieved independently of the properties of the 
coupling flow, while in a 
non-perturbative setting both aspects are linked. Only if the 
coupling flow computed from the perturbative beta functions meets certain 
conditions is it reasonable to assume that there exists an underlying 
non-perturbative framework to whose results the perturbative series is 
asymptotic. Specifically we formulate the

{\bf Criterion:} (Continuum limit via perturbation theory)
(PTC1) Existence of a formal continuum limit, i.e.~termwise removal of the 
UV cutoff is possible and the renormalized physical quantities 
are independent of the scheme and of the choice of interpolating fields, 
all in the sense of formal power series in the loop counting parameter. 
(PTC2) The perturbative beta functions have a Gaussian or a non-Gaussian 
fixed point and the dimension of its unstable manifold (as computed from the 
perturbative beta functions) equals the number of independent essential 
couplings. Equivalently, all essential couplings are asymptotically 
safe in perturbation theory. 

\subsection{Functional flow equations and UV renormalization}

The technique of functional renormalization group equations (FRGEs)
does not rely on a perturbative expansion and has been widely used 
for the computation of critical exponents and the 
flow of generalized couplings. For a systematic exposition 
of this technique and its applications we refer to the reviews 
\cite{Morris1, Pawlowski1, Wett2, Berges,Mitter}. 
Here we shall mainly use the effective average action $\Gamma_k$ 
and its `exact' FRGE. We refer to \cite{Gravirept} for a summary of this
formulation, and discuss in this Section how the UV renormalization 
problem presents itself in an FRGE. 

In typical applications of the FRG the ultraviolet renormalization problem 
does not have to be addressed. In the context of the asymptotic 
safety scenario this is different. By definition the perturbative series 
in a field theory based on an asymptotically safe functional measure 
has a  dependence on the UV cutoff 
which is not strictly renormalizable, see Section 1.3. The perturbative 
expansion of an FRGE must reproduce the structure of these divergencies. 
On the other hand in an exact treatment or based on 
different approximation techniques a reshuffling of the cutoff dependence 
is meant to occur which allows for a genuine continuum limit. 
We therefore outline here 
how the UV renormalization problem manifests itself in the framework 
of the functional flow equations. The goal will be to formulate a
criterion for the plausible existence of a genuine continuum 
limit in parallel to the one above based on perturbative indicators.

Again we illustrate the relevant issues for a scalar quantum field 
theory on flat space. For definiteness we consider here the flow 
equation for the {\it effective average action} $\Gamma_{\Lambda,k}[\phi]$,
for other types of FRGEs the discussion is similar though. 
The effective average action interpolates between the 
bare action $S_{\Lambda}[\phi]$ and the 
above, initially regulated, effective action $\Gamma_{\Lambda}$, according to 
\be 
S_{\Lambda}[\phi] \;\;\stackrel{k \ra \Lambda}{ \longleftarrow} \;\;
\Gamma_{\Lambda,k}[\phi] \;\; \stackrel{k \ra 0}{ \longrightarrow} \;\;
\Gamma_{\Lambda}[\phi]\,. 
\label{E2}
\end{equation}
Roughly speaking one should think of $\Gamma_{\Lambda,k}[\phi]$ as the 
conventional effective action but with only the momentum modes in 
the range $k^2 < p^2 < \Lambda^2$ integrated out. The $k \ra \Lambda$ limit 
in (\ref{E2}) will in fact differ from $S_{\Lambda}$ by a $1$-loop determinant 
$\ln \det [S^{(2)}_{\Lambda} + \cR_{\Lambda}]$. For the following discussion 
the difference is inessential and for (notational) simplicity we will identify
$\Gamma_{\Lambda,\Lambda}$ with $S_{\Lambda}$. 
Equation~(\ref{E2}) also presupposes that for fixed UV cutoff $\Lambda$ 
the limit $k \ra 0$ exists, which for theories with massless degrees of 
freedom is nontrivial.

The conventional effective action obeys a well-known functional 
integro-differential equation which implicitly defines it. Its counterpart 
for $\Gamma_{\Lambda,k}[\phi]$ reads 
\be
\exp\{-\Gamma_{\Lambda,k}[\phi]\} = 
\int \! [\cD \chi]_{\Lambda,k} 
\exp\Big\{ - S_{\Lambda}[\chi] + 
\int \! dx \, (\chi - \phi)(x) 
\frac{\delta \Gamma_{\Lambda,k}[\phi]}{\delta \phi(x)} \Big\}\,, 
\label{E3}
\end{equation}
where the functional measure $[\cD \chi]_{\Lambda,k}$ 
includes mostly momentum modes in the range $k^2 < p^2 < \Lambda^2$. 
This can be done by multiplying the kinematical measure by a suitable 
mode suppression factor 
\be 
[\cD \chi]_{\Lambda,k} = \cD \chi \exp\{ - C_{\Lambda,k}[\chi -\phi]\}\,,
\label{E3b}
\end{equation}
with a suitable quadratic form $C_{\Lambda,k}$. 
From (\ref{E3}) one can directly verify the alternative 
characterization in Eq.~(\ref{iGamma}).

The precise form of the mode suppression is inessential. In the following 
we outline a variant which is technically convenient. Here 
$C_{\Lambda,k}$ is a quadratic form in the fields defined in 
terms of a kernel $\cR_{\Lambda,k}$ chosen such that both 
$\cR_{\Lambda,k}$ and $k \dd_k \cR_{\Lambda,k}$ define 
integral operators of trace-class on the function space 
considered. We write $[\cR_{\Lambda,k} \chi](x) := \int \! dy \,
\cR_{\Lambda,k}(x,y) \chi(y)$ for the integral operator and 
${\mathrm{Tr}}[\cR_{\Lambda,k}]:= \int \! dx \, \cR_{\Lambda,k}(x,x) < \infty$
for its trace. The other properties of the kernel are best described 
in Fourier space, where $\cR_{k,\Lambda}$ acts as 
$[\cR_{\Lambda,k}\widehat{\chi}](p) = 
\int \! \frac{dq}{(2\pi)^d} \,\cR_{\Lambda,k}(p,q) \,
\widehat{\chi}(q)$, with $\widehat{\chi}(p) = \int \! dx\, \chi(x) 
\exp( - i p x)$, the Fourier transform of $\chi$ and similarly 
for the kernel (where we omit the hat for notational simplicity). 
The UV cutoff $0 \leq p^2 < \Lambda$ renders Euclidean momentum 
space compact and Mercer's theorem then provides simple sufficient 
conditions for an integral operator to be trace-class. 
We thus take the kernel $\cR_{\Lambda,k}(p,q)$ to be smooth, 
symmetric in $p,q$, and such that 
\be 
C_{\Lambda,k}[\chi] := \frac{1}{2} \int \! 
\frac{dp}{(2\pi)^d}\frac{dq}{(2\pi)^d} \,\widehat{\chi}(q)^*\, 
\cR_{\Lambda,k}(p,q) \,\widehat{\chi}(p) = 
\frac{1}{2} \int \!dx \, \chi(x)^*[\cR_{\Lambda,k} \chi](x) \geq 0\,,
\label{E4}
\end{equation}
for all continuous functions $\chi$. Similarly for 
$k \dd_k \cR_{\Lambda,k}(p,q)$. 
The trace-class condition is then satisfied and one can adjust the 
other features of the kernel to account for the mode suppression. 
These features are arbitrary to some extent; what matters is the 
limiting behavior for $p^2, q^2 \gg k^2$ and (with foresight) 
$\Lambda \ra \infty$.

The presence of the extra scale $k$ allows one to convert (\ref{E3}) 
into a functional differential equation~\cite{Wett1, Wett2, Berges}, 
\ba 
\label{CutFRGE}
\dk \Gamma_{\Lambda,k}[\phi] \is \frac{1}{2} {\mathrm{Tr}}\Big[ 
\dk \cR_{\Lambda,k} \, 
\Big( \Gamma^{(2)}_{\Lambda,k}[\phi] + \cR_{\Lambda,k}\Big)^{-1} \Big]
\\
\is \frac{1}{2} \int \! \frac{dq_1}{(2\pi)^d} \frac{dq_2}{(2\pi)^d}
k \dd_k \cR_{\Lambda,k}(q_1,q_2) 
\Big( \Gamma^{(2)}_{\Lambda,k}[\phi] + \cR_{\Lambda,k}\Big)^{-1}\!\!(q_2,q_1)\,. 
\nonumber
\end{eqnarray}
known as the {\it functional renormalization group equation} (FRGE) 
for the {\it effective average action}. For convenience we 
included a quick derivation of (\ref{CutFRGE}) in Appendix C of \cite{Gravirept}. 
In the second line of (\ref{CutFRGE}) we spelled out the 
trace using that $k\dd_k \cR_{\Lambda,k}$ 
is trace-class. Further $\Gamma_{\Lambda,k}^{(2)}[\phi]$ is the
integral operator whose kernel is the Hessian of the effective average action, 
i.e.\ $\Gamma_{\Lambda,k}^{(2)}(x,y) := 
\delta^2 \Gamma_{\Lambda,k}[\phi]/\delta \phi(x) \delta \phi(y)$, 
and $\cR_{\Lambda,k}$ is the integral operator in (\ref{E3}).

For finite cutoffs $(\Lambda,k)$ the trace of the right hand side 
of (\ref{CutFRGE}) will exist as the potentially problematic high momentum 
parts are cut off. In slightly more technical terms, since the product of a 
trace-class operator with a bounded operator is again trace-class, the 
trace in (\ref{CutFRGE}) is finite as long as the inverse of 
$\Gamma^{(2)}_{\Lambda,k}[\phi] + \cR_{\Lambda,k}$ 
defines a bounded operator. For finite UV cutoff one sees 
from the momentum space version of it that this will 
normally be the case. The trace-class property of the mode cutoff 
operator (for which (\ref{E4}) is a sufficient condition) also ensures
that the trace in (\ref{CutFRGE}) can be evaluated in any basis, the 
momentum space variant displayed in the second line is just 
one convenient choice.

Importantly the FRGE (\ref{CutFRGE}) is independent of the bare 
action $S_{\Lambda}$, which enters only via the initial condition 
$\Gamma_{\Lambda,\Lambda} = S_{\Lambda}$ (for large $\Lambda$).
In the FRGE approach the calculation of the 
functional integral for $\Gamma_{\Lambda,k}$ is replaced with the task of 
integrating this RG equation from $k =\Lambda$, where the initial condition 
$\Gamma_{\Lambda,\Lambda} = S_{\Lambda}$ is imposed, down to 
$k=0$, where the effective average action equals the ordinary effective 
action $\Gamma_{\Lambda}$.

All this has been for a fixed UV cutoff $\Lambda$. The removal of 
the cutoff is of course the central theme of UV renormalization.
In the FRG formulation one has to distinguish between two 
aspects. First, removal of the explicit $\Lambda$ dependence in the 
trace on the right hand side of (\ref{CutFRGE}), and second removal of the 
UV cutoff in $\Gamma_{\Lambda,k}$ itself, which was needed in order to make 
the original functional integral well-defined. 

The first aspect is unproblematic: the trace is manifestly finite 
as long as the inverse of $\Gamma^{(2)}_{\Lambda,k}[\phi] + \cR_{\Lambda,k}$ 
defines a bounded operator. {\it If} now $\Gamma^{(2)}_{\Lambda,k}[\phi]$ 
is {\it independently} known to have a finite and nontrivial limit 
as $\Lambda \ra \infty$, the explicit $\Lambda$ dependence carried 
by the $\cR_{\Lambda,k}$ term is harmless and the trace always exists.
Roughly this is because the derivative kernel $k \dd_k \cR_{\Lambda,k}$ 
has support mostly on a thin shell around $p^2 \approx k^2$, so that the 
(potentially problematic) large $p$ behavior of the other factor is 
irrelevant; cf.~\cite{Berges}.

The second aspect of course relates to the traditional UV renormalization 
problem. Since $\Gamma_{\Lambda,k}$ came from a regularized functional 
integral it will develop the usual UV divergencies 
as one attempts to send $\Lambda$ to infinity. The remedy is 
to carefully adjust the bare action $S_{\Lambda}[\phi]$ --
that is, the initial condition for the FRGE (\ref{CutFRGE}) --
in such a way that functional integral -- viz, the solution 
of the FRGE -- is asymptotically independent of $\Lambda$. 
Concretely this could be done by fine-tuning the way how the 
parameters $u_{\alpha}(\Lambda)$ in the expansion 
$S_{\Lambda}[\chi] = \sum_{\alpha} u_{\alpha}(\Lambda) 
P_{\alpha}[\chi]$ depends on $\Lambda$. However the 
FRGE method in itself provides {\it no means} to find the proper 
initial functional $S_{\Lambda}[\chi]$. 
Identification of the fine-tuned $S_{\Lambda}[\chi]$ lies at the 
core of the UV renormalization problem, irrespective of whether 
$\Gamma_{\Lambda,k}$ is defined directly via the functional 
integral or via the FRGE. Beyond perturbation theory the only 
known techniques to identify the proper $S_{\Lambda}$ start 
directly from the functional integral and are `constructive' 
in spirit. Unfortunately four dimensional quantum field theories of 
interest are still beyond constructive control.

One may also ask whether perhaps the cutoff-dependent FRGE (\ref{CutFRGE}) 
itself can be used to show that a limit $\lim_{\Lambda \ra \infty} 
\Gamma_{\Lambda,k}[\phi]$ exists. Indeed using other 
FRGEs and a perturbative ansatz for the solution has lead 
to economic proofs of perturbative renormalizability, 
i.e.\ of the existence of a formal continuum limit in the 
sense (PTC1) discussed before. Unfortunately 
so far this could not be extended to construct a 
nonperturbative continuum limit of fully fledged 
quantum field theories, see~\cite{Mitter} for a recent review
of such constructive uses of FRGEs. For the time being 
one has to be content with the following {\it if} ... {\it then}
statement:

{\it If} there exists a sequence of initial 
actions $S_{n \Lambda_0}[\chi]$, $n \in \N$, such that 
the solution $\Gamma_{\n \Lambda_0,k}[\phi]$ of the FRGE 
(\ref{CutFRGE}) remains finite as $n \ra \infty$, {\it then} 
the limit $\Gamma_k[\phi]:= \lim_{n \ra \infty} 
\Gamma_{n \Lambda_0,k}[\phi]$ has to obey the 
cut-off independent FRGE
\be 
\dk \Gamma_{k}[\phi] = \frac{1}{2} {\mathrm{Tr}}\Big[ 
\dk \cR_{k,\infty} \, 
\Big( \Gamma^{(2)}_{k}[\phi] + \cR_{k,\infty}\Big)^{-1} \Big]\,.
\label{FRGE}
\end{equation} 
Conversely, under the above premise, this equation should have 
at least one solution with a finite limit $\lim_{k \ra \infty} 
\Gamma_k[\phi]$. This limit can now be identified with the 
renormalized fixed point action $S_*[\chi]$. It is 
renormalized because by construction the cutoff dependencies have 
been eaten up by the ones produced by the trace in (\ref{CutFRGE}). 
It can be identified with a fixed point action because lowering 
$k$ amounts to coarse graining and $S_*[\chi]$ is the `inverse
limit' of a sequence of such coarse graining steps.

So far the positivity or unitarity requirement has not been discussed.
From the (Oster\-wal\-der--Schrader or Wightman) 
reconstruction theorems it is known how the unitarity of a 
quantum field theory on a flat spacetime translates into 
nonlinear conditions on the multipoint functions. Since the latter
can be expressed in terms of the functional derivatives of 
$\Gamma_k$, unitarity can in principle be tested retroactively,
and is expected to hold only in the limit $k \ra 0$. 
Unfortunately this is a very indirect and retroactive criterion.
One of the roles of the bare action $S_{\Lambda}[\chi]= 
\Gamma_{\Lambda,\Lambda}[\chi]$ is to encode properties which 
are likely to ensure the desired properties of $\lim_{k \ra 0} 
\Gamma_k[\phi]$. In theories with massless degrees of freedom 
the $k \ra 0$ limit is nontrivial and 
the problem of gaining computational control over the infrared physics 
should be separated from the UV aspects of the continuum limit as 
much as possible. However the $k \ra 0$ limit is essential to probe 
stability/positivity/unitarity.

One aspect of positivity is the convexity of the effective action.
The functional equations (\ref{CutFRGE}) and (\ref{FRGE}) do in 
itself ``not know'' that $\Gamma_k$ is the Legendre transform 
of a convex functional and hence must be itself convex. 
Convexity must therefore enter through the inital data
and it will also put constraints on the choice of the 
mode cutoffs. Good mode cutoffs are characterized by the fact that 
$\Gamma_k^{(2)} + \cR_k$ has positive spectral values for all $k$. 
If no blow-up occurs in the 
flow the limit $\lim_{k \ra 0}\Gamma_k^{(2)}$ will then also have 
non-negative spectrum. Of course this presupposes 
again that the proper initial conditions have been identified and 
the role of the bare action is as above.

For flat space quantum field theories 
one expects that $S_{\Lambda}[\chi]$ must be local, i.e.\ a 
differential polynomial of finite order in the fields so as to end up with 
an effective action $\lim_{k \ra \infty} \lim_{\Lambda \ra \infty} 
\Gamma_{\Lambda,k}[\phi]$ describing a local/microcausal unitary 
quantum field theory.

For convenient reference we summarize these conclusions in the 

{\bf Criterion:} (Continuum limit in the functional RG approach)
(FRGC1) A solution of the cutoff independent FRGE (\ref{FRGE}) which 
exists globally in $k$ (for all $0 \leq k \leq \infty$) 
can reasonably be identified with the continuum limit 
of the effective average action $\lim_{\Lambda \ra \infty} 
\Gamma_{\Lambda,k}[\phi]$ constructed by other means. 
For such a solution $\lim_{k \ra 0} \Gamma_k[\phi]$ 
is the full quantum effective action and $\lim_{k \ra \infty} 
\Gamma_k[\phi] = S_*[\phi]$ is the fixed point action.
(FRGC2) For a unitary relativistic quantum field theory
positivity/unitarity must be tested retroactively from 
the functional derivatives of $\lim_{k \ra 0} \Gamma_k[\phi]$.

We add some comments: 

Since the FRGE (\ref{FRGE}) is a differential equation in $k$,
an initial functional $\Gamma_{\mathrm{initial}}[\phi]$ has to be specified
for some $0<k_{\mathrm{initial}} \leq \infty$, to generate a local solution 
near $k=k_{\mathrm{initial}}$. The point is that for `almost all' choices of 
$\Gamma_{\mathrm{initial}}[\phi]$ the local solution {\it cannot} be 
extended to all values of $k$. Finding the rare initial 
functionals for which this is possible is the FRGE 
counterpart of the UV renormalization problem. 
The existence of the $k \ra 0$ limit is itself not part 
of the UV problem, in conventional quantum field theories
the $k \ra 0$ limit is however essential to probe 
unitarity/positivity/stability.

It is presently not known whether the above criterion can be 
converted into a theorem. Suppose for a quantum field theory on 
the lattice (with lattice spacing $\Lambda^{-1}$) the effective action 
$\Gamma_{\Lambda,k}^{\mathrm{latt}}[\phi]$ has been constructed 
nonperturbatively from a transfer operator satisfying
reflection positivity and that a continuum limit $\lim_{\Lambda \ra \infty}
\Gamma_{\Lambda,k}^{\mathrm{latt}}$ is assumed to exist. Does it 
coincide with a solution $\Gamma_k[\phi]$ of (\ref{FRGE}) 
satisfying the criteria (FRGC1), (FRGC2)? Note that this is 
`only' a matter of controlling the limit, for finite $\Lambda$ also 
$\Gamma_{\Lambda,k}^{\mathrm{latt}}$ will satisfy the 
flow equation (\ref{CutFRGE}). 

For an application to quantum gravity one will initially 
only ask for (FRGC1), perhaps with even only a partial 
understanding of the $k\ra 0$ limit. As mentioned, the 
$k \ra 0$ limit should also be related to positivity issues.  
The proper positivity requirement 
replacing (FRGC2) yet has to be found, however some 
constraint will certainly be needed. Concerning (FRGC1) 
the premise in the {\it if} ... {\it then} statement 
preceeding (\ref{FRGE}) has to be justified by external means
or taken as a working hypothesis. In principle one 
can also adopt the viewpoint that the quantum gravity 
counterpart of (\ref{FRGE}) introduced in \cite{Reuter} simply 
defines the effective action for quantum 
gravity whenever a solution meets (FRGC1). The main drawbacks 
of this proposal are: first, beyond perturbation theory 
little is known about the structure of $\Gamma_k$, e.g.~which 
type of nonlocalities are expected to occur. Second, the 
proposal would make it difficult to include information concerning 
(FRGC2). However difficult 
and roundabout a functional integral construction is, 
it allows one to incorporate `other' desirable features
of the system in a relatively transparent way. 

Also in the application to quantum gravity we shall therefore   
presuppose that a solution $\Gamma_k$ of the cutoff independent FRGE 
(\ref{FRGE}) satisfying (FRGC1) comes from an underlying functional 
integral. This amounts to the assumption that the renormalization 
problem for $\Gamma_{k,\Lambda}$ defined in terms of a functional 
integral can be solved and that the limit $\lim_{\Lambda \ra \infty} 
\Gamma_{k,\Lambda}$ can be identified with $\Gamma_k$. 
This is of course a rather strong hypothesis, however 
its selfconsistency can be tested within the FRG framework. 

To this end one truncates
the space of candidate continuum functionals $\Gamma^{\mathrm{trunc}}_k[\phi]$
to one where the initial value problem for the flow equation 
(\ref{FRGE}) can be solved in reasonably closed form. One can 
then by `direct inspection' determine the initial data for which 
a global solution exists. Convexity of the truncated $\lim_{k \ra 0} 
\Gamma^{\mathrm{trunc}}_k[\phi]$ can serve as guideline to identify 
good truncations. If the set of these initial data forms
a nontrivial unstable manifold of the fixed point $S_*^{\mathrm{trunc}}[\phi]=
\lim_{k\ra \infty} \Gamma^{\mathrm{trunc}}_k[\phi]$ application of the 
above criterion suggests that $\Gamma^{\mathrm{trunc}}_k$ can 
approximately be identified with the projection of the 
continuum limit $(\lim_{\Lambda \ra \infty}
\Gamma_{\Lambda,k})^{\mathrm{trunc}}$ of some $\Gamma_{\Lambda,k}$
computed by other means. The identification can only be an approximate 
one because in the $\Gamma_{k}^{\mathrm{trunc}}$ evolution one first 
truncates and then evolves in $k$, while in 
$(\lim_{\Lambda \ra \infty} \Gamma_{\Lambda,k}[\phi])^{\mathrm{trunc}}$ 
one first evolves in $k$ and then truncates. Alternatively one can 
imagine to have replaced the original dynamics by some `hierarchical' 
approximation implicitly defined by the property 
that $(\lim_{\Lambda \ra \infty} 
\Gamma^{\mathrm{hier}}_{\Lambda,k}[\phi])^{\mathrm{trunc}}
=\lim_{\Lambda \ra \infty} \Gamma^{\mathrm{hier}}_{\Lambda,k}[\phi]$. 
(See \cite{Felderhier}) for the relation between hierachical dynamics 
and the  local potential approximation).
The existence of an UV fixed point with a nontrivial unstable manifold 
for $\Gamma_k^{\mathrm{trunc}}$ can then be taken as witnessing the 
renormalizability of the `hierarchical' dynamics.

\newpage

\end{document}